\documentclass[12pt]{svmult}
\makeatletter
\def\section{\@startsection {section}{1}{\z@}{-3.5ex plus -1ex minus
 -.2ex}{2.3ex plus .2ex}{\large\bf}}
\def\subsection{\@startsection{subsection}{2}{\z@}{-3.25ex plus -1ex
minus -.2ex}{1.5ex plus .2ex}{\normalsize\bf}}

\newcommand{\bra}[1]{\langle{#1}|}
\newcommand{\ket}[1]{|{#1}\rangle}

\newcommand{\eq}[1]{Eq.~(\ref{#1})}

\newcommand{\nl}{\nonumber \\}

\def\beq{\begin{equation}}
\def\eeq{\end{equation}}
\def\beqa{\begin{eqnarray}}
\def\eeqa{\end{eqnarray}}
\newcommand{\sect}[1]{\setcounter{equation}{0}\section{#1}}

\newcommand{\EQ}{\begin{equation}}
\newcommand{\EN}{\end{equation}}
\newcommand{\bea}{\begin{eqnarray}}
\newcommand{\ena}{\end{eqnarray}}

\renewcommand{\a}{\alpha}
\renewcommand{\b}{\beta}

\newcommand{\dpb}{D$p$ brane}

\def\one{{\hbox{ 1\kern-.8mm l}}}

\def\sgh{{\rm sgh}}
\def\NS{{\rm NS}}
\def\R{{\rm R}}
\def\ii{{\rm i}}

\newlength{\bredde}
\def\slash#1{\settowidth{\bredde}{$#1$}\ifmmode\,\raisebox{.15ex}{/}
\hspace*{-\bredde} #1\else$\,\raisebox{.15ex}{/}\hspace*{-\bredde} #1$\fi}

\begin{document}
\begin{titlepage}
\rightline{DSF-23/2003} \rightline{NORDITA-2003-43 HE}
 \vskip
1.8cm \centerline{ \Large \bf Gauge Theories from D Branes} \vskip
1.4cm \centerline{Paolo Di Vecchia $^a$ and Antonella
Liccardo$^b$} \vskip .8cm \centerline{\sl $^a$ NORDITA,
Blegdamsvej 17, DK-2100 Copenhagen \O, Denmark}
\centerline{e-mail:{\tt divecchi@alf.nbi.dk}} \vskip 0.4cm
\centerline{\sl $^d$ Dipartimento di Scienze Fisiche, Universit\`a
di Napoli} \centerline{\sl Complesso Universitario Monte S.
Angelo, Via Cintia, I-80126 Napoli, Italy} \centerline{e-mail:{\tt
liccardo@na.infn.it}} \vskip 1.8cm
\begin{abstract}
In these lectures we start with a pedagogical introduction of the
properties of open and closed superstrings and then, using the
open/closed string duality, we construct the boundary state that
provides the description of the maximally supersymmetric Dp branes
in terms of the perturbative string formalism. We then use it for
deriving the corresponding supergravity solution and the
Born-Infeld action and for studying the properties of the
maximally supersymmetric gauge theories living on their
worldvolume. In the last section of these lectures we extend these
results to less supersymmetric and non-conformal gauge theories by
considering fractional branes of orbifolds and wrapped branes.

Lectures given at the School "Frontiers in Number Theory, Physics
and Geometry", Les Houches, March 2003.
\end{abstract}
\end{titlepage}

\section{Introduction}
\label{sec:intro}

The discovery that superstring theories contain not only strings
but also other non-perturbative p-dimensional objects, called p
branes, has been a source of major progress not only in order to
arrive at the formulation of M theory, but also for studying
perturbative and non-perturbative properties of the gauge theories
living on the brane world-volume. In fact  the so-called Dirichlet
branes (Dp branes) of type II theories admit two distinct
descriptions. On the one hand they are classical solutions of the
low-energy string effective action and therefore may be described
in terms of closed strings. On the other hand their dynamics is
determined by the degrees of freedom of the open strings with
endpoints attached to their world-volume, satisfying Dirichlet
boundary conditions along the directions transverse to the branes.
Thus they may be described in terms of open strings, as well. This
twofold description of the Dp branes was at the basis of the
Maldacena conjecture~\cite{MALDA} providing  the equivalence
between a closed string theory, as the type IIB on $AdS_5 \times
S^5$,  and ${\cal{N}}=4$ super Yang-Mills whose degrees of freedom
correspond to the massless excitations of the open strings having
their endpoints attached to a D3 brane. It is also at the basis of
recent studies of the perturbative and non-perturbative properties
of the gauge theories living on less supersymmetric and
non-conformal branes by means of classical solutions of the
supergravity equations of motion that we will present in the last
section of this lectures.

The fact that a Dp brane admits two distinct but equivalent
descriptions in the open and closed string channel goes in the
literature under the name of gauge-gravity correspondence. This
correspondence is a direct consequence of the open/closed string
duality that states that the one-loop annulus diagram of open
strings can be equivalently written as a tree diagram of closed
strings. This equivalence is known since the early days of string
theory and was subsequently developed later~\footnote{For a set of
relevant references on this subject see for instance
Ref.~\cite{ISLANDA}.}. It is in fact the open/closed string
duality that allows to construct  the boundary state that is the
basic tool for describing the Dp branes in the framework of string
theory. In these lectures we show how both the open and closed
string properties of Dp branes can be conveniently studied using
the formalism of the boundary state and how from it one can reach
a non trivial understanding of the properties of the gauge theory
living on their world-volume. This will be done first for
maximally supersymmetric theories, as for instance ${\cal{N}}=4$
super Yang-Mills that lives on a D3 brane, and will be extended in
the last section to less supersymmetric and non-conformal gauge
theories~\footnote{Recent reviews on this subject can be found in
  Ref.s~\cite{MATTEO,MIL}.}.

The lectures are organized as follows. After a self-consistent
review of the main properties of perturbative open and closed
superstring theories done in sect.~\ref{sec1}, we present the
description of Dp branes in terms of closed strings in
sect.~\ref{sec2} and  the one in terms of open strings in
sect.~\ref{sec3}. Sections~\ref{sec4} and \ref{sec5} are devoted
to the construction of the boundary state and to the calculation
of brane interaction by means of the one-loop annulus diagram. In
sect.~\ref{sec6} we derive from the boundary state the large
distance behavior of the corresponding classical supergravity
solution and the Born-Infeld action and we show that the gauge
theory living on the maximally supersymmetric Dp branes is the
dimensional reduction of ${\cal{N}}=1$ super Yang-Mills in ten
dimensions to $(p+1)$ dimensions. Finally in sect.~\ref{sec7} we
extend this procedure to less supersymmetric and non-conformal
gauge theories by considering fractional and wrapped branes.

\sect{Perturbative String Theory}
\label{sec1}
The action that describes the space-time propagation of  a
supersymmetric string in the superconformal gauge is given by
\begin{equation}
\label{sstac}
S=-\frac{T}{2}\int_M d\tau d\sigma
\left(\eta^{\alpha\beta}
\partial_\alpha X^\mu\partial_\beta X_\mu-i{\bar
\psi}^\mu\rho^\alpha\partial_\alpha\psi_{\mu} \right),
\end{equation}
where $T$ is the string tension related to the Regge slope by
$T=(2\pi\alpha')^{-1}$, $M$ is the world-sheet of the string
described by the  coordinate $\xi^{\alpha} \equiv (\tau, \sigma)$,
$\psi$ is a world-sheet Majorana spinor and the matrices
$\rho^{\alpha}$ provide a representation of the Clifford algebra
$\{ \rho^{\alpha}, \rho^{\beta} \} = - 2\eta^{\alpha \beta}$ in
two dimensions.

The previous action is invariant under the  following
supersymmetry transformations
\begin{equation}
\label{susyt}
\delta X^\mu=\bar\alpha \psi^\mu~~,~~ \delta
\psi^\mu=-i\rho^\alpha\partial_\alpha X^\mu\alpha~~,~~
\delta{\bar{\psi}^\mu} = i {\bar{\alpha}} \rho^{\alpha}
\partial_{\alpha} X^{\mu}~~, \end{equation} where $\alpha$ is a Majorana
spinor that satisfies the equation:
\begin{equation}
\rho^{\beta} \rho^{\alpha} \partial_{\beta} \alpha =0
\label{consu85}
\end{equation}
The equation of motion  and boundary conditions for the string
coordinate $X^{\mu}$  following from the action in Eq.(\ref{sstac}) are
\begin{equation}
\label{steq}
(\partial_\sigma^2-\partial_\tau^2)X^\mu=0~~~~,~~~~
\mu = 0,..., d-1,
\end{equation}
\begin{equation}
\label{stbc}
 \left( \partial_\sigma X \cdot \delta X
|_{\sigma=\pi}-
\partial_\sigma X \cdot \delta X |_{\sigma=0}\right)=0,
\end{equation}
where $\sigma\in [0,{\pi}]$. Eq.(\ref{stbc}) can
be satisfied either by imposing the periodicity condition
\begin{equation}
\label{bccl}
X^\mu(\tau,0)=X^\mu(\tau,\pi),
\end{equation}
which leads to a
theory of closed strings, or by requiring
\begin{equation}
\label{bcop}
\partial_\sigma X_\mu\delta X^\mu|_{0,\pi}=0,
\end{equation}
separately at both $\sigma=0$ and $\sigma=\pi$ obtaining a
theory of open strings. In the latter case Eq.(\ref{bcop}) can be
satisfied in either of the two ways
\begin{equation}
\label{neudic}
\left\{\begin{array}{l}
\partial_\sigma X_\mu|_{0,\pi}=0
\rightarrow {\rm Neumann~boundary~conditions}\\
\delta X^\mu|_{0,\pi}=0\rightarrow {\rm Dirichlet~boundary~conditions}.
\end{array}
\right.
\end{equation}
If the open string satisfies Neumann boundary conditions at both
its endpoints (N-N boundary conditions) the general solution of
the Eq.s (\ref{steq}) and (\ref{stbc})  is equal to
\begin{equation}
\label {expo1}
X^\mu(\tau ,\sigma)= q^\mu+2\alpha'p^\mu \tau +i\sqrt{2\alpha' }\sum_{n\neq
0}
\left(\frac {\alpha^{\mu}_{n}}{n}e^{-in \tau}\cos{n\sigma}\right),
\end{equation}
where $n$ is an integer.
For D-D boundary conditions we have
instead
\begin{equation}
\label {expo2}
X^\mu(\tau
,\sigma)= \frac{c^\mu(\pi-\sigma)+d^\mu \sigma}{\pi}
-\sqrt{2\alpha' }\sum_{n\neq 0} \left(\frac
{\alpha^{\mu}_{n}}{n}e^{-in \tau}\sin{n\sigma}\right).
\end{equation}
One can also have mixed boundary conditions. The expression of the
string coordinates in this case can be found in
Ref.~\cite{ISLANDA}.

For a closed string the most general solution of the Eq.s of
motion and of the periodicity condition in Eq.(\ref{bccl}) can be
written as follows
\begin{equation}
\label {expc}
X^\mu(\tau, \sigma)=
q^\mu+2\alpha'p^\mu \tau +i\sqrt{\frac{\alpha'}{2} } \sum_{n\neq
0} \left(\frac {\alpha^{\mu}_{n}}{n}e^{-2in(\tau -\sigma)}+ \frac
{\widetilde\alpha^{\mu}_{n}}{n}e^{-2in(\tau+\sigma)}\right),
\end{equation}
In order to discuss the fermionic degrees of freedom, it is useful
to introduce light cone coordinates
\begin{equation}
\label{lccor}
\xi_+=\tau+\sigma
~~~~~~~;~~~~~~~\xi_-=\tau-\sigma~~~,~~~\partial_{\pm}\equiv
\frac{\partial}{\partial \xi_{\pm}} ,
\end{equation}
In terms of them the Eq. of motion for $\psi$ becomes
\begin{equation}
\label{ssteq}
\partial_+\psi^\mu_-=0~~~~~;~~~~~\partial_-\psi^\mu_+=0,
\end{equation}
where
\begin{equation}
\label{psipm}
\psi^\mu_\pm=\frac{1\mp\rho^3}{2}\psi^\mu~~~~;~~~~
\rho^3\equiv\rho^0\rho^1.
\end{equation}
The boundary conditions following
from Eq. (\ref{sstac}) are
\begin{equation}
\label{sstbc}
\left(
\psi_+\delta\psi_+-\psi_-\delta\psi_-\right)|_{\sigma=0}^{\sigma=\pi}=0.
\end{equation} which, in the case of an open string are satisfied if
\begin{equation}
\label{fbcu1}
\left\{
\begin{array}{l}
\psi_-(0,\tau)=\eta_1{\psi_+}(0,\tau)\\
\psi_-(\pi,\tau)=\eta_2{\psi_+}(\pi,\tau)
\end{array}
\right. ,
\end{equation}
where $\eta_1$ and $\eta_2$ can take the values $\pm 1$.
In particular if $\eta_1 =\eta_2$
we get what is called the Ramond (R) sector of the open string,
while if $\eta_1 =- \eta_2$ we get  the
Neveu-Schwarz (NS) sector.
\noindent
In the case of a closed string the fermionic coordinates $\psi_{\pm}$ are
independent from each other and they can be either periodic or
anti-periodic.
This amounts to impose the following  conditions:
\begin{equation}
\label {ucbc}
\psi^\mu_-(0,\tau)=\eta_3\psi^\mu_-(\pi,\tau)~~~~~
{\psi^\mu_+}(0,\tau)=\eta_4{\psi^\mu_+}(\pi,\tau),
\end{equation}
that satisfy the boundary conditions in Eq.(\ref{sstbc}).
In this case we  have four different sectors
according to the two values that $\eta_3$ and $\eta_4$ take
\begin{equation}
\left\{
\begin{array}{l}
\eta_3=\eta_4=1\Rightarrow {\rm (R-R)}\\
\eta_3=\eta_4=-1\Rightarrow {\rm (NS-NS)}\\
\eta_3=-\eta_4=1\Rightarrow {\rm (R-NS)}\\
\eta_3=-\eta_4=-1\Rightarrow {\rm(NS-R)}\\
\end{array}\right. .
\label {rns2}
\end{equation}
The general solution of Eq.(\ref{ssteq}) satisfying the boundary
conditions in Eq.s (\ref{fbcu1}) is given by
\begin{equation}
\label{modpsil}
\psi^\mu_\mp \sim \sum_{t} \psi^\mu_t
e^{- it(\tau\mp\sigma)}~~~~~~~{\rm where}~~~~~
\left\{ \begin{array}{l}
t\in Z+\frac{1}{2}\rightarrow {\rm NS \,\,\,sector}\\
t\in Z\rightarrow {\rm R \,\,\, sector}
\end{array}
\right. ,
\end{equation}
while the ones satisfying the boundary conditions in Eq.(\ref{ucbc}) are
given
by
\begin{equation}
\label{modpsil2}
\psi^\mu_- \sim \sum_{t} \psi^\mu_t
e^{-2it(\tau-\sigma)}~~~~~~~
{\rm where}~~~~~
\left\{
\begin{array}{l}
t\in Z+\frac{1}{2}\rightarrow {\rm NS \,\,\, sector}\\
t\in Z\rightarrow {\rm R \,\,\,sector}
\end{array}
\right. ,
\end{equation}
\begin{equation}
\label{modpsir}
\psi^\mu_+ \sim \sum_{t} \widetilde\psi^\mu_t
e^{-2it(\tau +\sigma)}~~~~~~~
{\rm where}~~~~~
\left\{
\begin{array}{l}
t\in Z+\frac{1}{2}\rightarrow {\widetilde{\rm {NS}} \,\,\,{\rm sector}}\\
t\in Z\rightarrow {\widetilde{\rm {R}} \,\,\,{\rm sector}}
\end{array}
\right. .
\end{equation}
In the quantum string theory the oscillators $\alpha_n$ and
${\widetilde\alpha}_n$ play the role of creation and annihilation
operators acting on a Fock space  and satisfying the commutation
relations
\begin{equation}
\label{oscqua1}
[\alpha^\mu_m,\alpha^\nu_n]=[\widetilde\alpha^\mu_m,\widetilde\alpha^\nu_n]=
m\delta_{m+n,0}\eta^{\mu\nu}~~~~~;~~~~~ [{\hat q}^\mu,{\hat
p}^\nu]=i\eta^{\mu\nu}~,
\end{equation}
\begin{equation}
\label{oscqua2}
[\alpha^\mu_m,\widetilde\alpha^\nu_n]=[\hat q^\mu,\hat q^\nu]=
[\hat p^\mu,\hat p^\nu]=0~.
\end{equation}
which can be obtained by imposing the standard equal time
commutators between the bosonic string coordinates. Analogously
the fermionic oscillators satisfy the anticommutation relations
\begin{equation}
\label{ferquosc}
\{\psi^\mu_t,\psi^\nu_v\}=\{\tilde\psi^\mu_t,\tilde\psi^\nu_v\}=
\eta^{\mu\nu}\delta_{v+t,0}\,\,\,\,\,\,\,\,\,\,\,\,
\{\psi^\mu_t,\tilde\psi^\nu_v\}= 0
\end{equation}
following from the canonical anticommutation relations between the
fermionic coordinates.
 The closed string vacuum state
$|0\rangle|\tilde 0\rangle|p\rangle$ with momentum $p$ is defined
by the conditions
\[
\alpha^{\mu}_{n}|0\rangle |\tilde 0\rangle|p \rangle =
\tilde\alpha^{\mu}_{n} |0\rangle |\tilde 0\rangle |p \rangle
=0~~\forall n >0 ~~,~~
\]
\[
\psi^{\mu}_{t}|0\rangle|\tilde 0\rangle|p \rangle =
\tilde\psi^{\mu}_{t} |0\rangle|\tilde 0\rangle |p \rangle =0~~
{\rm with}~~~~~ \left\{
\begin{array}{l}
t\geq\frac{1}{2}\rightarrow {\rm NS \,\,\, sector}\\
t\geq 1\rightarrow {\rm R \,\,\,sector}
\end{array}
\right. ,
\]
\begin{equation}
\label{vacdef}
\hat p^\mu |0\rangle |\tilde 0\rangle|p\rangle
=p^\mu |0\rangle |\tilde 0\rangle|p\rangle~.
\end{equation}
For open strings the vacuum state is $|0\rangle |p\rangle$, and its definition
involves only one kind of oscillators.

Notice that in the R sector there are fermionic zero modes
satisfying the Dirac algebra that follows from Eq.(\ref{ferquosc})
for $t,v =0$:
\begin{equation}
\{ \psi_{0}^{\mu} , \psi_{0}^{\nu} \} =
\eta^{\mu \nu}
\label{diraca}
\end{equation}
and therefore they can be represented as Dirac $\Gamma$-matrices.
This implies that the ground state of the Ramond sector transforms
as a Dirac spinor and  we can  label it with a spinor index
$A$. It is annihilated by all annihilation operators $\psi_{t}^{\mu}$
with $ t > 0$. The action of  $\psi_0 $ on the open string vacuum is given
by
\begin{equation}
\psi_0^\mu\, |A\rangle =
\frac{1}{\sqrt{2}} \left(\Gamma^\mu\right)^A_{~C} |C\rangle~~~,
\label{psi0op}
\end{equation}
while that on the closed string ground state is given
by
\begin{eqnarray} \psi_0^\mu\, |A\rangle |{\widetilde B}\rangle &=&
\frac{1}{\sqrt{2}} \left(\Gamma^\mu\right)^A_{~C} \,\left(\!\one\,
\right)^B_{~D}|C\rangle\, |{\widetilde D}\rangle \nl {\widetilde
\psi}_0^\mu \,|A\rangle |{\widetilde B}\rangle &=&
\frac{1}{\sqrt{2}} \left(\Gamma_{11}\right)^A_{~C}
\,\left(\Gamma^\mu\right)^B_{~D}\, |C\rangle |{\widetilde
D}\rangle
\label{psi0}
\end{eqnarray}
Because of the Lorentz metric the Fock space defined by the
relations in Eq.s (\ref {oscqua1}) - (\ref{ferquosc}) contains
states with negative norm. To select only physical states one has
to introduce  the energy momentum tensor and the supercurrent and
to look at the action that these two operators  have on the
states. The energy-momentum tensor, in the light cone coordinates,
has two non zero components \begin{equation} \label{sliccoem}
T_{++}=\partial_+ X \cdot
\partial_+ X +\frac{i}{2}\psi_+ \cdot \partial_+\psi_{+} ~~~~;~~~~
T_{--}=\partial_- X \cdot \partial_- X +\frac{i}{2}\psi_- \cdot
\partial_-\psi_{-}~~, \end{equation}
while the supercurrent, which is the N{\"{o}}ther current
associated to the supersymmetry transformation in
Eq.(\ref{susyt}), is \begin{equation} \label{scurpm} J_-=\psi_-
\cdot
\partial_-X ~~~~;~~~~J_+=\psi_+ \cdot \partial_+X ~~. \end{equation} The
Fourier components of $T_{--}$ and $T_{++}$ are called Virasoro
generators and are given by
\begin{equation}
\label{svgeno}
 L_{n}=\frac{1}{2}\sum_{m\in Z}\alpha_{-m}\cdot
\alpha_{n+m} +
~\frac{1}{2}\sum_t\left(\frac{n}{2}+t\right)\psi_{-t} \cdot
\psi_{t+n}, ~~\forall n>0
\end{equation}
with an analogous expression for
$\tilde L_m$ in the closed string case, while
\begin{equation} L_0 = \alpha '
\hat p^2 +\sum_{n=1}^{\infty} \alpha_{-n} \cdot \alpha_n +\sum_{t
>0} t \psi_{-t} \cdot \psi_t. \label{sL0on}
\end{equation}
in the open string case and
\begin{equation} L_0 = \frac{\alpha '}{4} \hat p^2
+\sum_{n=1}^{\infty} \alpha_{-n} \cdot \alpha_n +\sum_{t>0} t
\psi_{-t} \cdot \psi_t \label{L0v} \,\,\,\,,\,\,\,\, {\tilde{L}}_0
= \frac{\alpha '}{4} \hat p^2 +\sum_{n=1}^{\infty}
{\tilde{\alpha}}_{-n} \cdot {\tilde{\alpha}}_n +\sum_{t >0} t
{\tilde{\psi}}_{-t} \cdot {\tilde{\psi}}_t
\end{equation} for a closed
string. The conditions which select the physical states involve
also the Fourier components of the supercurrent, denoted with
$G_t$ (and $\widetilde G_t$ for closed strings). The operator
$G_t$ is given by
\begin{equation}
\label{sucurfu}
G_t=\sum_{n=-\infty}^\infty\alpha_{-n} \cdot \psi_{t+n}~~~~;~~~~
\end{equation}
with an analogous expression for
${\widetilde G}_t$ in the closed string case.

The physical states are those satisfying  the conditions \begin{equation}
\label {sphcond1} \left\{
\begin{array}{l}
L_m|\psi_{\rm phys}\rangle =0 ~~~~~m > 0\\
(L_0-a_0 )|\psi_{\rm phys}\rangle =0\\
G_t|\psi_{\rm phys}\rangle =0 ~~~~~\forall t \geq 0\\
\end{array}
\right. , \end{equation}  where $a_0 =\frac{1}{2}$ for  the NS sector and
$a_0 =0$ for  the R sector. In the closed string case one must also
impose analogous conditions involving
$\tilde L_m$ and $\tilde G_t$.

The spectrum of the theory is given in the open string case by
\begin{equation}
\label{smassao} M^2=\frac{1}{\alpha'}\left(\sum_{n=1}^{\infty}
\alpha_{-n} \cdot \alpha_n +\sum_{t>0} t\psi_{-t} \cdot \psi_t-a_0
\right)
\end{equation}
while in the closed case by
 \begin{equation}
\label{smassacl} M^2= \frac{1}{2} \left( M^2_+ +M^2_- \right)~~,
\end{equation} where
\begin{equation}
\label{smassacl1}
M^2_-=\frac{4}{\alpha'}\left(\sum_{n=1}^{\infty} \alpha_{-n} \cdot
\alpha_n +\sum_t t\psi_{-t} \cdot \psi_t-a_0 \right),\,\,\,\,\,
 M^2_+=\frac{4}{\alpha'}\left(
\sum_{n=1}^{\infty}  \widetilde\alpha_{-n} \cdot
\widetilde\alpha_n +\sum_t t\widetilde\psi_{-t} \cdot
\widetilde\psi_t-{\tilde{a}}_0 \right).
\end{equation}
For closed strings  we should also add the level matching
condition
\begin{equation} (L_0 - {\widetilde{L}}_0 -a_0  +{\tilde{a}}_0 ) |
\psi_{\rm phys} \rangle =0
\label{levmat4}.
\end{equation}
Let us now concentrate on the massless spectrum of superstring
theories. Imposing the physical states conditions given in Eq.
(\ref{sphcond1}) for  open strings, we get in the NS sector the
following massless state:
\begin{equation}
\epsilon_{\mu}(k)
\psi_{-1/2}^{\mu} | 0, k
\rangle  \hspace{1cm} k \cdot \epsilon =0 \hspace{1cm};\hspace{1cm} k^2
=0
\label{mlsNS}
\end{equation}
that corresponds to a gauge vector field,
with transverse polarization, while in the R sector the massless
state is given by:
\begin{equation} u_A (k) | A , k \rangle  \hspace{1cm};
\hspace{1cm} u_A ( k \cdot \Gamma )^{A}_{\,\,\,B} =0  \hspace{1cm}
k^2 =0
\label{mlsR}
\end{equation}
that in general corresponds to a spinor field in $d$ dimensions. A
necessary condition for having space-time supersymmetry is that
the number of physical bosonic degrees of freedom be equal to that
of physical fermionic degrees of freedom. In the NS sector we have
found a vector field, which in $10$ dimensions has  $8$ degrees of
freedom (i.e. $(d-2)$). In the R sector instead we got a spinor
field. The number of degrees of freedom of a spinor  in $d$
dimension is $2^{d/2}$,  if it is a Dirac spinor, $2^{d/2}/2$ if
it is a Majorana or Weyl spinor and $2^{d/2}/4$ in the case of
Majorana-Weyl spinor (for $d$ even). In $d=10$ the only spinor
having the same number of degrees of freedom of a vector is the
Majorana-Weyl (with 8 d.o.f.) . Thus in order to have
supersymmetry we must impose the ground state of the Ramond sector
to be a Majorana-Weyl spinor. Since the fermionic coordinates
$\psi^\mu$ are real we expect the spinor to be Majorana. In order
to get also a Weyl ground state we must impose an additional
condition that goes under the name of GSO projection. In the
Ramond sector the GSO projection implies that we must restrict
ourselves to the states that are not annihilated by one of the two
following operators (for instance the one with the sign $+$):
\begin{equation}
P_{R} =\frac{1 \pm
\psi_{11}(-1)^{F_{R}}}{2} \hspace{.5cm} {\rm where} \hspace{.5cm}
F_R = \sum_{n=1}^{\infty} \psi_{-n} \cdot \psi_{n} \hspace{1cm}
\psi_{11} \equiv 2^5 \psi_{0}^{0} \psi_{0}^{1} \dots \psi_{0}^{9}
\label{gsof}
\end{equation}
On the other hand, in order to eliminate the
states with half-integer squared mass in units of $\alpha '$ that
are present in the spectrum of the NS but not in the R sector, we must
perform a similar projection also in the NS sector introducing
the operator
\begin{equation}
\label{ferno}
P_{NS}=\frac{1+(-1)^{F_{NS}}}{2} \hspace{.5cm} {\rm
where} \hspace{.5cm} F_{NS}= \sum_{t=1/2}^\infty \psi_{-t} \cdot
\psi_t -1
\end{equation}
Because of the previous projections,  the tachyon with mass
$M^2=-1/(2\alpha')$, appearing in the NS-sector of the spectrum is
projected out and the ground state fermion is a Majorana-Weyl
spinor in ten dimensions with only $8$ physical degrees of
freedom.

In the closed string case we have four different sectors and we
have to perform the GSO projection in each sector. Then one needs
to define  analogous quantities $\widetilde F_{NS}, \widetilde
F_R, \widetilde P_{NS}$ and $\widetilde P_R$ in terms of the right
handed oscillators. In so doing one can choose ${\widetilde P}_R$
to have the same or the opposite sign $(\pm)$ with respect to the
one appearing in the definition of $P_R.$ Then if $P_R$ and
${\widetilde P}_R$ are defined with the same sign ($+$ or $-$) the
two Majorana-Weyl spinors $u_A$ and $\widetilde u_A$ of the left
and right sectors have the same chirality ($+$ or $-$). Choosing
instead the opposite sign they have opposite chirality. These two
situations corresponds to two different superstring models. Indeed
the first case corresponds to the type IIB (chiral) theory and the
second case to the type IIA (non chiral) theory. In the NS-NS
sector the massless states are given by:
\begin{equation}
\psi_{-1/2}^{\mu} {\tilde{\psi}}_{-1/2}^{\nu} | 0, \frac{k}{2}\rangle
|\widetilde{0, \frac{k}{2}} \rangle
\hspace{1cm} k^2 =0
\label{msst}
\end{equation}
Those corresponding to a graviton are
obtained by saturating the state in the previous equation with the
symmetric and traceless tensor:
\begin{equation} \epsilon_{\mu \nu}^{(h)} =
\epsilon_{\nu \mu}^{(h)} \hspace{1cm} \epsilon_{\mu \nu}^{(h)}
\eta^{\mu \nu} =0
\label{grav0}
\end{equation}
Those corresponding to an
antisymmetric $2$-form tensor are obtained by saturating the state
in Eq.(\ref{msst}) with an antisymmetric polarization tensor:
\begin{equation}
\epsilon_{\mu \nu}^{(A)} = - \epsilon_{\nu \mu}^{(A)}
\label{kr}
\end{equation}
Finally the dilaton is obtained by saturating the state in
Eq.(\ref{msst}) with the following tensor:
\begin{equation} \epsilon_{\mu
\nu}^{(\phi)} = \frac{1}{\sqrt{8}} \left[ \eta_{\mu \nu} - k_{\mu}
\ell_{\nu} - k_{\nu} \ell_{\mu} \right]
\label{dila0}
\end{equation}
where
$\ell^2 = k^2 =0$ and $\ell \cdot k =1$. The physical conditions imply
that the polarization tensors for both the graviton and the
antisymmetric tensor satisfy the condition:
\begin{equation} k^{\mu}
\epsilon_{\mu \nu}^{(h)} = 0 \hspace{1cm}   k^{\nu} \epsilon_{\mu \nu}^{(A)}
=0
\label{conseq}
\end{equation}
Then we have a R-NS sector whose massless state is given by:
\begin{equation}
u_A (k) |A, k/2 \rangle  \,\, \epsilon_{\mu} (k) \psi^{\mu}_{-1/2}
|\widetilde{0, k/2} \rangle
\label{msset}
\end{equation}
By introducing a spinorial quantity with a vector index $ \chi^{\mu} (k)
\equiv
u (k) \epsilon^{\mu} (k)$ it is easy to check that the physical conditions
imply that $\chi$ satisfies the two equations:
\begin{equation}
(\chi)_B (\Gamma^{\mu})^{B}_{\,\,A} k_{\mu}  =0  \hspace{2cm} k \cdot \chi =0
\label{phyco}
\end{equation}
The vector spinor is reducible under the action of the Lorentz group. It can
be decomposed in the following way:
\begin{equation}
({\hat{\chi}}_{\mu})^{A} = \left[({\hat{\chi}}_{\mu})^{A} - \frac{1}{D}
\Gamma_{\mu} ( \Gamma^{\nu} {\hat{\chi}}_{\nu} )^{A}  \right] +
\frac{1}{D} \Gamma_{\mu} ( \Gamma^{\nu} {\hat{\chi}}_{\nu} )^{A}
\label{deco54}
\end{equation}
The first term corresponds to a gravitino with spin $3/2$, while the second
one to the dilatino with spin $1/2$. They have opposite chirality.
The same considerations apply also to the R-NS sector that provides also a
gravitino and a dilatino as in the case of the NS-R sector. In both these
sectors one gets space-time fermions.

Finally we have the R-R sector that as the NS-NS sector contains
bosonic states. The massless states of the R-R sector are given
by:
\begin{equation}
u_A (k) {\tilde{u}}_{B} (k) |A, k/2 \rangle |{\widetilde
{B , k/2}} \rangle
\label{rrstates}
\end{equation}
They are physical states
(annihilated by $G_0$ and ${\tilde{G}}_0$) if $u$ and $\tilde{u}$
satisfy the Dirac equation:
\begin{equation} u_A (k) ( k \cdot \Gamma
)^{A}_{\,\,B} = {\tilde{u}}_A (k) ( k \cdot \Gamma )^{A}_{\,\,B}=0
\label{direq23}
\end{equation}

In order to investigate further aspects of the R-R spectrum and also
for discussing vertex operators that are needed to write scattering
amplitudes among string states, it is
useful to use the conformal properties of string theory. Indeed
string theory in the conformal gauge is a two-dimensional
conformal field theory. Thus, instead of the operatorial analysis
that we have discussed until now, one could give an equivalent
description by using the language of conformal field theory in
which one works with OPEs rather then commutators or
anticommutators. We are not going to discuss here this alternative
description in detail (for a careful discussion see \cite
{FMS},\cite{ISLANDA}), but we will limit our conformal discussion
only to those string aspects that get an easier formulation in
terms of conformal field theory.

In the conformal formulation one introduces the variables $z$ and
${\bar{z}}$ that are related to the world sheet variables $\tau$
and $\sigma$ through the conformal transformation:
\begin{equation} z = e^{
2i( \tau - \sigma)}~~~~~~~~~~;~~~~~~~~~~ {\bar{z}} = e^{ 2 i( \tau
+ \sigma)}~,
\label{zbarz}
\end{equation}
in the closed string case and
\begin{equation} z = e^{i (\tau-\sigma)}~~~;~~~ \bar z=e^{i (\tau+\sigma)}
\label{zopen}
\end{equation}
in the open string case and then express the bosonic and fermionic
coordinates, given in Eq.s (\ref{expo1}), (\ref{expo2}) and
(\ref{modpsil}) for open strings and in Eq.s  (\ref{expc}),
(\ref{modpsil2}) and (\ref{modpsir}) for closed strings, in terms
of $z$ and $\bar z$.

Using the conformal language the R-R vacuum state $| A \rangle |
\tilde B \rangle$ can be written in terms of the NS - NS vacuum by
introducing the spin fields $S^{A} (z),{\tilde S}^{B} (\bar z)$
satisfying the equation:
\begin{equation} \lim_{z \rightarrow 0} S^{A} (z) {\tilde
S}^{B} (\bar z)|0\rangle |\tilde 0\rangle  = |A\rangle\tilde B
\rangle~~,
\label{spfi86}
\end{equation}
where $|0\rangle $ is the NS
vacuum. Inserting this equation in
Eq.(\ref{rrstates}) one can expand it as follows
\[
\lim_{z \rightarrow 0} u_{A} (k) S^{A} (z) {\tilde{u}}_B
{\tilde{S}}^{B} ({\bar{z}})
|0 \rangle | {\widetilde 0} \rangle =
\]
\begin{equation}
 = \frac{1}{2^5} \sum_{n=0}^{10}
\frac{(-1)^{n+1}}{n!} u_A (k) (\Gamma_{\mu_1 \dots \mu_n}   C^{-1} )^{AB}
{\tilde{u}}_{B} \lim_{z \rightarrow 0} S^{C} (z)
( C \Gamma^{\mu_1 \dots \mu_{n}})_{CD} {\tilde{S}}^D ( {\bar{z}}) |0 \rangle
|{\widetilde 0} \rangle
\label{deco64}
\end{equation}
where $\Gamma_{\mu_1 \dots \mu_n }$ is the completely antisymmetrized product
of $n$ $\Gamma$-matrices.

Also in this case the two spinors can be taken with the same or
opposite chirality. Let us assume that they have the same
chirality corresponding to IIB superstring theory. In this case
they both satisfy the two following Eq.s
\begin{equation}
u_A \left(\frac{1+
\Gamma_{11}}{2} \right)^{A}_{\,\,B} =0 \hspace{.5cm};
\hspace{.5cm} \left(\frac{1 - \Gamma_{11}}{2} \right)^{A}_{\,\,B}
(C^{-1})^{BC} {\tilde{u}}_C =0
\label{chi74}
\end{equation}
where the second is obtained from the first by using that
$\Gamma_{11}^{T} = - C \Gamma_{11} C^{-1}$. With the help of the
two previous Eq.s it is straightforward to show that:
\begin{equation}
u_A ( \Gamma_{\mu_1 \dots \mu_n
} C^{-1} )^{AB} {\tilde{u}}_B = (-1)^{n+1} u_A ( \Gamma_{\mu_1
\dots \mu_n } C^{-1} )^{AB} {\tilde{u}}_B
\label{qea23}
\end{equation}
This
means that only the terms with $n$ odd contribute in the sum in
Eq.(\ref{deco64}). If we had spinors with opposite chirality then
only the terms with $n$ even will be contributing. It remains to
show that the even or the odd values of $n$ are actually not
independent. In fact the first Eq. in (\ref{chi74}) implies that
\begin{equation}
u_E \left( \frac{1 + \Gamma_{11}}{2} \right)^{E}_{\,\,F} (
\Gamma_{\mu_1 \dots \mu_n } C^{-1} )^{FG}{\tilde{u}}_G =0
\label{ide38}
\end{equation}
But, by using that
\begin{equation} \Gamma_{11}
\Gamma_{\mu_1 \dots \mu_n} = \frac{(-1)^{(n+2)(n-1)/2}}{(10-n)!}
\epsilon_{\mu_1 \dots \mu_n \mu_{n+1} \dots \mu_{10}}
\Gamma^{\mu_{n+1} \dots \mu_{10}}
\label{rel85}
\end{equation}
in
Eq.(\ref{ide38}) we get
\begin{equation}
u \Gamma_{\mu_1 \dots \mu_n } C^{-1}
{\tilde{u}}  + \frac{(-1)^{(n+2)(n-1)/2}}{(10-n)!} \epsilon_{\mu_1
\dots \mu_n \mu_{n+1} \dots \mu_{10}} u \Gamma^{\mu_{n+1} \dots
\mu_{10}} {\tilde{u}} =0
\label{rela749}
\end{equation}
that shows that the
terms with $n > 5 $ are related to those with $n< 5$. For $n=5$ we
get the following self-duality relation:
\begin{equation}
u \Gamma_{\mu_1
\dots \mu_5 } C^{-1} {\tilde{u}}  + \frac{1}{5!} \epsilon_{\mu_1
\dots \mu_{10}} u \Gamma^{\mu_{6} \dots \mu_{10}} {\tilde{u}} =0
\label{rela654}
\end{equation}
This means that in type IIB we can limit
ourselves to $n=1,3,5$, while in type IIA to $n=2,4$, being the
other values related to those.

It can also be shown that the quantities in Eq.(\ref{rela749})
correspond to field strengths and not to  potentials:
\begin{equation}
F_{\mu_1 \dots \mu_n } \equiv  u\Gamma_{\mu_1 \dots \mu_n} C^{-1}
{\tilde{u}} \label{fipo67}
\end{equation}
This follows from the fact
that $F_{\mu_1 \dots \mu_n } $ satisfies both the Eq. of motion
and the Bianchi identity that in form notation are given by:
\begin{equation}
d F_n = d ({}^* F)_{10-n} =0
\label{bia32}
\end{equation}
They can be obtained from Eq.(\ref{fipo67}) remembering that the
two spinors appearing in it, in order to be physical, must satisfy
the Dirac equation given in Eq.(\ref{direq23}). Therefore we have
the following R-R potentials
\begin{equation}
C_0  \,\, , C_2 \,\, C_4 \hspace{2cm} {\rm
in~~type~~IIB}
\label{pote34}
\end{equation}
\begin{equation} C_1 \,\, , C_3
\hspace{2cm} {\rm in~~type~~ IIA}
\label{pote43}
\end{equation}
where the subindex indicates the rank of the form.

In conclusion the bosonic spectrum of the two closed superstring IIA
and IIB consists of a graviton $G_{\mu \nu}$, a dilaton $\phi$ and
a two-form potential $B_2$ in the NS-NS sector and of the RR
fields given in Eq.(\ref{pote34}) for the type IIB and in
Eq.(\ref{pote43}) for the type IIA theory. The number of physical
degrees of freedom of the previous fields is given in Table 1.
\begin{table}
\caption{Degrees of Freedom}
\begin{center}
\begin{tabular}{|c|c|c|}
\hline
STATE            & d-dims                 & 10-dims   \\ \hline
$G_{\mu \nu}$     & (d-2)(d-1)/2 -1         &  35       \\ \hline
$\phi$            &  1                      &   1       \\ \hline
$B_2$             &  (d-2)(d-3)/2           &   28      \\ \hline
$C_0$             &  1                      &  1        \\ \hline
$C_1$             &  d-2                    &   8       \\ \hline
$C_2$             &  (d-2)(d-3)/2           &   28      \\ \hline
$C_3$             &  (d-2)(d-3)(d-4)/6      &   56      \\ \hline
$C_4 $  &  $(d-2)(d-3)(d-4)(d-5)/(2\cdot 4!) $ & 35     \\ \hline
$\chi_{\mu}^{A}$  & $ (d-3) 2^{d/2}/4$        &  56     \\ \hline
$\psi^A$          &  $ 2 ^{d/2}/4 $             &   8     \\ \hline
\end{tabular}
\end{center}
\end{table}
We have assumed that $d$ is even. If $d$ is odd the fermionic
degrees of freedom are given by $2^{(d-1)/2}$ instead of
$2^{d/2}$. By counting the number of physical degrees of freedom
for both type II theories it is easy to check that the number of
bosonic degrees of freedom $(128)$ equals that of fermionic ones
$(128)$ as expected in a supersymmetric theory. It turns out that
the actions describing the low-energy degrees of freedom in the
two closed superstring theories are the type IIB and type IIA
supergravities that we will write down in the next section.

Until now we have analyzed the string spectrum in the operatorial
framework, in which the physical states are constructed by acting
with the oscillators $\alpha$ and $\psi$ on the vacuum state.
However, in order to compute string scattering amplitudes, the
conformal description of string theory is much more suitable than
the operatorial one. Indeed in the conformal framework one defines
a vertex operator for each string state and then express a
scattering amplitude among string states in terms of a
correlator between the corresponding vertex operators.

In a conformal field theory one introduces the concept of conformal
or  primary field ${\cal V}(z)$ of dimension $h$ as an object
that satisfies the following OPE with the energy momentum tensor
$T(z)$:
\begin{equation}
\label{confi}
T(z){\cal V}(w)= \frac{\partial_w {\cal
V}(w)}{z-w}+ \frac{h {\cal V}(w)}{(z-w)^2} +...~~.
\end{equation}
where the dots indicate non singular terms when $z \rightarrow w$.
In order to perform a  covariant quantization of string theory we
have to consider also the ghost and superghost degrees of freedom
that
 we have completely disregarded until now. They
arise from the exponentiation of the Faddev-Popov determinant that
is obtained when the string is quantized through the path-integral
quantization. In particular choosing the conformal gauge, one gets
the following action \cite{FMS}
\begin{equation}
\label{Sbc}
S_{\rm gh-sgh}
\sim \int d^2 z[(b\bar\partial c+{\rm c.c.})+
\left(\beta\bar\partial\gamma+c.c\right)]~~,
\end{equation}
where $b$ and
$c$ are fermionic fields with conformal dimension equal
respectively to $2$ and $-1$ while  $\beta$ and $\gamma$ are
bosonic fields with conformal dimensions equal respectively to
$3/2$ and $-1/2$.

With the introduction of  ghosts the string action  in the
conformal gauge becomes invariant under the BRST transformations
and the physical states are  characterized by the fact that they
are annihilated by the BRST charge that is given by
\begin{equation}
\label{carS}
Q \equiv \oint dz J_{BRST} (z) = Q_0+Q_1+Q_2~~,
\end{equation}
where
\begin{equation}
\label{cars2}
Q_0= \oint\frac{dz}{2\pi
i}c(z)\left[T(z) +T^{\beta\gamma}(z)+
\partial c(z)b(z)
\right]
\end{equation}
and
\begin{equation}
\label{cars3}
Q_1=\frac{1}{2}\oint\frac{dz}{2\pi i}
\gamma(z)\psi(z)\cdot\partial X(z)~~~; ~~~
Q_2=-\frac{1}{4}\oint\frac{dz}{2\pi i}\gamma^2(z)b(z)
\end{equation}
where
\begin{equation} T^{bc}(z)=[-2 b \partial c -
\partial b  c]: \,\,\,\,\,\,\,\,\,
T^{\beta\gamma}(z)=[-\frac{3}{2} \beta \partial \gamma-\frac{1}{2}
\partial \beta  \gamma]
\label{2.6b}
\end{equation}
It is convenient to use the bosonized variables:
\begin{equation}
\gamma (z) = {\rm e}^{\varphi (z)} \eta (z)~~~~~,~~~~ \beta (z) =
\partial \xi (z) {\rm e}^{- \varphi (z)}
\label{bosvar56}
\end{equation}
In terms of them $Q_1$ and $Q_2$ become:
\begin{equation}
Q_1=\frac{1}{2}\oint\frac{dz}{2\pi i} {\rm e}^{\varphi (z)} \eta
(z) \psi(z)\cdot\partial X(z)~~~; ~~~ Q_2=
\frac{1}{4}\oint\frac{dz}{2\pi i}b(z)\eta (z) \partial \eta (z)
{\rm e}^{2 \varphi (z)}
\label{bosq62}
\end{equation}
A vertex operator
corresponding to a physical state must be BRST invariant, i.e.
\begin{equation} \label{verstsu}
 [Q,{\cal W} (z)]_\eta=0
\end{equation}
 where
$[,]_\eta$ means  commutator ($\eta =-1$) [anticommutator ($\eta
=1$)] when the vertex operator is a bosonic [fermionic] quantity.

In the massless NS sector of open strings the correct BRST
invariant vertex operator with the inclusion of the ghosts and
superghosts contribution turns out to be
\begin{equation}
\label{vert3}
{\cal W}_{-1}(z)=c(z) e^{-\varphi(z)}
\epsilon\cdot\psi(z) e^{i{\sqrt{2\alpha'}}k\cdot X(z)}
\end{equation}
Here and in the following we use dimensionless string fields $X$
and $\psi$, and also the momentum $k$ always appear in the
dimensionless combination ${\sqrt{2\alpha'}}k$. The previous
vertex is BRST invariant if $k^2= \epsilon\cdot k=0.$ This can be
shown in the following way. Let us define for convenience:
\begin{equation}
{\cal
W}_{-1}(z) \equiv c(z) F(z)
\label{def49}
\end{equation}
and let us compute:
\[
[ Q_0 , {\cal W}_{-1}(w)] = \oint \frac{dz}{2 \pi i}
c(z)\left[T(z) +T^{\beta\gamma}(z)+ \partial c(z)b(z) \right]
{\cal W}_{-1}(w) =
\]
\begin{equation}
= \oint \frac{dz}{2 \pi i} \left\{c (z) c(w) \left[
    \frac{\partial_w  F(w)}{z-w} +
    \frac{F(w)}{(z-w)^2}
\right] + \frac{ (c \partial c F) (w)}{z-w}   \right\}
\label{ope39}
\end{equation}
where we have used the fact that $F(w)$ is a conformal field with
dimension equal to $1$, which implies that the four-momentum of
the vertex must be light-like $k^2 =0$. Expanding around the point
$z =w$ and keeping only the terms that are singular i.e. the only
ones that give a non-vanishing contribution in the previous
equation we get:
\begin{equation}
\oint \frac{dz}{2 \pi i}
\left\{ \left[ c(w) + (z-w) \partial c (w)\right] \left[
\frac{c(w) \partial F (w)}{z-w} + \frac{{\cal W}_{-1}(w)
}{(z-w)^2} \right] - \frac{\partial c(w){\cal
W}_{-1}(w)}{(z-w)}\right\} =0
\label{ope493}
\end{equation}
because the term
$c(w)$ in the  square bracket gives no contribution ( $c^2 =0$ )
and the other singular terms just trivially cancel. Following the
same procedure it can also be checked that
\begin{equation}
[Q_1 ,{\cal
W}_{-1}(w) ] = [Q_2 ,{\cal W}_{-1}(w)]=0
\label{ope43}
\end{equation}
The second equation is valid in general, while the first one is
only valid if $\epsilon \cdot k =0$. In conclusion we have seen
that the vertex operator in Eq.(\ref{vert3}) is BRST invariant if
$k^2 = \epsilon \cdot k  =0$.

We can proceed in an analogous way in the R sector and obtain the
following BRST invariant vertex operator for the massless
fermionic state of  open superstring~\cite{FMS}:
\begin{equation}
\label{vert7}
{\cal W}_{-1/2}(z)=u_A(k) c(z)
S^A(z)e^{-\frac{1}{2}\varphi(z)} e^{i{\sqrt{2\alpha'}}k\cdot X (z)}
\end{equation}
It is BRST invariant if $k^2 =0$ and $u_A \left( \Gamma^{\mu}
\right)^{A}_{\,\,B} k_{\mu} =0$. Both  vertices in Eq.s
(\ref{vert3}) and (\ref{vert7}) have conformal dimension equal to
zero.

Moreover in superstring for each physical state we can construct
an infinite tower of equivalent physical vertex operators all
(anti)commuting with the BRST charge and characterized  according
to their superghost picture $P$ that is equal to the total ghost
number of the scalar field $\varphi$ and of the $\eta\xi$ system
that appear in the "bosonization" of the $\beta\gamma$ system
according to Eq.(\ref{bosvar56}) (for more details see
\cite{ISLANDA}):
\begin{equation} P =
\oint \frac{dz}{2 \pi i} \left(- \partial \varphi + \xi \eta
\right)
\label{pict}
\end{equation}
For example the vertex in Eq.(\ref{vert3}), which does not contain  the
fields $\eta$ and $\xi$ but only the field $\varphi$ appearing
in the exponent with a factor $-1$, is in the picture $-1$.
Analogously the vertex in Eq.(\ref{vert7}) is in the picture
$-1/2$. Vertex operators in different pictures are related through
the picture changing procedure that we are now going to describe
shortly. Starting from a BRST invariant vertex ${\cal{W}}_{t}$ in
the picture $t$ (characterized by a value of $P$ equal to $t$),
where $t$ is integer (half-integer) in the NS (R) sector, one can
construct another BRST invariant vertex operator ${\cal W}_{t+1}$
in the picture $t+1$ through the following operation~\cite{FMS}
\begin{equation}
\label{vert4}
{\cal W}_{t+1}(w)= [Q,2\xi(w){\cal W}_{t}(w)]_{\eta}
=\oint_{w} \frac{dz}{2\pi i}J_{BRST}(z)~2\xi(w){\cal W}_{t}(w)~~.
\end{equation}
Using the Jacobi identity and the fact that $Q^2 =0$ one can
easily show that the vertex ${\cal W}_{t+1} (w)$ is BRST
invariant:
\begin{equation}
[Q , {\cal W}_{t+1} ]_{\eta} =0
\label{brsinv}
\end{equation}
On the other hand the vertex ${\cal W}_{t+1} (w) $ obtained
through the construction in Eq.(\ref{vert4}) is not BRST trivial
because the corresponding state contains the zero mode $\xi_0$
that is not contained in the Hilbert space of the
$\beta\gamma$-system as it can be seen by looking at the
expressions of $\beta$ and $\gamma$ in terms of $\xi$ given in Eq.
(\ref{bosvar56}).

In conclusion all the vertices constructed through the procedure
given in Eq.(\ref{vert4}) are BRST invariant and non trivial in
the sense that all give a non-vanishing result when inserted for
instance in a tree-diagram correlator  provided that the total
picture number is equal to $-2$ (see \cite{ISLANDA}for more
details). Using the picture changing procedure from the vertex
operator in Eq.(\ref{vert3}) we can construct the vertex operator
in the $0$ superghost picture which is given by~\cite{PETTO}
\begin{equation}
\label{vert6}
{\cal W}_{0}(z)=c(z)
{\cal V}_1(z)-\frac{1}{2}\gamma(z){\cal V}_0(z)~~.
\end{equation}
with
\begin{equation}
\label{ver2} {\cal V}_0(z)=\epsilon\cdot\psi(z) e^{i{\sqrt{2\alpha'}}k\cdot
X(z)}~~~{\rm and }~~~ {\cal V}_1(z)=(\epsilon\cdot\partial X(z)
+i{\sqrt{2\alpha'}}k\cdot\psi \epsilon\cdot\psi) e^{i{\sqrt{2\alpha'}}k\cdot X(z)}~~.
\end{equation}

Analogously starting from the massless vertex in the R sector in
Eq.(\ref{vert7}) one can construct the corresponding vertex in an
arbitrary superghost picture $t$.

In the closed string case the vertex operators are given by the
product of two vertex operators of the open string. Thus for the
massless NS-NS sector in the superghost picture $(-1,-1)$ we have
\begin{equation}
\label{vert7b}
{\cal W}_{(-1,-1)}=\epsilon_{\mu\nu}{\cal
V}^\mu_{-1}(k/2,z) \widetilde{\cal V}^\nu_{-1}(k/2,\bar z)~~,
\end{equation}
where ${\cal V}^\mu_{-1}(k/2,z)=c(z)\psi^\mu(z)e^{-\varphi(z)}
e^{i\frac{{\sqrt{2\alpha'}}k}{2}\cdot X(z)}$ and  $\widetilde{\cal V}^\nu_{-1}$ is
equal to an analogous expression in terms of the tilded modes.
This vertex is BRST invariant if $k^2=0$ and
$\epsilon_{\mu\nu}k^\nu=k^\mu\epsilon_{\mu\nu}=0.$

In the R-R sector the vertex operator for  massless states in the
$(- \frac{1}{2},- \frac{1}{2})$ superghost picture is
\begin{equation}
\label{vert8}
{\cal W}_{(-1/2,-1/2)}=
\frac{\left(C\Gamma^{\mu_1...\mu_{m+1}}\right)_{AB}
F_{\mu_1...\mu_{m+1}}}{2\sqrt{2} (m+1)!}{\cal V}^A_{-1/2}(k/2,z)
\widetilde{\cal V}^B_{-1/2}(k/2,\bar z)
\end{equation}
where ${\cal
V}^A_{-1/2}(k/2,z)=c(z)S^A(z)e^{-\frac{1}{2}\varphi(z)}
e^{i\frac{{\sqrt{2\alpha'}}k}{2}\cdot X(z)}$ and
\begin{equation}
\label{effe}
F_{\mu_1...\mu_{m+1}}=\frac{(-1)^{m+1}}{2^5}u_D(k)(\Gamma_{\mu_1...\mu_{m+1}}
C^{-1})^{DE}\widetilde u_E(k)~~.
\end{equation}
It is BRST invariant if
$k^2=0$ and $F_{\mu_1...\mu_m}$ is a field strength satisfying
both the Maxwell equation $(dF=0)$ and the Bianchi identity
$(d\,^*F=0)$.

For future purposes it is useful to give also the vertex operator
of a physical R-R state in the asymmetric picture $(-1/2,-3/2)$.
Indicating with $A_{\mu_1...\mu_m}$  the gauge potential corresponding to
the field strength $F_{\mu_1...\mu_{m+1}}$, this vertex is given
by~\cite{BILLO}:
\begin{equation}
\label{vert9}
{\cal
W}_{(-1/2,-3/2)}=\sum_{M=0}^{\infty} \frac{a_M}{2{\sqrt 2}}
\left(C {\cal{A}}^{(m)} \Pi_M\right)_{AB} {\cal
V}^A_{-1/2+M}(k/2,z)\widetilde{\cal V}^B_{-3/2-M}(k/2,\bar z)
\end{equation}
where
\begin{equation}
\left(C {\cal{A}}^{(m)} \right)_{AB} = \frac{\left(C
\Gamma^{\mu_1...\mu_m} \right)}{m!}
A_{\mu_1...\mu_m}~~,~~\Pi_q=\frac{1+(-1)^q\Gamma_{11}}{2}
\label{aaa}
\end{equation}
 and
\begin{equation}
\label{vert10}
{\cal
V}^A_{-1/2+M}(k/2,z)=\partial^{M-1}\eta(z)...\eta(z) c(z)
S^A(z)e^{\left(-\frac{1}{2}+M\right)\varphi(z)}
e^{i\frac{\sqrt{2\alpha'}k}{2}\cdot X(z)}
\end{equation}
\begin{equation}
\label{vert11}
\widetilde{\cal V}^B_{-3/2-M}(k/2,\bar z)=
\bar\partial^{M}\widetilde\xi(\bar
z)...\bar\partial\widetilde\xi(\bar z) \widetilde c(\bar z)
\widetilde S^A(\bar z)e^{\left(-\frac{3}{2}-M\right)
\widetilde\varphi(\bar z)} e^{i\frac{\sqrt{2\alpha'}k}{2}\cdot \widetilde X(\bar
z)}
\end{equation}
It can be shown that the vertex operator in
Eq.(\ref{vert9}) is BRST invariant if $k^2=0$ and the following
two conditions are satisfied
\begin{equation} a_M=
\frac{(-1)^{M(M+1)}}{[M!(M-1)!...1]^2}~~~~,~~~~d {}^* A^{(m)}
=0~~~.
\label{condi8}
\end{equation}
By acting with the picture changing operator on the vertex in
Eq.(\ref{vert9}) it can be shown that one obtains the vertex in
the symmetric picture in Eq.(\ref{vert8}). In particular one can
show that only the first term in the sum in Eq.(\ref{vert9})
reproduces the symmetric vertex, while all the other terms  give
BRST trivial contributions. Notice that by changing the superghost
picture of the vertex operator one also changes the physical
content of the specific vertex. Indeed while the vertex operator
in the symmetric picture is proportional to the field strength,
the one in the asymmetric picture depends on the potential.
Obviously the picture changing procedure does not affect the
amplitudes, which always depend on the field strength.

Having constructed the BRST invariant vertex operators we can use
them to compute scattering amplitudes between string states. In
order to get a non-vanishing result, we must use three
BRST-invariant vertices of the form $c(z) V(z)$ ($V(z)$ is a
primary field with conformal dimension equal to $1$ that does not
contain the ghosts $b$ and $c$) corresponding to the states for
which we fix the corresponding Koba-Nielsen variables to $0,1$ and
$\infty$ and $(N-3)$ vertices without the factor $c(z)$. The last
ones are also BRST-invariant because we integrate over the
corresponding Koba-Nielsen variable. In this way the product of
the $N$ vertices has ghost number equal to $3$ and when it is
taken between the BRST and projective-invariant vacuum
characterized by ghost number $q=0$ we get a non-zero result.
Moreover one must also require the product of the $N$ vertex
operator to have  superghost number equal to $-2$.

As an example let us calculate the amplitude among three gluon
states with momentum $k_i$ and polarization $\epsilon_i$ at tree
level in perturbation theory. This is given by \cite{DLMMR}
\begin{equation}
\label{ampiezza}
A^0(\epsilon_1,k_1;\epsilon_2,k_2;\epsilon_3,k_3)=
C_0 (i N_{\rm op})^3 {\rm Tr}(\lambda^{a_1}\lambda^{a_2}\lambda^{a_3})
\langle W_{-1}(z_1)W_{-1}(z_2)W_{0}(z_3)\rangle
\end{equation}
where the normalization of the tree level amplitude and of the
states in  $d$ dimensions are
\begin{equation}
\label{costanti}
C_{0}= \frac{1}{g_{\rm
op}^2(2\alpha')^{\frac{d}{2}}}
\,\,\,\,\,\,\,\,\,\,\,\,\,\,\,\,\,\,\,\,\,\,\,\,\, N_{\rm op}=2
g_{\rm op}(2\alpha')^{\frac{d-2}{4}}~,
\end{equation}
the matrices $\lambda$ are the generators of the gauge group
$SU(N)$ in the fundamental representation, normalized as
\begin{equation}
{\rm Tr}(\lambda^a\lambda^b)=\frac{\delta^{ab}}{2} \label{trac}
\end{equation}
and finally
$W_{-1}$ is the gluon vertex in the superghost picture $-1$
given in Eq.(\ref{vert3}) and $W_{0}$ is the one in the picture
$0$ defined in Eq.(\ref{vert6}).
Let us evaluate the correlator
\[
\langle W_{-1}(z_1)W_{-1}(z_2)W_{0}(z_3)\rangle=\epsilon^1_\mu\epsilon^2_\nu
\epsilon^3_\rho
\langle c(z_1)c(z_2)c(z_3)\rangle\langle e^{-\phi(z_1)}e^{-\phi(z_2)}\rangle
\times
\]
\begin{equation}
\left[
\langle \psi^{\mu}(z_1) \psi^{\nu} (z_2)\rangle
\langle \prod_{i=1}^3e^{ik_i\cdot X(z_i)}  \partial X^\rho(z_3)
\rangle + \langle \prod_{i=1}^3 e^{ik_i\cdot X(z_i)}\rangle
\langle \psi^{\mu} (z_1) \psi^{\nu} (z_2)i{\sqrt{2\alpha'}}k_3\cdot
\psi(z_3) \psi^{\rho} (z_3)\rangle
\right]
\label{correlat}
\end{equation}
Using the correlators \cite{FMS}
\begin{equation}
\label{3e} \langle e^{i{\sqrt{2\alpha'}}k_i\cdot
X(z_i)}e^{i{\sqrt{2\alpha'}}k_j\cdot
X(z_j)}\rangle=(z_{ij})^{{2\alpha'}k_i\cdot k_j}
\end{equation}
\begin{equation}
\langle e^{i{\sqrt{2\alpha'}}k_i\cdot
X(z_i)} \partial X^\mu(z_j) \rangle= \frac{i{\sqrt{2\alpha'}}k_i^\mu}{z_{ij}}
\,\,\,\,\,\,\,\,\,\,\,\,\,\,\,\,\,\,\,\,\,\,
\langle \psi^{\mu} (z_i) \psi^{\nu} (z_j)\rangle=-
\frac{\eta^{\mu \nu}}{z_{ij}}
\label{3ed}
\end{equation}
\begin{equation}
\label{ccc}
\langle c(z_i)c(z_j)c(z_k)\rangle=z_{ij}z_{ik}z_{jk}~~,
\,\,\,\,\,\,\,\,\,\,\,\,\,\,\,\,\,\,\,\,\,\,
\langle e^{a\phi(z_i)}e^{b\phi(z_j)}\rangle=\frac{1}{(z_{ij})^{ab}}
\end{equation}
with $z_{ij}=(z_i-z_j)$,
Eq. (\ref{correlat}) becomes
\begin{equation}
\langle W_{-1}(z_1)W_{-1}(z_2)W_{0}(z_3)\rangle= {\sqrt{2\alpha'}}
\epsilon^1_\mu\epsilon^2_\nu\epsilon^3_\rho z_{23}z_{13} \left\{
-\frac{\eta^{\mu\nu}}{z_{12}}\left[\frac{k_1^\rho}{z_{13}}
+\frac{k_2^\rho}{z_{23}}\right]+\frac{k_3^\nu\eta^{\mu\rho}
-k_3^\mu\eta^{\nu\rho}}{z_{13}z_{23}} \right\} \label{corre2}
\end{equation}
where we have used the on-shell condition to eliminate the factors
$(z_{ij})^{{2\alpha'}k_i\cdot k_j}$. Substituting Eq.s
(\ref{costanti}) and (\ref{corre2}) in Eq.(\ref{ampiezza}), using
momentum conservation and transversality one gets:
\begin{equation}
\label{ampiezza1}
A^0(\epsilon_1,k_1;\epsilon_2,k_2;\epsilon_3,k_3)=  8g_{\rm
op}(2\alpha')^{\frac{d-4}{4}} {\rm
Tr}(\lambda^{a_1}\lambda^{a_2}\lambda^{a_3})(
\epsilon_1\cdot\epsilon_2
k_1\cdot\epsilon_3+\epsilon_3\cdot\epsilon_1
k_3\cdot\epsilon_2+\epsilon_2\cdot\epsilon_3 k_2 \cdot\epsilon_1)
\end{equation}
In order to compare this result with the field theory $3$-gluon
amplitude we need to add to the previous expression corresponding to
the permutation $(1,2,3)$ also the contribution of the
anticyclic permutation $(3,2,1)$ that can be easily obtained from
eq.(\ref{ampiezza1}). In so doing one gets
\begin{equation}
\label{ampiezzaf}
A^0(\epsilon_1,k_1;\epsilon_2,k_2;\epsilon_3,k_3)= 4 i g_{\rm
op}(2\alpha')^{\frac{d-4}{4}} {\rm f^{a_1 a_2 a_3}} (
\epsilon_1\cdot\epsilon_2
k_1\cdot\epsilon_3+\epsilon_3\cdot\epsilon_1
k_3\cdot\epsilon_2+\epsilon_2\cdot\epsilon_3 k_2 \cdot\epsilon_1)
\end{equation}
where we have used that ${\rm
Tr}([\lambda^{a_1}\lambda^{a_2}]\lambda^{a_3})= \frac{i}{2} {\rm f^{a_1
a_2 a_3}}$. Comparing  Eq.(\ref{ampiezzaf}) with the $3$-gluon
scattering amplitude in $d$ dimension,  we get the following
relation between the $d$-dimensional gauge coupling
constant and the string parameters $g_{\rm op}$ and $\alpha'$
\begin{equation}
g_d=2 g_{\rm op}(2\alpha')^{\frac{d-4}{4}}
\end{equation}

\sect{Classical $p$-brane: the closed string perspective}
\label{sec2} At semiclassical level, in addition to strings,
closed string theories naturally contain other extended objects
that are called $p$ branes which are solutions of the low energy
string effective action. They act as sources of the massless R-R
closed string fields and saturate the BPS bound between mass and
charge. In order to see how they appear in the theory, let us
briefly discuss the low energy limit of string theory. The low
energy effective action of type IIB string theory is given by the
so-called  type IIB supergravity. Its bosonic part in the Einstein
frame is given by: ~\footnote{Our conventions for curved indices
and forms are the following: $\varepsilon^{0\dots9}=+1$; signature
$(-,+^9)$; $\mu,\nu=0,\dots,9$; $\alpha,\beta=0,\ldots,3$;
$i,j=4,5$; $\ell,m=6,\ldots,9$; $\omega_{(n)}={1\over n!}
\,\omega_{\mu_1 \dots \mu_n} dx^{\mu_1}\wedge\dots\wedge
dx^{\mu_n}$, and $*\omega_{(n)}={\sqrt{-\det G}\over
n!\,(10-n)!}\, \varepsilon_{\nu_1\dots\nu_{10-n}\mu_1 \dots \mu_n}
\,\omega^{\mu_1 \dots \mu_n}  dx^{\nu_1}\wedge\dots\wedge
dx^{\nu_{10-n}}.$}
\[
S_{\rm IIB} = \frac{1}{2 \kappa^2} \Bigg\{ \int d^{10} x~
 \sqrt{-\det G}~ R - \frac{1}{2} \int \Big[ d \phi \wedge {}^* d \phi
 \,+\, {\rm e}^{- \phi} H_{3}  \wedge {}^* H_{3}\,+\, {\rm e}^{2
 \phi}\, F_{1} \wedge {}^* F_{1}
\]
\begin{equation}
 + \,\,{\rm e}^{\phi} \,{\widetilde{F}}_{3} \wedge {}^*
 {\widetilde{F}}_{3} \,+\, \frac{1}{2}\, {\widetilde{F}}_{5}
\wedge {}^* {\widetilde{F}}_{5}  \, -\,  C_{4} \wedge H_{3}
 \wedge F_{3} \Big] \Bigg\}
\label{tendim3}
\end{equation}
where
\begin{equation}
H_{3} = d B_{2}~~~,~~~F_{1}=d C_{0}~~~,~~~ F_{3} =
d C_{2}~~~,~~~F_{5} = d C_{4}
\label{form2}
\end{equation}
are, respectively, the field strengths of the NS-NS 2-form and of
the 0-, 2- and 4-form potentials of the R-R sector and
\begin{equation}
{\widetilde{F}}_{3} = F_{3}
+ C_{0} \wedge H_{3}~~~~ ,~~~~{\widetilde{F}}_{5} = F_{5} +
C_{2} \wedge H_{3}~~.
\label{form3}
\end{equation}
Moreover, $2 \kappa^2 \equiv 16 \pi G_{N}^{10}=
 (2\pi)^{7}\,g_s^2\,\alpha'^4 $ where $g_s$
is the string coupling  constant, and the self-duality constraint
${}^* {\widetilde{F}}_{5}={\widetilde{F}}_{5}$  has to be
implemented on shell.

The low energy type IIA string effective action, corresponding to type
IIA supergravity has, in our conventions, the following expression:
\[
S_{{IIA}}= \frac{1}{2\kappa^2}\Bigg\{
            \int d^{10}x\sqrt{-G}\, R \Bigg. \,\, +
\]
\begin{equation}
\label{IIAAction}
- \Bigg.
           \frac{1}{2}
            \int\Big( d\phi\wedge  {}^*d\phi
                 + e^{-\phi}H_3\wedge {}^* H_3
                 - e^{3\phi/2}F_2\wedge {}^* F_2
                 - e^{\phi/2}\tilde{F}_4\wedge {}^*\tilde{F}_4
                 + B_2 \wedge F_4\wedge F_4
            \Big)\Bigg\}  \, ,
\end{equation}
where we have considered only the bosonic degrees of freedom.
The field strengths appearing in the last equation are given by:
\begin{equation}
  H_3=dB_2\, , \qquad F_2=dC_1\, , \qquad F_4=dC_3\, , \qquad
  \tilde{F}_4=F_4-C_1\wedge H_3\, ,
\end{equation}
and $2 \kappa^2 =(2\pi)^{7 }g_{s}^{2}{\alpha'}^4$.

The previous supergravity actions have been written in the
so-called Einstein frame. It is also useful to give their
expression in a different frame which is called the string frame,
the two being related by the following rescaling of the metric:
\begin{equation}
g_{\mu \nu} = e^{(\phi - \phi_0)/2} G_{\mu
  \nu}~~~,{\rm with}~~~e^{\phi_0} = g_s
\label{str87}
\end{equation}
where with $ g_{\mu \nu}$ we have indicated the string frame
metric. In general under a rescaling of the metric
\begin{equation}
{\hat{g}}_{\mu \nu} = e^{- 2 s \gamma (x)}  g_{\mu \nu}
\label{resca}
\end{equation}
the various terms that appear in the supergravity Lagrangian are
modified as follows:
\begin{equation}
\sqrt{- {\hat{g}}} = e^{- s d \gamma (x)} \sqrt{- g}
\label{det}
\end{equation}
\begin{equation}
{\hat{R}} = e^{2s \gamma (x)} \left[ R + 2 s (d-1) \frac{1}{\sqrt{-g}}
\partial_{\mu} \left( \sqrt{-g} g^{\mu \nu} \partial_{\nu}
 \gamma (x) \right) - s^2
(d-1)(d-2) g^{\mu \nu} \partial_{\mu} \gamma (x) \partial_{\nu} \gamma (x)
\right]
\label{curva}
\end{equation}
\[
e^{- 2 a \gamma (x)} \sqrt{- {\hat{g}}}\left[ \hat{R} - b {\hat{g}}^{\mu \nu}
\partial_{\mu} \gamma \partial_{\nu} \gamma \right] =
e^{- \left[ 2a + s(d-2) \right]\gamma (x)} \times
\]
\begin{equation}
\times \sqrt{-g} \left\{R + \left[
s^2 (d-1)(d-2)+ 4 as (d-1) -b \right] g^{\mu \nu} \partial_{\mu} \gamma
\partial_{\nu} \gamma \right\}
\label{equa}
\end{equation}
\begin{eqnarray}
- \frac{\sqrt{- {\hat{g}}}}{2 n!} e^{a_n \gamma (x)}{\hat{g}}^{\mu_{1} \nu_{1}}
\dots {\hat{g}}^{\mu_{n} \nu_{n}} H_{\mu_1 \dots \mu_{n}} H_{\nu_1 \dots \nu_n}
& = & \nonumber \\
=- \frac{\sqrt{- {{g}}}}{2 n!} e^{[ a_n - (d-2n)s] \gamma (x)}
{{g}}^{\mu_{1} \nu_{1}}
\dots {{g}}^{\mu_{n} \nu_{n}} H_{\mu_1 \dots \mu_{n}} H_{\nu_1 \dots \nu_n}
\label{antisy}
\end{eqnarray}
where for the sake of generality we have kept the dimension of the
space-time $d$ arbitrary.

Applying the previous formulas in Eq.
(\ref{tendim3}) we get the purely bosonic part of the IIB
supergravity action in the string frame:
\[
S_{IIB} = \frac{1}{2 \kappa^{2}} \Bigg\{ \int d^{10} x \sqrt{-g}
e^{-2 (\phi - \phi_0)} \Big[ R + 4 g^{\mu \nu} \partial_{\mu} \phi
    \partial_{\nu} \phi - \frac{1}{12}
    H_{\mu \nu \rho} H^{\mu \nu \rho} \Big] +  \Bigg.
\]
\begin{equation}
\Bigg.
- \frac{1}{2} \int \Big[ {\widetilde{F}}_3 \wedge {}^* {\widetilde{F}}_3 +
 F_1 {}^* \wedge F_1   + \frac{1}{2} {\widetilde{F}}_{5}
 \wedge {}^* {\widetilde{F}}_{5}  \, -\,  C_{4} \wedge H_{3}
 \wedge F_{3} \Big] \Bigg\}
\label{stri84}
\end{equation}
while for the type IIA effective action in the string frame one gets:
\[
S_{IIA} = \frac{1}{2 \kappa^{2}} \Bigg\{ \int d^{10} x \sqrt{-g}
e^{-2 (\phi -\phi_0)} \Big[ R + 4 g^{\mu \nu} \partial_{\mu} \phi
    \partial_{\nu} \phi - \frac{1}{12}
    H_{\mu \nu \rho} H^{\mu \nu \rho} \Big] +  \Bigg.
\]
\begin{equation}
\Bigg.
- \frac{1}{12} \int \Big[ {\widetilde{F}}_4 \wedge {}^* {\widetilde{F}}_4 +
 F_2 {}^* \wedge F_2   +   \, -\,  B_{2} \wedge F_{4}
 \wedge F_{4} \Big] \Bigg\}
\label{stri85}
\end{equation}
The classical equations of motion derived from the previous
low-energy actions admit solutions corresponding to
$p$-dimensional objects called $p$-brane. In the following we want
just to remind their main properties. The starting point is either
Eq.(\ref{tendim3}) or (\ref{IIAAction}) in which we neglect the
NS-NS two-form potential and we keep only one R-R field, namely:
\begin{equation}
S = \frac{1}{2 \kappa^2} \int d^d x \sqrt{-g}  \left[ R  -
\frac{1}{2} \left(\nabla \phi \right)^2 -
\frac{1}{2 (n+1)!} {\rm e}^{- a \phi }
\left( F_{p+1} \right)^{2} \right]~~,
\label{action}
\end{equation}
where we have indicated with $d$ the space-time dimension. The $p$ brane
solution is obtained by making the following ansatz for the metric:
\begin{equation}
d {{s}}^2 = \left[ H(r) \right]^{2 A}
\left(\eta_{\alpha\beta}dx^\a
dx^\b \right) + \left[ H(r) \right]^{2B} (\delta_{ij}d x^idx^j)~~,
\label{metri}
\end{equation}
with $\alpha,\,\beta\in\{0,...p\} $,  $i,\,j\in\{p+1,...d-1\} $ , and
the ansatz
\begin{equation}
{\rm e}^{- {{\phi}}(x)} = \left[ H(r) \right]^{\tau}
~~~~,~~~~{\cal{C}}_{01...p}(x) = \pm  [ H(r) ]^{-1}~~,
\label{dil}
\end{equation}
for the dilaton $ \phi$ and for the R-R $(p+1)$-form potential
${\cal C}$ respectively. $H (r)$ is assumed to be only a function
of the square of the transverse coordinates $r = x_{\bot}^{2}= x_i
x^i$.  If the parameters are chosen as
\begin{equation}
A= - \frac{d - p -3}{2(d-2)}~~~,~~~B= \frac{p+1}{2(d-2)}~~~,~~~
\tau= \frac{{{a}}}{2}~~~,~~~,
\label{expo}
\end{equation}
with ${{a}}$ obeying the equation
\begin{equation}
\frac{2 (p+1)( d-p-3)}{d-2}  + {{a}}^2 = 4~~,
\label{equ}
\end{equation}
then the function $H(r)$ satisfies the flat space Laplace
equation. A BPS solution corresponding to a $p$ dimensional
extended object, namely an extremal $p$-brane solution, is
constructed by introducing in the right hand side of the Eq.s of
motion, following from the action (\ref{action}), a
$\delta$-function source term in the transverse directions which
can be obtained from the following boundary action:
\begin{equation}
S_{bound} = - \tau_p \int d^{p+1} \xi e^{a \phi/2} \sqrt{-
 \det G_{\alpha \beta}} + \tau_p \int {{C}}_{p+1}
\label{sbound56}
\end{equation}
where the constant $a$  in $d=10$ is given by:
\begin{equation}
a = \frac{p-3}{2}
\label{aaa95}
\end{equation}
If we restrict ourselves to the simplest case of just one
$p$-brane, we obtain for $H (r)$
\begin{equation}
H(r) = 1 + 2 \kappa T_p G(r) ~~,
\label{sol}
\end{equation}
where
\begin{equation}
G(r) = \left\{ \begin{array}{cc}
 \left[ (d-p -3) r^{(d-p-3)}
\Omega_{d-p-2}\right]^{-1} &  ~~~p < d-3 ~~,\\
  - \frac{1}{2 \pi} \log r  &   ~~~p= d-3 ~~,
\end{array} \right.
\label{solu}
\end{equation}
with
\begin{equation}
\Omega_{q}={ (2 \pi)^{(q+1)/2}}/{ \Gamma \Big((q+1)/2\Big)}
\label{in8}
\end{equation}
being the area of a unit $q$-dimensional sphere $S_q$.
For future use it is convenient to introduce the quantity:
\begin{equation}
Q_p = \mu_p\,\frac{\sqrt{2}\,\kappa\,}{(d-p-3)\,\Omega_{d-p-2}}~~;~~
\mu_p \equiv \sqrt{2} T_p~~;~~\tau_p = \frac{\mu_p}{\sqrt{2} \kappa}.
\label{Qp}
\end{equation}
and (if $p < d-3$) to rewrite $H(r)$ in Eq.(\ref{sol}) as follows:
\begin{equation}
H(x) = 1 +  \frac{Q_p}{r^{d-3-p}}~~,
\label{conv43}
\end{equation}
The classical solution has a mass per unit $p$-volume, $M_p$ and an electric
charge with respect to the R-R field, $\mu_p$, given respectively by
\begin{equation}
M_p = \frac{T_p}{\kappa}~,~ \hspace{1cm} \mu_p = \pm \sqrt{2} T_p~~.
\label{char}
\end{equation}

\sect{D$p$ branes: the open string perspective}
\label{sec3}

The $p$-branes solutions of the low-energy string effective
actions discussed in the previous section have a complementary
description, in the open string framework, as D$p$ branes that is
as hyperplanes on which open string attach their endpoints. The
existence of such hyperplanes is required by the extension of
T-duality, which is a symmetry of (bosonic) closed string theory,
to the open string case. The fundamental observation made by
Polchinski~\cite{POL95} has been to identify the D$p$ branes
appearing in open string theory with the $p$-branes solutions of
the supergravity Eq.s of motion. Let us briefly review how
T-duality enforces the existence of D$p$ branes (for a detailed
discussion see \cite{ISLANDA}). T-duality is the transformation
that interchanges  the winding states with the Kaluza-Klein states
appearing in the  closed string spectrum when the theory is
compactified on a torus. It turns out that the bosonic closed
string theory is invariant under T-duality, while in the
supersymmetric case, this transformation is in general not a
symmetry but brings from a certain string theory to another string
theory.

Let us first discuss the bosonic case. When one space coordinate is
compactified on a circle of
radius $R$, the bosonic closed string mass operator acquires the
following form
\begin{equation}
\label{cmassa}
{ M}^2=\frac{2}{\alpha'}
\left[\sum_{n=1}^{\infty}  \left(\alpha_{-n} \cdot \alpha_n+
\widetilde\alpha_{-n} \cdot \widetilde \alpha_n
\right)-2 \right]
+ \left(\frac{n}{R}\right)^2+ \left(\frac{w R}{\alpha'}\right)^2 ~.
\end{equation}
from which  we see that  the  spectrum of the closed string has been enriched with
respect to the non-compact case by the appearance of two kinds of
particles: the usual K-K modes - coming from the
standard
quantization of the momentum conjugate to the compact direction -
which contribute to the energy as
$ \frac{n}{R}$,
 together with some new excitations that are called
winding modes because they can be thought of as generated by the
winding of the closed string around the compact direction, which in
fact contributes to the energy of the system as
\begin{equation}
\label{winen}
T~ 2\pi R w=\frac{wR}{\alpha'}~,
\end{equation}
where $T = 1/(2 \pi \alpha')$ is the string tension. All previous
formulas can be trivially generalized to the case of a toroidal
compactified theory in which a number of coordinates $X^\ell$ are
compactified on circles with radii $R^{(\ell)}.$

{From} Eq.(\ref{cmassa}) we see that the spectrum of the theory is
invariant under the exchange of KK modes with winding modes together with
an inversion of the radius of compactification:
\begin{equation}
\label{tdua}
w \leftrightarrow n~~~~~~~~~~;~~~~~~~~~~R\leftrightarrow
{\hat R }\equiv\frac{\alpha'}{R}~.
\end{equation}
This is called a T-duality transformation and $\hat R$ is the
compactification radius of the T-dual theory.  It can also be shown that both
the partition function and the correlators are invariant under T-duality.
This means that
 T-duality  is a symmetry of the bosonic closed string theory.
As a consequence, whenever we have to consider compactified
theories, we can limit ourselves to  the case
$R\geq\sqrt{\alpha'}.$ That is the reason why   $\sqrt{\alpha'}$
is often called the  minimal length of string theory.

Consistently with Eq.(\ref{cmassa}) one can also define the action
of T-duality on the string coordinate $X$ as follows (see
\cite{ISLANDA} for details)
\begin{equation}
X_-\rightarrow X_- \,\,\,\,\,\,\,\,\, X_+\rightarrow -X_+
\label{dut}
\end{equation}
where we have written
\begin{equation}
\label{embleri}
X =\frac{1}{2}\left( X_- + X_+ \right),
\end{equation}
with
\begin{equation}
\label{xmur}
X_- = q + 2 \sqrt{2\alpha'}(\tau-\sigma)\alpha_0
+i\sqrt{2 \alpha'}\sum_{n\neq 0}\frac{\alpha_n}{n}
~e^{-2in(\tau-\sigma)}~~,
\end{equation}
and
\begin{equation}
\label{xmul}
X_+ = q + 2\sqrt{2\alpha'}(\tau+\sigma)\widetilde\alpha_0
+i\sqrt{ 2 \alpha'}\sum_{n\neq 0}\frac{\widetilde\alpha_n}{n}
~e^{-2in(\tau+\sigma)}~~,
\end{equation}
Therefore the T-duality transformation acts on the right sector as a parity
operator changing sign of the right moving coordinate $X_+$
and leaving unchanged the left moving one $X_-$.

In an open string theory, even in its compactified version, there
are only K-K modes, while the winding modes are absent. This could
suggest that T-duality is not a symmetry of the open string
theory. Such a conclusion, however, is not satisfactory because
theories with open strings also contain closed strings! Therefore
if $d-p-1$ directions are compactified on circles with radii
$R^{\ell}$, performing the limits $R^{\ell}\rightarrow 0$ one
would end with  open strings living in a $p+1$-dimensional
subspace of the entire space-time, and closed strings in the
entire $d$-dimensional target space. Indeed in that limit the open
string sector would keep no trace of the compact dimensions,
losing effectively $d-p-1$ directions, because all the K-K modes
become infinitely massive and decouple from the spectrum. Instead
in the closed sector, even if the K-K modes decouple, the winding
modes would not disappear giving a continuum of states and,
through a T-duality transformation, one could completely restore
all the $d$ space-time dimensions, as a consequence of the fact
that in the limit $R^{\ell}\rightarrow 0$ the T-dual radii go to
infinity. This mismatch can be solved by requiring that, in the
T-dual picture,  open string still can oscillate in $d$
dimensions, while their endpoints are fixed on a $p+1$-dimensional
hyperplane that we call {\dpb}. In this scenario open strings
satisfy Dirichlet boundary conditions in the $d-p-1$ transverse
directions. These are allowed boundary conditions, as we have
already seen in Eq.(\ref{neudic}), although they destroy the
Poincar{\`{e}} invariance of the theory. In conclusion, in order
to avoid a different behavior between the closed and the open
sector of  string theory, we must require the action of T-duality
on an open string theory to be that of transforming Neumann
boundary conditions into Dirichlet ones. This can, in fact, be
very naturally obtained if we extend the definition of the T-dual
coordinate given in Eq.(\ref{dut}) to the open string case. In
this way we obtain the following T-dual open string coordinate:
\begin{equation}
\label{ducorr}
{\hat X}^{\ell} = \frac{1}{2}\left[ X_{-}^{\ell} -X_{+}^{\ell} \right]~,
\end{equation}
where now the left and right movers contain the same set of oscillators
\begin{equation}
\label{xmuro}
X_{-}^{\ell} = q^\ell +  c^\ell
+\sqrt{ 2 \alpha'}(\tau-\sigma)\alpha^\ell_0
+i\sqrt{2\alpha'}\sum_{n\neq 0}\frac{\alpha^\ell_n}{n}
~e^{-in(\tau-\sigma)}~~,
\end{equation}
and
\begin{equation}
\label{xmulo}
X_{+}^{\ell} = q^\ell - c^\ell +
\sqrt{2 \alpha'}(\tau+\sigma)\alpha^\ell_0
+i\sqrt{2\alpha'}\sum_{n\neq 0}\frac{\alpha^\ell_n}{n}
~e^{-in(\tau+\sigma)}~~,
\end{equation}
{From} Eq.s (\ref{ducorr}), (\ref{xmuro}) and (\ref{xmulo}) one
can immediately see that T-duality has transformed a string
coordinate satisfying Neumann boundary conditions and given by
$1/2 \left[X_{-}^{\ell} + X_{+}^{\ell}
 \right]$ into a  T-dual one satisfying Dirichlet boundary conditions and given
by Eq.(\ref{ducorr}).

In superstring theory the effect of  T-duality on
the bosonic coordinates is exactly the same as discussed for the bosonic
string, namely  T-duality acts as a parity transformation over the tilded
 sector,
while for the fermionic coordinates the transformations under T-duality
can be fixed by requiring the superconformal invariance of the theory
which imposes
\begin{equation}
\label{tduafer}
\psi_+\rightarrow -\psi_+~~~~;~~~~\psi_-\rightarrow \psi_-~,
\end{equation}
or in terms of the oscillators
\begin{equation}
\label{tduafero}
{\widetilde{\psi}}_t\rightarrow -\widetilde\psi_t~~~~;~~~~\psi_t\rightarrow
\psi_t~,
\end{equation}

We end this section by giving the spectrum of open superstrings having
their end-points attached to a Dp-brane. This is given by the
following formula:
\begin{equation}
\alpha' k_{||}^2 + \sum_{n=1}^{\infty} n a^{\dagger}_{n} \cdot a_n +
\sum_t t \psi^{\dagger}_{t} \cdot \psi_t -a =0
\label{speope}
\end{equation}
where $a = \frac{1}{2} [0]$ in the NS [R] sector and $k_{||}$ is the
momentum of the string parallel to the brane. In particular the
massless states in the NS sector are given by $ ( \psi_{-1/2}^{\alpha},
\psi_{-1/2}^{i}) | 0, k \rangle$ corresponding to a gauge boson $A_\alpha$  and
to $(9-p)$  Higgs scalars $\Phi^i$ related to the translational modes
of the brane along the directions transverse to its
world-volume. These gauge and scalar fields living on the world-volume
of a Dp-brane become non-abelian transforming all of them according to
the adjoint representation of the gauge group if instead of a single
Dp-brane we have a bunch of $N$ coincident Dp-branes. In this case in
fact we get  $N^2$ massless states
corresponding to the fact that the open strings can have their
end-points on each of the $N$ branes.

\sect{Boundary State}
\label{sec4}

The interaction between two D$p$ branes is given by the vacuum
fluctuation of an open string stretching between them. This is
similar to what happens in the Casimir effect, where the
interaction between two superconducting plates is obtained by
computing the vacuum fluctuation of the electromagnetic field, due
to the presence of the boundary plates. Thus D$p$ brane
interaction is simply given by the one-loop open string
"free-energy" which is usually represented by the annulus.
Furthermore the same interaction admits a complementary
description in the closed string language. Indeed, by exchanging
the variables $\sigma$ and $\tau$, the one-loop open string
amplitude can also be viewed as a tree diagram of a closed string
created from the vacuum, propagating for a while and then
annihilating again into the vacuum. These two equivalent
descriptions of the same diagram are called respectively the
`open-channel' and the `closed-channel' and the relation between
the two description is called open/closed string duality. We want
to stress that the physical content of the two descriptions is a
priori completely different. In the first case we describe the
interaction between two Dp branes as a one-loop amplitude of open
strings, which is the amplitude of a quantum theory of open
strings, while in the second case we describe the same interaction
as a tree-level amplitude of closed strings, which is instead a
classical amplitude in a theory of closed strings. The fact that
these two descriptions are equivalent is a consequence of the
conformal symmetry of string theory that allows one to connect the
two apriori different descriptions.

To show that, let us consider a one-loop diagram with  an open
string circulating in it and stretching between two parallel D$p$
branes with coordinates respectively $(y^{p+1},...,y^{d-1})$ and
$(w^{p+1},...,w^{d-1})$. The open string satisfies Neumann
boundary conditions  along the directions longitudinal to the
branes both at $\sigma=0 $ and $\sigma = \pi$
\begin{equation}
\partial_{\sigma} X^{\alpha}|_{\sigma=0,\pi} =0 \hspace{2cm} \alpha=0, 1, ...., p \label{neu1}~,
\end{equation}
while along the transverse directions one has
\begin{equation}
\label {bc2} X^i|_{\sigma=0}=y^i~~~~~~~~~~
X^i|_{\sigma=\pi}=w^i~~~~~ i = p+1,..., d-1~,
\end{equation}
where we take $\sigma$ and $\tau$ in the two intervals
$\sigma\in[0,\pi]$ and $\tau\in [0, T].$ There is a conformal
transformation acting on the previous open string boundary
conditions which transforms them into the boundary conditions for
a closed string propagating between the two D$p$ branes. In terms
of the complex coordinate $\zeta\equiv\sigma+i\tau,$ this
transformation reads \beq \label {ctz} \zeta = \sigma + i \tau
\rightarrow -i\zeta=\tau - i\sigma~, \eeq or equivalently \beq
\label{ctst} (\sigma, \tau)\rightarrow (\tau,-\sigma)~. \eeq In
order to have the closed string variables $\sigma$ and $\tau$ to
vary in the intervals $\sigma\in[0, \pi]$ and  $\tau \in [0,
\hat{T}]$ one can exploit conformal invariance, once more,
performing the following rescaling \beq \label{res} \sigma
\rightarrow \frac{\pi}{T}\sigma ~~~~~~~~~~ \tau\rightarrow
\frac{\pi}{T}\tau~, \eeq where we have defined \beq \label{res2}
\hat{T}=-\pi^2/T\,\,. \eeq From the previous equations it follows
that {\it a loop of an open string propagating through the proper
time $T$ is conformally equivalent to a tree-level amplitude of a
closed string which propagates through the proper time $\hat T\sim
1/T$}.

In the closed string channel we need to construct the two boundary
states $|B_X \rangle $ that describe the two D$p$ branes
respectively at $\tau=0$ and $\tau ={\hat{T}}$. The equations that
characterize these states are obtained by applying the conformal
transformation previously constructed to the boundary conditions
for the open string given in Eq.s (\ref{neu1}) and (\ref{bc2}). At
$\tau =0$ we get the following conditions:
\begin{equation}
\label {bc1c}
\partial_{\tau}X^\alpha|_{\tau=0}|B_X \rangle =0 ~~~~~~~~~~\alpha =0,...,p~,
\end{equation}
\begin{equation}
\label {bc2c} X^i|_{\tau=0}|B_X \rangle =y^i ~~~~~~~~~~i =
p+1,..., d-1~.
\end{equation}
Analogous conditions can be obtained for the D$p$ brane at $\tau =
{\hat{T}}$.

The previous equations can be easily written in terms of the
closed string oscillators by making use of the expansion in
Eq.(\ref{expc}), obtaining
\begin{equation}
\label{over1} (\alpha_n^\alpha+\widetilde\alpha_{-n}^\alpha)|B_X
\rangle =0~~;~~ (\alpha_n^i-\widetilde\alpha_{-n}^i)|B_X \rangle
=0~~\forall n\neq 0
\,\,\,\,\,\,
{\hat{p}}^\alpha|B_X \rangle  = 0 ~~~~~({\hat{q}}^i-y^i)|B_X
\rangle =0~.
\end{equation}
Introducing the matrix
\begin{equation}
\label{matS} S^{\mu\nu}=(\eta^{\alpha\beta},-\delta^{ij})~,
\end{equation}
it is easy to see that the state satisfying the previous equations
is
\begin{equation}
\label{b1} |B_X \rangle  =\frac{T_p}{2} \delta^{d-p-1}({\hat
q}^i-y^i) \left(\prod_{n=1}^\infty e^{-\frac{1}{n} \alpha_{-n}
S\cdot\widetilde\alpha_{-n}}\right)|0\rangle _{\alpha}|0\rangle
_{\widetilde\alpha} |p=0\rangle ~,
\end{equation}
where
\begin{equation}
\label {Tp} T_p=\frac {{\sqrt \pi}}{2^{\frac{d-10}{4}}}
(2\pi\sqrt\alpha')^{\frac{d}{2}-2-p} ~.
\end{equation}
The normalization of the boundary state $T_p/2$ can be fixed by
computing the interaction between two parallel D$p$ branes both in
the open and in the closed string channel and comparing the two
results. See Ref.~\cite{ISLANDA} for details.

In the superstring case, together with the bosonic boundary state
$|B_X \rangle $ one also has a fermionic component $|B_{\psi}
\rangle $ which can be constructed by performing the conformal
transformation, which brings from the open to the closed channel,
on the boundary conditions for an open superstring stretching
between two D$p$ branes .

In Eq. (\ref{fbcu1}) we have given the fermionic boundary conditions
of an open superstring corresponding to the case in which the
bosonic degrees of freedom satisfy Neumann boundary conditions in
all directions. If the bosonic coordinate satisfies Dirichlet
boundary conditions in some of the directions, those boundary
conditions are changed as follows:
\begin{equation}
\label{fbcuS} \left\{
\begin{array}{l}
\psi^\mu_-(0,\tau)=\eta_1 S^{\mu}_{\,\,\,\nu} {{\psi}_+}^{\nu} (0,\tau)\\
\psi^\mu_-(\pi,\tau)=\eta_2 S^{\mu}_{\,\,\,
\nu}{{\psi}_+}^{\nu}(\pi, \tau)
\end{array}
\right.
\end{equation}
where the matrix $S$ has been defined in Eq.(\ref{matS}). This can
be easily understood using T-duality. Indeed  T-duality transforms
Neumann into Dirichlet boundary conditions for the bosonic
coordinate and, as discussed in sect.~\ref{sec3}, changes the sign
of the fermionic coordinate in the right sector leaving that of
the left sector unchanged. Moreover we must also give the
periodicity or anti periodicity conditions for the fermionic
degrees of freedom in going around the loop. These are chosen to
be
\begin{equation}
\label{fbcp} \left\{
\begin{array}{l}
\psi_-(\sigma,0)=\eta_3 \psi_-(\sigma, T)\\
{\psi}_+(\sigma,0)=\eta_4{\psi}_+(\sigma, T)
\end{array}
\right.
\end{equation}
where $\eta_3$ and $\eta_4$ can take the values $\pm 1$. From the
boundary conditions in Eq.s (\ref{fbcuS}) and (\ref{fbcp}) we get
\begin{equation}
\label {eta34} \psi^\mu_-(0,0)=\eta_1
S^{\mu}\,_{\nu}\psi^\nu_+(0,0)= \eta_1 \eta_4 S^{\mu}_{\,\,\, \nu}
{{\psi}_+}^\nu (0, T)
\end{equation}
and
\begin{equation}
\label{eta342} \psi^\mu_-(0,0)=\eta_3  \psi^\mu _-(0, T)= \eta_3
\eta_1 S^{\mu}_{\,\,\, \nu} {{\psi}_+}^\nu(0, T)
\end{equation}
The two set of boundary conditions in Eq.s (\ref{fbcuS}) and
(\ref{fbcp}) must be consistent with each other, thus
$\eta_3=\eta_4$.

In order to pass to the closed string channel, one has to take
into account that the right and left fermionic coordinates
$\psi_-$ and ${\psi}_+$ are two-dimensional conformal fields with
conformal weight $h=\frac{1}{2}$ with respect to the
variables $\zeta$ and $\bar\zeta $ respectively and then, under
the conformal transformation (\ref{ctz}), they  transform as
\begin{equation}
\psi_-(\zeta)\rightarrow\psi'_-(\zeta)= (-i)^{\frac{1}{2}} \psi_-
(f(\zeta)) \label{ii}
\end{equation}
and
\begin{equation}
\label {iii} \psi_+(\bar\zeta)\rightarrow\psi'_+(\bar\zeta)
=(i)^{\frac{1}{2}} \psi_+ (\bar f(\bar\zeta))
\end{equation}
This implies that, performing the previous transformation on
Eq.(\ref{fbcuS}), there is  a relative factor $i$ appearing
between the right and left modes, which transforms the boundary
conditions (\ref{fbcuS}) and (\ref{fbcp}) in
\begin{equation}
\label{fbcucl} \left\{
\begin{array}{l}
\psi^\mu_-(0,\sigma)=i\eta_1S^{\mu}_{\,\,\,\nu}{\psi^\nu_+}(0,\sigma)\\
\psi^\mu_-({\hat{T}}, \sigma)=i\eta_2S^{\mu}_{\,\,\, \nu}
{\psi^\nu_+}({\hat{T}}, \sigma)
\end{array}
\right.
\end{equation}
and
\begin{equation}
\label{fbcpcl} \left\{
\begin{array}{l}
\psi^\mu_-(0,\tau )=\eta_3\psi^\mu_-(\pi,\tau)\\
{\psi^\mu_+}(0,\tau)=\eta_3{\psi^\mu_+}(\pi,\tau)
\end{array}
\right.
\end{equation}
where we have explicitly put $\eta_4=\eta_3$. The identity between
$\eta_3$ and $\eta_4,$ implies that the fermionic boundary state
has only the R-R  and the NS-NS sectors. {From} the first equation
in (\ref{fbcucl}), one can derive the overlap Eq.s for the
fermionic boundary state:
\begin{equation}
\label {psib} (\psi^\mu_-(0,\sigma)-i\eta S^{\mu}_{\,\,\, \nu}
{\psi^\nu_+}(0,\sigma)) |B_{\psi}, \eta \rangle  =0
\end{equation}
where $\eta = \pm 1$, which, in the case of the NS-NS-sector, are
satisfied by
\begin{equation}
\label{bnsns} |B_{\psi} , \eta \rangle  = -i \prod_{t=1/2}^\infty
\left(e^{i\eta\psi_{-t}\cdot S\cdot \widetilde \psi_{-t}} \right)
|0\rangle
\end{equation}
In the R-R sector the boundary state has the same form as in the
NS-NS sector for what the non-zero modes is concerned, but with
integer instead of half-integer modes. We get therefore
~\footnote{The unusual phases introduced in Eq.s (\ref{bnsns}) and
(\ref{brr}) will turn out to be convenient to study the couplings
of the massless closed string states with a D-brane and to find
the correspondence with the classical D-brane solutions obtained
from supergravity. Note that these phases are instead irrelevant
when one computes the interactions between two D-branes.}
\begin{equation}
\label{brr} |B_{\psi} , \eta \rangle  =-\prod_{t=1}^\infty
e^{i\eta\psi_{-t} \cdot S\cdot\widetilde \psi_{-t}} |B_{\psi} ,
\eta \rangle ^{(0)}
\end{equation}
where the zero mode contribution $|B_{\psi}, \eta \rangle ^{(0)}$
is given by
\begin{equation}
\label{solover} |B_\psi , \eta\rangle ^{(0)} = {\cal
M}_{AB}|A\rangle |\widetilde B\rangle
\end{equation}
with
\begin{equation}
\label{soloverM} {\cal
M}_{AB}=\left(C\Gamma^0...\Gamma^p\frac{1+i\eta\Gamma^{11}}{1+i\eta}
\right)_{AB}
\end{equation}
$C$ is the charge conjugation matrix and $\Gamma^\mu$ are the
Dirac $\Gamma$ matrices in the $10$-dimensional space (see
Ref.~\cite{cpb} for some detail about the derivation of Eq.s
(\ref{solover}) and (\ref{soloverM})).

The boundary state discussed until now describes only the degrees
of freedom corresponding to the string coordinate $X$ and $\psi$.
In order to have a BRST invariant object we have to
supplement it with a component describing  the ghost and
superghosts degrees of freedom.
The complete boundary state for both the NS-NS and R-R sectors is
given by:
\begin{equation}
|B, \eta \rangle _{R,NS} =  |B_{mat}, \eta \rangle  | B_{g}, \eta
\rangle \label{bounda3}
\end{equation}
where
\begin{equation}
|B_{mat} \rangle  = |B_X \rangle  |B_{\psi}, \eta \rangle
\hspace{1cm};\hspace{1cm} |B_{g}\rangle  = |B_{gh} \rangle  |
B_{sgh}, \eta \rangle \label{bounda4}
\end{equation}
The matter part of the boundary state consists of the bosonic
component given in Eq.(\ref{b1}) and of the fermionic one  given
in Eq.(\ref{bnsns}) for the NS-NS sector and in Eq.(\ref{brr}) for
the R-R sector. The ghost part $|B_g \rangle $ contains the
boundary state corresponding to the ghosts $(b,c)$ and the one
corresponding to the superghosts $(\beta, \gamma )$. BRST
invariance requires that the total boundary state satisfies the
equation
\begin{equation}
\label{BRST} (Q+\widetilde Q)|B,\eta \rangle =0~,
\end{equation}
where the BRST charge has been given in Eq.s
(\ref{carS})-(\ref{bosq62}). With some algebra one can show that
the ghosts and superghosts boundary states satisfying
Eq.(\ref{BRST}) are
\begin{equation}
\label{bgh} |B_{gh}\rangle= e^{\sum_{n=1}^\infty
(c_{-n}{\widetilde b}_{-n}- b_{-n}{\widetilde c}_{-n})}
\left(\frac {c_0+{\widetilde c}_0}{2}\right)|q=1\rangle
|{\widetilde{q=1}} \rangle
\end{equation}
where $|q=1\rangle $ is the state that is annihilated by the
following oscillators
\begin{equation}
\label{q1} c_n|q=1\rangle =0~~~~~\forall n\geq 1;~~~~~;~~~~~~
b_m|q=1\rangle =0~~~~~\forall m\geq 0~.
\end{equation}
and
\begin{equation}
\label{bs8} \ket{B_\sgh,\eta}_{\rm NS} =
\exp\biggl[\ii\eta\sum_{t=1/2}^\infty(\gamma_{-t} \tilde\beta_{-t}
- \beta_{-t}
  \tilde\gamma_{-t})\biggr]\,
  \ket{P=-1}\,\ket{\tilde P=-1}~,
\end{equation}
in the NS sector in the picture $(-1,-1)$ and
\begin{equation}
\label{bs10} \ket{B_\sgh,\eta}_\R = \exp\biggl[
\ii\eta\sum_{t=1}^\infty(\gamma_{-t} \tilde\beta_{-t} - \beta_{-t}
\tilde\gamma_{-t})\biggr]\,
 \ket{B_\sgh,\eta}_\R^{(0)}~~,
\end{equation}
in the R sector in the $(-1/2,-3/2)$ picture. The superscript
$^{(0)}$ denotes the zero-mode contribution that, if $\ket{P=-{1/
2}}\,\ket{\tilde P=-{3/ 2}}$ denotes the superghost vacuum  that
is annihilated by $\beta_0$ and $\tilde \gamma_0$, and is given
by~\cite{YOST}
\begin{equation}
\label{bsrsg0} \ket{B_\sgh,\eta}_\R^{(0)} =
\exp\left[\ii\eta\gamma_0\tilde\beta_0\right]\,
  \ket{P=-{1/ 2}}\,\ket{\tilde P=-{3/ 2}}~~.
\end{equation}
We would like to stress that the boundary states
$\ket{B}_{\NS,\R}$ are written in a definite picture $(P,\tilde
P)$ of the superghost system, where $P$ is given in
Eq.(\ref{pict}) and $\tilde P= -2 -P$ in order to soak up the
anomaly in the superghost number. In particular we have chosen
$P=-1$ in the NS sector and $P=-1/2$ in the R sector, even if
other choices would have been in principle possible \cite{YOST}.
Since $P$ is half-integer in the R sector, the boundary state
$\ket{B}_{\R}$ has always $P\not=\tilde P$, and thus it can couple
only to R-R states in the asymmetric picture $(P,\tilde P)$.
Notice that, as we have seen in section~\ref{sec1} the massless
R-R states in the $(-1/2,-3/2)$ picture contain a part that is
proportional to the R-R {\it potentials}~\cite{sagnotti,BILLO}, as
opposed to the standard massless R-R states in the symmetric
picture $(-1/2,-1/2)$ that are always proportional to the R-R
field strengths. This implies that the coupling of the boundary
state with the R-R fields is expressed in terms of the potentials.

The boundary state in Eq.(\ref{bounda3}) depends on the value of
$\eta$. Actually the GSO projection selects a specific combination
of the two values of $\eta=\pm 1$. In the
NS-NS sector the GSO projected boundary state is
\begin{equation}
\label{bs22a} \ket{B}_\NS  \equiv  {1 +(-1)^{F+G}\over 2}\,\, {1
+(-1)^{\widetilde F+\widetilde G}\over 2}\, \ket{B,+}_\NS ~~,
\end{equation}
where $F$ and $G$ are the fermion and superghost number operators
\begin{equation}
{F} = {\sum_{m=1/2}^{\infty}\psi_{-m} \cdot \psi_m}-1 ~~,~~ {G} =
{- \sum_{m=1/2}^{\infty} \left( \gamma_{-m}  \beta_m + \beta_{-m}
\gamma_m \right)} ~~. \label{fergh}
\end{equation}
Their action on the boundary state corresponding to the fermionic
coordinate $\psi$ and to the superghosts can easily be computed
and one gets:
\begin{equation}
(-1 )^{F} |B_{\psi} , \eta\rangle  = - |B_{\psi} , -\eta\rangle
~~~~;~~~~ (-1 )^{{\widetilde{F}}} |B_{\psi} , \eta\rangle  = -
|B_{\psi} , -\eta\rangle \label{acti3}
\end{equation}
\begin{equation}
(-1 )^{G} |B_{sgh} , \eta\rangle  =  |B_{sgh} , -\eta\rangle ~~~~;
~~~~ (-1 )^{{\widetilde{G}}} |B_{sgh} , \eta\rangle  =  |B_{sgh} ,
-\eta\rangle \label{acti4}
\end{equation}
Using the previous expressions after some simple algebra we get
\begin{equation}
\label{bs22ab} \ket{B}_\NS = {1\over 2} \Big( \ket{B,+}_\NS -
\ket{B,-}_\NS \Big)
\end{equation}
Passing to the R-R sector the GSO projected boundary state is
\begin{equation}
\label{bs22ba} \ket{B}_\R  \equiv  {1 +(-1)^{p} (-1)^{F+G}\over
2}\,\, {1 - (-1)^{\widetilde F+\widetilde G}\over 2}\,
\ket{B,+}_\R ~~.
\end{equation}
where $p$ is even for Type IIA and odd for Type IIB, and
\begin{equation}
(-1)^{F} = \psi_{11}(-1)^{\sum\limits_{m=1}^{\infty}\psi_{-m}
\cdot \psi_m} ~~,~~ {G} = - \gamma_0 \beta_0 -
\sum\limits_{m=1}^{\infty} \left[ \gamma_{-m} \beta_m +
\beta_{-m}\gamma_m \right]~~. \label{fermion}
\end{equation}
{From} the previous expressions it is easy to see after some calculation that
the action of the fermion number operators is given by:
\begin{equation}
(-1)^{F} |B_{\psi}, \eta \rangle  = (-1)^p |B_{\psi}, - \eta
\rangle \hspace{.5cm}; \hspace{.5cm} (-1)^{{\widetilde{F}}}
|B_{\psi}, \eta \rangle  =  |B_{\psi}, - \eta \rangle \label{}
\end{equation}
and
\begin{equation}
(-1)^{G} |B_{sgh}, \eta \rangle  =   |B_{sgh}, - \eta \rangle
\hspace{.5cm};\hspace{.5cm} (-1)^{{\widetilde{G}}} |B_{sgh}, \eta
\rangle  = -  |B_{sgh}, - \eta \rangle \label{fernnu3}
\end{equation}
Using the previous expressions after some straightforward
manipulations, one gets
\begin{equation}
\label{bs22bb} \ket{B}_\R  =
    {1\over 2} \Big( \ket{B,+}_\R + \ket{B,-}_\R\Big)~~.
\end{equation}

\sect{Interaction Between D$p$ branes}
\label{sec5}

In this section we study the static interaction between two
Dp branes located at a distance $y$ from each other. Then
we will generalize it to the interaction between a D$p$ and a D$p'$
brane, with $NN\equiv{\rm min} \{p,p'\} +1$ directions common to
the brane world-volumes, $DD\equiv {\rm min}\{d-p-1,d-p'-1\}$
directions transverse to both, and $\nu = (d -NN -DD)$ directions
of mixed type. We will not consider instantonic D-branes, hence
also $NN\geq 1$.
The two D-branes simply interact via tree-level
exchange of closed strings with propagator
\begin{equation}
D= \frac{\alpha '}{4 \pi} \int d^2 z z^{L_0 - a} {\bar{z}}^{{\tilde{L}}_0 -a}
\label{DDD}
\end{equation}
where $a=1/2$ $(0)$ in the NS-NS (R-R) sector. Introducing the two
boundary states $\ket{B_1}$ and $\ket{B_2}$ describing the two
D-branes, the static amplitude is given by
\begin{equation}
  \label{bs32}
 A = \bra{B_1}~D~\ket{B_2}=\frac{T_p^2}{4}
\frac{\alpha'}{4\pi}\int_{|z|\langle1}
\frac{d^2z}{|z|^2}{\cal A}\,{\cal A}^{(0)}~~,
\end{equation}
where we have indicated with ${\cal A}$ and  ${\cal A}^{(0)}$
respectively the non zero mode and the zero mode contribution in
which the previous amplitude can be factorized. Starting from the
non-zero modes we have to evaluate amplitudes of the form
\[
\langle 0| \langle \tilde 0|~\prod_{m=1}^\infty \left( e^{-
\frac{1}{m}\alpha_m\cdot S\cdot\widetilde\alpha_m} \right)
z^{N_\alpha} \bar z^{\widetilde N_\alpha}\prod_{n=1}^\infty \left(
e^{-\frac{1}{n} \alpha_{-n} \cdot S \cdot
{\widetilde{\alpha}}_{-n}} \right) |0\rangle|\tilde 0\rangle=
\]
with
\begin{equation}
\label{N} N_\alpha\equiv\sum_{n=1}^{\infty}  \alpha_{-n} \cdot
\alpha_n~~~~~;~~~~~ \widetilde N_\alpha\equiv\sum_{n=1}^{\infty}
{\widetilde\alpha}_{-n} \cdot {\widetilde\alpha}_n~,
\end{equation}
for the bosonic degrees of freedom and
\[
\langle 0|\langle \tilde 0|~\prod_{t}^\infty \left(
e^{i\eta_1\psi_t\cdot S\cdot\widetilde\psi_t} \right)z^{N_\psi}
\bar z^{\widetilde N_\psi}\prod_{t}^\infty \left(
e^{-i\eta_2\psi_{-t} \cdot S \cdot {\widetilde{\psi}}_{-t}}
\right) |0\rangle|\tilde 0\rangle
\]
with
\begin{equation}
\label{Npsi} N_\psi\equiv\sum_{t}  t\psi_{-t} \cdot
\psi_t~~~~~;~~~~~ \widetilde N_\psi\equiv\sum_t
{t\widetilde\psi}_{-t} \cdot {\widetilde\psi}_t~,
\end{equation}
for the fermionic ones, where $t\geq 1/2$ $(1)$ in the NS-NS (R-R)
sector. Using that
\begin{equation}
\label{zN} z^{N_\alpha}
e^{\alpha_{-n}}z^{-N_\alpha}=e^{\alpha_{-n} z^n}~~~~~{\rm
and}~~~~~ {\bar z}^{N_\alpha} e^{\alpha_{-n}}{\bar z}^{-N_\alpha}=
e^{\alpha_{-n} {\bar z}^n}~~~~~\forall n \neq 0~.
\end{equation}
and analogous expressions involving fermionic operators, one can
explicitly evaluate the contractions among the oscillators getting
the following contributions (for $d=10$)
\begin{equation}
\label{serb} {\rm bosons}\longrightarrow
\prod_{n=1}^\infty\left(\frac{1}{1-q^{2n}}\right)^{8}
\end{equation}
\begin{equation}
{\rm fermions}\longrightarrow
\label{serf}
\prod_{n=1}^\infty\left({1+\eta_1\eta_2 q^{2n-1}}\right)^{8}~.
\end{equation}
with $q = |z|={\it e}^{- \pi t}$. Introducing the functions $f_i$
defined as
\begin{equation}
\label{f12} f_1\equiv q^{{1\over 12}} \prod_{n=1}^\infty (1 -
q^{2n})~~~~;~~~~ f_2\equiv \sqrt{2}q^{{1\over 12}}
\prod_{n=1}^\infty (1 + q^{2n})~~;
\end{equation}
\begin{equation}
\label{f34} f_3\equiv q^{-{1\over 24}} \prod_{n=1}^\infty (1 +
q^{2n -1}) ~~~~;~~~~ f_4\equiv q^{-{1\over 24}} \prod_{n=1}^\infty
(1 - q^{2n -1}) ~~,
\end{equation}
which under the modular transformation $t\rightarrow 1/t$
transform as
\begin{equation}
\label{mtfi} f_1(e^{-\frac{\pi}{t}})=\sqrt{t}f_1(e^{-\pi t})~~;~~
f_2(e^{-\frac{\pi} {t}})=f_4(e^{-\pi t})~~; f_3(e^{-\pi
t})=f_3(e^{-\frac{\pi}{ t}})~,
\end{equation}
the GSO projected NS-NS amplitude turns out to be
\begin{equation}
\label{ansnz0} {\cal A}_{\rm{NS-NS}}= \frac{1}{2} \left[
\left({f_3\over f_1}\right)^{8} - \left({f_4\over
f_1}\right)^{8}\right]~~,
\end{equation}
while in the R-R sector, before the GSO projection, we get
\begin{equation}
\label{arrnz} {\cal A}_{\rm{R-R}}(\eta_1,\eta_2) =
 \left[ \frac{1}{16}
\left(\frac{f_2}{f_1}\right)^{8}
\delta_{\eta_1\eta_2,1}+\delta_{\eta_1\eta_2,-1}\right]~,
\end{equation}
Let us discuss now the zero modes contribution. In the NS-NS
sector there are zero modes only in the bosonic sector and they
contribute as follows:
\begin{equation}
\label{zp} \langle p=0|\delta^{d_\perp}(\hat
q_i)|z|^{\frac{\alpha'}{2}{\hat p}^2} \delta^{d_\perp}(\hat
q_i-y_i)|p=0\rangle=V_{p+1}\int\frac{d^{d_\perp}Q}{(2\pi)^{d_\perp}}|z|^{
\frac{\alpha'}{2}{Q}^ 2} e^{ iQ \cdot y}~,
\end{equation}
where the normalization for the momentum  has been chosen as
\begin{equation}
\label{knorm} \langle k|k'\rangle=2\pi \delta(k-k')~,~~~~~ {\rm
with}~~~~~~~~ (2\pi)^d\delta^d(0)\equiv V_d~.
\end{equation}
Performing the gaussian integral, Eq. (\ref{zp}) becomes
\begin{equation}
\label{zpp} V_{p+1} e^{-y^{2}/(2 \pi \alpha' t)} \left( 2 \pi^2 t
\alpha' \right)^{-d_{\perp}/2}~~~ .
\end{equation}
Inserting  it, together with Eq.s (\ref{ansnz0}) in
Eq.(\ref{bs32}) we get the total NS-NS contribution to the
interaction between two D$p$ branes
\begin{equation}
\label{ansns1} A_{\rm{NS-NS}} = \frac{V_{p+1}}{2}\,
(8\pi^2\alpha')^{-{(p+1)\over 2}} \int_0^\infty
\frac{dt}{{t}^{(9-p)\over 2}}\,{\rm e}^{- y^2 /(2\alpha' \pi
t)}\left[ \left({f_3\over f_1}\right)^{8}
  - \left({f_4\over f_1}\right)^{8}\right]~~,
\end{equation}
The evaluation of the zero mode contribution in the R-R sector
requires more care due to the presence of zero modes both in the
fermionic matter fields and the bosonic superghosts. Inserting Eq.
(\ref{arrnz}) into Eq. (\ref{bs32}) we can write the total R-R
contribution as
\[
A_{\rm{R-R}}(\eta_1,\eta_2) =
 V_{p+1}\, (8\pi^2\alpha')^{-{(p+1)\over 2}}\int_0^\infty {dt} \left( 1\over t \right)^{(9-p)\over 2}\, {\rm
e}^{- y^2 /(2\pi \alpha' t)}
\]
\begin{equation}
\times \left[\frac{1}{16} \left(\frac{f_2}{f_1}\right)^{8}
\delta_{\eta_1\eta_2,+1}+\delta_{\eta_1\eta_2,-1}\right]
~{}_\R^{(0)}\!\langle B^1,\eta_1 | B^2,\eta_2\rangle_\R^{(0)}~~,
\label{arr1}
\end{equation}
where
\begin{equation}
\ket{B,\eta}_\R^{(0)} =
\ket{B_\psi,\eta}_\R^{(0)}~\ket{B_\sgh,\eta}_\R^{(0)}~~.
\label{bsr00}
\end{equation}
Notice that in \eq{arr1} it is essential {\it not} to separate the
matter and the superghost zero-modes. In fact, a na{\"{\i}}ve
evaluation of ${}_\R^{(0)}\!\langle B^1,\eta_1 |
B^2,\eta_2\rangle_\R^{(0)}$ would lead to a divergent or ill
defined result: after expanding the exponentials in
${}_\R^{(0)}\!\langle B^1_\sgh,\eta_1 |
B^2_\sgh,\eta_2\rangle_\R^{(0)}$, all the infinite terms with any
superghost number contribute, and yield the divergent sum
$1+1+1+...$ if $\eta_1\eta_2=-1$, or the alternating sum
$1-1+1-...$ if $\eta_1\eta_2=1$. This problem has been addressed
in Ref.~\cite{YOST} and solved by introducing a regularization
scheme for the pure Neumann case ($NN=10$). This method has then
been extended to the most general case with D-branes in
Ref.~\cite{BILLO}. Here, we give the final result for the
fermionic zero mode part of the R-R sector:
\begin{equation}
{}_\R^{(0)}\!\langle B^1,\eta_1 | B^2,\eta_2\rangle_\R^{(0)}  =
-16 \delta_{\eta_1\eta_2,1} \label{result3n}
\end{equation}
which generalizes to the case of $\nu\neq 0$ as follows
\begin{equation}
{}_\R^{(0)}\!\langle B^1,\eta_1 |
B^2,\eta_2\rangle_\R^{(0)}  =
-16 \,\delta_{\nu,0}\,\delta_{\eta_1\eta_2,1} + 16 \,
\delta_{\nu,8}\,\delta_{\eta_1\eta_2,-1}~~.
\label{result3}
\end{equation}
Then we get the following expression for the R-R contribution
\begin{equation}
 A_{\rm{R-R}}
= V_{p+1}  (8\pi^2\alpha')^{-{(p+1)\over 2}} \cdot \int_0^\infty
{dt} \left(1\over t \right)^{(9-p)\over 2}\, {\rm e}^{-  y^2 /(2
\pi \alpha' t)} ~\frac{1}{2} \left[
-\left(\frac{f_2}{f_1}\right)^{8}\right]~~. \label{arr30}
\end{equation}
Finally the previous amplitudes can be generalized to the
interaction between a D$p$ and a D$p'$ brane as follows (details
on this generalization can be found in Ref.~\cite{BILLO})
\begin{eqnarray}
\label{ansns} A_{\rm{NS-NS}} &=& V_{NN}\,
(8\pi^2\alpha')^{-{NN\over 2}} \int_0^\infty {dt} \left(1\over t
\right)^{DD\over
2}\,{\rm e}^{- y^2 /(2\alpha' \pi t)}\,\nonumber\\
& & \times \frac{1}{2}\left[ \left({f_3\over f_1}\right)^{8 -\nu}
  \left({f_4\over f_2}\right)^{\nu}
  - \left({f_4\over f_1}\right)^{8 -\nu}
  \left({f_3\over f_2}\right)^{\nu}\right]~~,
\end{eqnarray}
for the NS-NS sector and
\begin{equation}
 A_{\rm{R-R}}
= V_{NN}  (8\pi^2\alpha')^{-{NN\over 2}} \cdot \int_0^\infty {dt}
\left(1\over t \right)^{DD\over 2}\, {\rm e}^{-  y^2 /(2 \pi
\alpha' t)} ~\frac{1}{2} \left[
-\left(\frac{f_2}{f_1}\right)^{8}\,
\delta_{\nu,0}+\delta_{\nu,8}\right]~~. \label{arr3}
\end{equation}
for the R-R sector, where $V_{NN}$ is the common world-volume of
the two D-branes.

Due to the ``abstruse identity'', the total D-brane amplitude
\begin{equation}
A=A_{\rm{NS-NS}}+A_{\rm{R-R}}
\label{total}
\end{equation}
vanishes if $\nu=0,4,8$; these are precisely the configurations of
two D-branes which break half of the supersymmetries of the Type II
theory and satisfy the BPS no-force condition.

As we said before, open/closed string duality allows to evaluate
the interaction between Dp branes either in the closed,
as we have done before, or in the open channel. The expressions
are of course equal, as one can check performing the modular
transformation $t\rightarrow \tau = \frac{1}{t}$ which brings from one channel
to the other. In this duality it is
particulary interesting to understand the spin structure
correspondence between the two computations. Indeed for each spin
structure of, for instance, the open string channel one can find a
correspondent spin structure in the closed string one that gives
exactly the same contribution to the free energy. In the open
string language the free energy corresponding to the various spin
structures is given by:
\begin{equation}
F_i = \int_{0}^{\infty} \frac{d \tau}{\tau} Tr_i \left[ {\it e}^{- 2 \pi \tau
(L_0-a)}  \right]
\label{Fi}
\end{equation}
where $a =1/2 (a =0)$ in the NS(R) sector, the index $i$ runs over the four
open string spin structures:
\begin{equation}
i = NS, NS (-1)^F , R , R (-1)^F
\label{spst}
\end{equation}
and the factor $(-1)^F$ comes from the open string GSO projectors
defined in Eq.s (\ref{gsof}) and  (\ref{ferno}) for the R and NS
sectors respectively, which must be inserted in the trace in
Eq.(\ref{Fi}). On the other hand the various spin structures in
the closed string channel are given by:
\begin{equation}
F_{\eta \eta'} = \bra{B,\eta} D \ket{B, \eta'}_{NS-NS, R-R}
\label{feta}
\end{equation}
where $\eta, \eta' = \pm 1$.

By explicit calculation one gets the contribution of the four spin
structures in the open string channel to be given by:
\begin{equation}
\int_{0}^{\infty} \frac{d \tau}{\tau} Tr_{NS} \left[{\it e}^{- 2 \pi \tau L_0
} \right] =\frac{V_{NN}} {( 8 \pi^2 \alpha' )^{NN/2}}
\int_{0}^{\infty} \frac{d \tau}{\tau} \tau^{-NN/2} {\it e}^{b^2 \tau/(2 \pi
\alpha')} \left( \frac{f_2 (k}{f_4 (k)}\right)^{\nu}
\left( \frac{f_3 (k)}{f_1 (k)}\right)^{8- \nu}
\label{b11}
\end{equation}

\begin{equation}
\int_{0}^{\infty} \frac{d \tau}{\tau} Tr_{R} \left[{\it e}^{- 2 \pi \tau L_0
} \right] =\frac{V_{NN}} {( 8 \pi^2 \alpha' )^{NN/2}}
\int_{0}^{\infty} \frac{d \tau}{\tau} \tau^{-NN/2} {\it e}^{b^2 \tau/(2 \pi
\alpha')} \left( \frac{f_3 (k)}{f_4 (k)}\right)^{\nu}
\left( \frac{f_2 (k)}{f_1 (k)}\right)^{8- \nu}
\label{b2}
\end{equation}

\begin{equation}
\int_{0}^{\infty} \frac{d \tau}{\tau} Tr_{NS} \left[{\it e}^{- 2 \pi \tau L_0
} (-1)^F \right] = - \frac{V_{NN}} {( 8 \pi^2 \alpha' )^{NN/2}}
\int_{0}^{\infty} \frac{d \tau}{\tau} \tau^{-NN/2} {\it e}^{-b^2 \tau/(2 \pi
\alpha')} \left( \frac{f_4 (k)}{f_1 (k)}\right)^{8} \delta_{\nu 0}
\label{b3}
\end{equation}

\begin{equation}
\int_{0}^{\infty} \frac{d \tau}{\tau} Tr_{R} \left[{\it e}^{- 2 \pi \tau L_0
} (-1)^F \right] = - \frac{V_{NN}} {( 8 \pi^2 \alpha' )^{NN/2}}
\int_{0}^{\infty} \frac{d \tau}{\tau} \tau^{-NN/2} {\it e}^{-b^2 \tau/(2 \pi
\alpha')} \delta_{\nu 8}
\label{b4}
\end{equation}
where
\begin{equation}
k = {\it e}^{- \pi \tau}
\label{kkk}
\end{equation}
On the other hand in the closed channel we get
\begin{equation}
\bra{B, \eta} D \ket{B, \eta}_{NS-NS} = \frac{V_{NN}} {( 8 \pi^2 \alpha')^{NN/2}}
\int_{0}^{\infty} \frac{dt}{t} t^{1 -DD/2} {\it e}^{-b^2/(2 \pi \alpha' t)}
\left( \frac{f_4 (q)}{f_2 (q)}\right)^{\nu} \left(\frac{f_3 (q)}{f_1 (q)}
\right)^{8 -\nu}
\label{c1}
\end{equation}
\begin{equation}
\bra{B, \eta} D \ket{B, - \eta}_{NS-NS} = \frac{V_{NN}} {( 8 \pi^2 \alpha')^{NN/2}}
\int_{0}^{\infty} \frac{dt}{t} t^{1 -DD/2} {\it e}^{-b^2/(2 \pi \alpha' t)}
\left( \frac{f_3 (q)}{f_2 (q)}\right)^{\nu} \left(\frac{f_4 (q)}{f_1 (q)}
\right)^{8 -\nu}
\label{c2}
\end{equation}
\begin{equation}
\bra{B, \eta} D \ket{B, \eta}_{R-R} = -  \frac{V_{NN}} {( 8 \pi^2 \alpha')^{NN/2}}
\int_{0}^{\infty} \frac{dt}{t} t^{1 -DD/2} {\it e}^{-b^2/(2 \pi \alpha' t)}
\left( \frac{f_2 (q)}{f_1 (q)}\right)^{8} \delta_{\nu 0}
\label{c3}
\end{equation}
\begin{equation}
\bra{B, \eta} D \ket{B,- \eta}_{R-R} =  \frac{V_{NN}} {( 8 \pi^2 \alpha')^{NN/2}}
\int_{0}^{\infty} \frac{dt}{t} t^{1 -DD/2} {\it e}^{-b^2/(2 \pi \alpha' t)}
\delta_{\nu 8}
\label{c4}
\end{equation}

By introducing the variable $t=1/\tau$ and using the
transformations properties of the functions $f_i $, it is easy to
show that Eq.s (\ref{b11})-(\ref{b4}) are respectively equal
 to Eq.s (\ref{c1})-(\ref{c4}) and specifically  the various spin structures in the open and in the closed string
channel are related as follows
\begin{equation}
\bra{B, \eta} D \ket{B, \eta}_{NS-NS} = \int_{0}^{\infty} \frac{d \tau}{\tau}
Tr_{NS} \left[ {\it e}^{-2 \pi \tau L_0 }\right]
\label{a1}
\end{equation}
\begin{equation}
\bra{B, \eta} D \ket{B, -\eta}_{NS-NS} = \int_{0}^{\infty} \frac{d \tau}{\tau}
Tr_{R} \left[ {\it e}^{-2 \pi \tau L_0 }\right]
\label{a2}
\end{equation}
\begin{equation}
\bra{B, \eta} D \ket{B, \eta}_{R-R} = \int_{0}^{\infty} \frac{d \tau}{\tau}
Tr_{NS} \left[ {\it e}^{-2 \pi \tau L_0 } (-1)^{F}\right]
\label{a3}
\end{equation}
\begin{equation}
\bra{B, \eta} D \ket{B, -\eta}_{R-R} = -\int_{0}^{\infty} \frac{d
\tau}{\tau} Tr_{R} \left[ {\it e}^{-2 \pi \tau L_0} (-1)^{F}
\right] \label{a4}
\end{equation}

The GSO projection in the closed string channel is performed by taking
the following combination:
\begin{equation}
\ket{B}_{NS-NS} = \frac{1}{2} \left[ \ket{B, +}_{NS-NS} -
\ket{B, -}_{NS-NS}\right]
\label{gsons}
\end{equation}
in the NS-NS sector and by
\begin{equation}
 \ket{B}_{R-R} = \frac{1}{2} \left[ \ket{B, +}_{R-R} +
\ket{B, -}_{R-R}\right]
\label{gsoram}
\end{equation}
in the R-R sector. Using the previous Eq.s one can easily show
that the following identities are satisfied
\begin{equation}
{}_{NS-NS} \bra{B} D \ket{B}_{NS-NS} = \frac{1}{2} \int_{0}^{\infty}
\frac{d \tau}{\tau} \left\{ Tr_{NS} \left[ {\it e}^{- 2 \pi \tau L_0}
\right] - Tr_R \left[ {\it e}^{- 2 \pi \tau L_0}  \right] \right\}
\label{r23}
\end{equation}
and
\begin{equation}
{}_{R-R} \bra{B} D \ket{B}_{R-R} = \frac{1}{2} \int_{0}^{\infty}
\frac{d \tau}{\tau} \left\{ Tr_{NS} \left[ {\it e}^{- 2 \pi \tau L_0}
(-1)^F
\right] - Tr_R \left[ {\it e}^{- 2 \pi \tau L_0}  (-1)^F \right] \right\}
\label{r24}
\end{equation}

\sect{Classical Solutions and Born-Infeld Action From Boundary
State}
\label{sec6}
In this section we want to use the boundary
state introduced in Sect. (\ref{sec4}) and describing Dp branes in
string theory in order to obtain the classical solutions  of the
low-energy string effective action discussed in Sect.
(\ref{sec2}). In particular we will show that the large distance
behavior of the graviton, dilaton and R-R $p+1$-form fields that
one obtains from the boundary state exactly agrees with that
obtained from the classical solution in sect.~\ref{sec2}.
Afterwards we will use the boundary state for computing the
one-point couplings of the Dp branes with the massless closed
string states and show that they agree with those obtained from
the Born-Infeld action.

To understand how to use the boundary state in order to determine
the long distance behavior of the classical massless fields
generated by a Dp brane, we compare the boundary state description
of the interaction between two Dp branes with  its field theory
counterpart. In the boundary state  language two Dp branes
interact via the exchange of a closed string propagator as
$\bra{B_1}D\ket{B_2}$ where $D$ is the closed string propagator
which propagates {\it all} the closed string states. But,
 if the distance between the two branes is big enough,
the dominant contribution to this interaction  comes from the
exchange of the massless closed string states. Thus at large
distance we can factorize the previous amplitude as follows \beq
\label{factintb} \bra{B_1}D\ket{B_2}\sim \sum_\Psi
\bra{B_1}{P_\Psi}\rangle\bra{P_\Psi} D|{B_2}\rangle \eeq where
$P_\Psi$ runs only over the projectors of the closed superstring
massless states.

In the field theory language the interaction between two branes
can be expressed as the coupling of the field generated by one
brane with the corresponding current generated by the other
brane or equivalently as the coupling of two current terms, one
for each brane, through the appropriate propagator, summed over all
the states $\Psi$ propagating between them:
\beq
\label{factintf}
 \sum_\Psi
J_\Psi^1\cdot\Phi^2_\Psi\,\sim\, \sum_\Psi\,\int\, J^1_\Psi\cdot
D_\Psi\cdot J^2_\Psi
\eeq
where $\Phi^2_\Psi=\int\, D_\Psi\cdot
J_\Psi^2$ is the field corresponding to the state $\Psi$ and
generated by the brane $2,$ $D_\Psi$  is its propagator and
$J^1_\Psi$ is the current of the brane $1$ which is coupled to the
field $\Phi_{\Psi}^{1}$. In the field theory language $\langle
B^i\ket{P_\Psi}$ gives exactly the currents $J^i_\Psi$, as we will
discuss later. Thus comparing Eq. (\ref{factintb}) with Eq.
(\ref{factintf}) we deduce that the long distance behavior of the
classical massless fields generated by a {\dpb}, that we have
called $\Phi_\Psi$, can be determined by computing the projection
of the boundary state along the various fields $\Psi$ after having
inserted a closed string propagator as
\begin{equation}
\label{filflu}
\Phi_\Psi=\langle P_\Psi|D|B\rangle
\end{equation}
Let us apply this procedure for computing  the expression for the
generic NS-NS massless field which is given by
\begin{equation}
J^{\mu\nu}\equiv {}_{-1}
\langle {\widetilde{0}}| {}_{-1}\langle 0| \psi^{\nu}_{1/2}~
{\widetilde{\psi}}^{\mu}_{1/2}  |D|
B\rangle_{NS}= - \frac{T_p}{2k^2_\perp} V_{p+1}S^{\nu\mu}
\label{masscam}
\end{equation}
Specifying the different polarizations corresponding to the various fields
(see Refs.~\cite{cpb,antone} for details) we get
\begin{equation}
\label{dilatone}
\delta \varphi=\frac{1}{\sqrt{d - 2}}\,\left(\eta^{\mu\nu} -
k^{\mu} \ell^{\nu} - k^{\nu}\ell^{\mu}\right)
J_{\mu\nu}=
 \frac{d-2p-4}{2{\sqrt {2(d-2)}}}\,\mu_p\,
\frac{V_{p+1}}{k_{\bot}^{2}}
\end{equation}
for the dilaton,
\[
\delta {\tilde{h}}_{\mu \nu}(k) =\frac{1}{2}\Big(J_{\mu\nu}+J_{\nu\mu}\Big)
-\frac{\delta\varphi}{\sqrt{d-2}}\, \,\eta_{\mu\nu}=
\]
\begin{equation}
=\sqrt 2 \mu_p \frac{ V_{p+1}}{ k_{\bot}^2}\,
{\rm diag} \left(- A, A \dots A, B \dots B \right)~~,
\label{gravitone}
\end{equation}
for the graviton, where $A$ and $B$ are given in Eq. (\ref{expo})
and
\begin{equation}
\delta {\cal{B}}_{\mu\nu}(k) = \frac{1}{\sqrt 2}\Big( J_{\mu\nu} - J_{\nu\mu}
\Big)=0
\label{ant1}
\end{equation}
for the antisymmetric tensor. In the R-R sector we get instead
\begin{equation}
\delta {\cal{C}}_{01\dots p}(k) \equiv
\bra{P^{(C)}_{01\cdots p}} \,D\,\ket{B}_{\rm
R} =\mp\,{\mu_p}
\frac{V_{p+1}}{k_{\bot}^2}
\label{ra1}~~.
\end{equation}
Expressing the previous fields in configuration space using  the
following Fourier transform valid for $p < d-3$
\begin{equation}
\int d^{(p+1)}x\,d^{(d-p-1)} x \frac{{\rm e}^{i k_{\bot} \cdot x_{\bot}}}{
(d-p-3)\,
r^{d-p-3}\,\Omega_{d-p-2}}= \frac{V_{p+1}}{k^2_\perp}~,
\label{ft}
\end{equation}
remembering the expression $Q_p$ defined in Eq.(\ref{Qp})
and  rescaling the fields according to
\begin{equation}
{\phi} = { \sqrt{2}\kappa} {{\varphi}} ~~~~,~~~~
{ h}_{\mu \nu} = {2 \kappa} {{\tilde{h}}}_{\mu \nu} ~~~~,~~~~
{{ {{C}}}}_{01...p} = {\sqrt{2}}{\kappa}
{\cal{C}}_{01...p}~~, \label{newfi}
\end{equation}
we get the following large distance behavior
\begin{equation}
\label{dilatone2}
\delta\phi(r)=   \frac{d-2p-4}{{2\sqrt{2(d-2)}}}\,
\frac{Q_{p}}{r^{d-p-3}}
\end{equation}
for the dilaton,
\begin{equation}
\delta { h}_{\mu \nu}(r) = 2\frac{Q_{p}}{ r^{d-p-3}}\,
{\rm diag} \left(- A,  \dots A, B \dots B \right)~~,
\label{gravitone2}
\end{equation}
for the graviton and
\begin{equation}
\label{RR2}
{\delta{{ C}}}_{01...p} (r) = \mp\frac{Q_{p}}{r^{d-p-3}}
\end{equation}
for the R-R form potential.

The previous equations reproduce exactly the behavior for $r\to
\infty$ of the metric in Eq.(\ref{metri}) and of the R-R potential
given in Eq.(\ref{dil}). In fact at large distance their
fluctuations around the background values are exactly equal to
${\delta{ h}}_{\mu\nu}$ and ${\delta{{{ C}}}}_{01...p}$. In the
case of the dilaton, in order to find agreement between the
boundary state and the classical solution, we have to take $d=10$,
consistently with the fact that this is  the critical dimension
for superstrings.

Let us now discuss how to derive the Born-Infeld action, which
describes the low-energy dynamics of a Dp brane from the boundary
state. First we evaluate the couplings of a Dp brane with the
closed string massless states showing that the structure of those
couplings is the same as that obtained from the Born-Infeld action
and actually the comparison with what comes from the Born-Infeld
action allows us to fix the brane tension and charge in terms of
the string parameters $\alpha'$ and $g_s$.

The coupling of a D$p$ brane with a specific massless field $\Psi$
can be computed by saturating the boundary state with the
corresponding field $\langle \Psi | $ ($\langle \Psi | $ can be
$\langle \Psi_h|,\langle \Psi_B|, \langle \Psi_\varphi|$
corresponding respectively to the graviton, antisymmetric tensor
and dilaton or $\langle {\cal C}_{(n)} |$ corresponding to a R-R
state). By proceeding in this way we get the following couplings:
\begin{equation}
{\cal T}_{\varphi} \equiv  \frac{1}{2\sqrt{2}}\,J^{\mu\nu}\,
\left( \eta_{\mu\nu}-k_{\mu} \ell_{\nu} - k_{\nu}\ell_{\mu}\right)
\,\varphi =\frac{1}{2 \sqrt{2}}\, V_{p+1}\,\,T_p ~\frac{d-2p-4}{2}
\,\varphi~~; \label{adil2}
\end{equation}
for the dilaton,
\begin{equation}
{\cal T}_h \equiv J^{\mu \nu} \,{\tilde h}_{\mu\nu} =-
V_{p+1}\,T_p~ \eta^{\alpha \beta}{\tilde h}_{\beta\alpha}
\label{agra2}
\end{equation}
for the graviton,
\begin{equation}
{\cal T}_{\cal B} \equiv  \frac{1}{\sqrt{2}}\, J^{\mu \nu} \,{\cal
B}_{\mu\nu} = 0 \label{aas2}
\end{equation}
for the NS-NS $2$-form potential and
\begin{equation}
{\cal T}_{{\tilde C}_{n}} \equiv \langle {\tilde C}_{(n)}
\ket{B}_{R} = -\frac{{\tilde C}_{\mu_1 \dots \mu_{n}}}{16
\sqrt{2}(n)! } V_{p+1}
 \frac{T_p}{2} 2
Tr \left( \Gamma^{\mu_{n} \dots \mu_1} \Gamma^0
\dots \Gamma^p \right)
\label{wnplus12}
\end{equation}
for the R-R states. Computing the trace one gets
\begin{equation}
{\cal T}_{{\cal C}_{(p+1)}} = \frac{\sqrt{2}\,T_p}{(p+1)!}\,
V_{p+1}\,{\cal C}_{\alpha_0\ldots\alpha_p}
\,\varepsilon^{\alpha_0\ldots\alpha_p} \label{fite452}
\end{equation}
where we have used the fact that for $d=10$ the $\Gamma$ matrices
are $32\times 32$ dimensional matrices, and thus $Tr( \one) = 32$.
Here $\varepsilon^{\alpha_0\ldots\alpha_p}$ indicates the
completely antisymmetric tensor on the D brane
world-volume~\footnote{Our convention is that
$\varepsilon^{0\ldots p}=-\varepsilon_{0\ldots p}=1$.}. It can be
checked that the previous couplings can be obtained by extracting
the terms linear in the massless closed string states from the
following action:
\begin{equation}
S_{DBI} = - \tau_p \int d^{p+1} \xi~{\rm e}^{- \kappa {{\varphi}}
(3- p )/(2\sqrt{2})}
\sqrt{- \det\left[ g_{\alpha \beta} + \sqrt{2} \kappa {\tilde B}_{\alpha \beta}
  {e)^{- \kappa \phi/\sqrt{2}}} \right] } + \mu_p \int_{V_{p+1}}
{\tilde C}_{p+1}~~.
\label{borninfe22}
\end{equation}
provided that the brane tension and  charge are given by
\begin{equation}
\tau_p = \frac{T_p}{\kappa} = \frac{(2 \pi \sqrt{\alpha'})^{1-p}}{2
  \pi g_s \alpha'}~~,~~ \mu_p = \sqrt{2} T_p = \sqrt{2 \pi} ( 2 \pi
  \sqrt{\alpha'} )^{3 - p}~~,
\label{techa45}
\end{equation}
where $T_p$ is given in Eq.(\ref{Tp}) for $d=10$.

In the previous action the closed string fields are canonically normalized
while in the supergravity actions in Eq.s (\ref{tendim3}) and
(\ref{IIAAction}) they are not. In order to have in      the
Born-Infeld action the massless closed string fields normalized as in
Eq.s (\ref{tendim3}) and  (\ref{IIAAction}) we need to introduce the
following fields:
\begin{equation}
\sqrt{2} \kappa \varphi = \phi~~,~~\sqrt{2} \kappa {\cal C} = {{C}}~~,~~
\sqrt{2} \kappa {\cal B} = {{B}}
\label{nor98}
\end{equation}
In terms of these new fields the Born-Infeld action becomes:
\[
S_{BI} = - \tau_p \int d^{p+1} x \,\,{\rm e}^{-(3-p)\phi /4} \times
\sqrt{- \det \left[G_{\alpha \beta} + {\rm e}^{-\phi/2}
    \left({{B}}_{\alpha \beta} + 2 \pi \alpha' F_{\alpha \beta} \right)
    \right]} \,+
\]
\begin{equation}
+ \tau_p \int_{V_{p+1}} \sum_n {{C}}_{n} {\rm e}^{(2\pi \alpha' F +B)}
\label{BI12}
\end{equation}
which has been generalized in order to include also the brane
coupling with open string fields. Actually the dependence of the
Born-Infeld action on the open string fields can be explicitly obtained
by constructing the boundary state having a gauge field on it and
repeating the calculation just done (see Refs.~\cite{ISLANDA,antone}
for details). It is important at this point to stress that the
Born-Infeld action in Eq. (\ref{BI12}) contains two kinds of very
different fields. The first ones are the massless closed string
excitations of the NS-NS and R-R sectors that live in the entire
ten-dimensional space and that
enter in the Born-Infeld action through their pullback into the
world-volume of the Dp brane defined by
\begin{equation}
G_{\alpha \beta} = G_{\mu \nu} \partial_{\alpha} x^{\mu}
\partial_{\beta}
x^{\nu}~~,~~ B_{\alpha \beta} = B_{\mu \nu} \partial_{\alpha} x^{\mu}
\partial_{\beta}
x^{\nu}
\label{pullba}
\end{equation}
with a similar expression for the R-R fields. The second ones are
instead the fields corresponding to the massless open-string
states discussed at the end of Sect.~\ref{sec3}, that live on the
world-volume of the brane and that are the gauge field
$A_{\alpha}$ with field strength $F_{\alpha \beta}$ and the Higgs
fields  $\Phi^i$ related to the transverse brane coordinates $x^i$
by the relation $ x^i \equiv 2 \pi \alpha' \Phi^i$. They play the
role of longitudinal and transverse coordinates of the brane. This
is consistent with the fact that the dynamics of the brane is
determined by the open strings having their end-points on the
brane. In the case of a system of $N$ coincident branes the
Born-Infeld action gets modified by the fact that the coordinates
of the branes become non-abelian fields. The complete expression
of the non-abelian Born-Infeld action is not known. But for our
purpose it is sufficient to consider the non-abelian extension
given in Ref.~\cite{TSE} where the symmetrized trace is
introduced. Moreover it can be shown that a system of $N$
coincident Dp branes is a BPS state preserving $1/2$ supersymmetry
(corresponding to $16$ preserved supersymmetries) and as a
consequence they are not interacting. This can be easily seen by
plugging  the classical solution given in Eq.s (\ref{metri}) and
(\ref{dil}) with $d=10$ in the Born-Infeld action in
Eq.(\ref{BI12}) obtaining the interaction term of the Lagrangian.
In fact if we do that neglecting the coordinates of the brane we
get
\begin{equation}
\tau_p \int d^{p+1} x \left\{ - H^{ [(p- 7)(p+1) - (p-3)^2]/16}
+  \frac{1}{H} -1 \right\}  = -\tau_{p} \int d^{p+1} x
\label{noforce18}
\end{equation}
that is independent on the distance $r$ between the brane probe
described by the Born-Infeld action and the system of $N$ coincident
branes described by the classical solution.

A system of $N$ Dp branes has a $U(N)$ gauge theory living
on its world-volume with $16$ supersymmetries corresponding, in the
case of $p=3$, to ${\cal{N}}=4$ super Yang-Mills in four dimensions
that is a conformal invariant theory with vanishing
$\beta$-function. Its Lagrangian can be obtained by expanding the
first term of the Born-Infeld action up to the quadratic order in the
gauge fields living on the brane. Neglecting the term independent from
the gauge fields that we have already computed in Eq.(\ref{noforce18})
we get the following Lagrangian:
\begin{equation}
L = \frac{1}{g_{YM}^{2}} \left[- \frac{1}{4} F_{\alpha \beta} F^{\alpha
    \beta}  + \frac{1}{2} \partial_\alpha \Phi^i \partial^{\alpha}
    \Phi^i  \right] + \dots
\label{YMax45}
\end{equation}
where the gauge coupling constant is indeed a constant given by:
\begin{equation}
g_{YM}^{2} = \frac{2}{\tau_p (2 \pi \alpha ')^2} = \frac{2 g_s
  \sqrt{\alpha'} ( 2 \pi \sqrt{\alpha'})^p}{(2 \pi \alpha')^2}
\label{couco45}
\end{equation}
where the extra factor $2$ in the second term comes from the fact
that we have normalized the generators of the $SU(N)$ gauge group
according to Eq. (\ref{trac}).

In particular for $p=3$ we get $ g_{YM}^{2} = 4 \pi g_s$. The action in
Eq.(\ref{YMax45}) corresponds to the dimensional reduction of
the ${\cal{N}}=1$ super Yang-Mills in ten dimensions to $(p+1)$
dimensions.

The previous considerations imply that the low-energy dynamics of
branes can be used to determine the properties of gauge theories and
viceversa.

\section{Non conformal Branes}
\label{sec7}

\subsection{Generalities and general formulae}

In the previous sections we have seen that a D brane has the
twofold property of being a solution of the low-energy string
effective action and of having a gauge theory living on its
world-volume. These complementary descriptions of a D brane open
the way to study the quantum properties of the world-volume gauge
theory from the classical dynamics of the brane and viceversa.
This goes under the name of gauge/gravity correspondence. At the
end of the previous section we have already used this
correspondence to derive the gauge coupling constant  of the
${\cal{N}}=4$ super Yang-Mills from the supergravity solution. In
particular, we have seen that the gauge coupling constant is not
running consistently with  the fact that the ${\cal{N}}=4$
world-volume theory is a conformal invariant theory. In this
section we want to extend these results to more realistic theories
that are less supersymmetric and non conformal. Two kinds of
branes have been used to study those gauge theories, namely
fractional branes of some simple orbifold and branes wrapped on
some nontrivial two-cycle of a Calabi-Yau space. As we have done
for the D  branes previously discussed, we will first construct
the classical solution corresponding to a system of fractional and
wrapped D branes and then insert it in the expressions for the
gauge coupling constant and the vacuum angle $\theta_{YM}$ of the
gauge theory living on their world-volume, expressed in terms of
the supergravity fields.

It may at first sight look puzzling that the supergravity solution
involving the massless closed string fields can provide  perturbative
information on the gauge theory living in the world-volume of a D
brane. In fact, by computing for instance the one-loop annulus diagram
in the full string theory
we expect that the perturbative information on the gauge theory living
on a D brane be given by the contribution of the massless open string
states that is in general totally different from that of the massless
closed string fields. It turns out, however, as shown explicitly in
Ref.~\cite{DLMP} for the case of fractional branes, that in certain cases the
contribution of the massless open string states to the annulus diagram
transform under open/closed string duality exactly into that of the
massless closed string states without any mixing with the massive states.
This procedure shows that in the case of fractional branes the
gauge/gravity correspondence follows
from open/closed string duality.

Let us now briefly illustrate how to derive these
gauge-gravity
relations for the gauge theory living on fractional D3
and wrapped D5 branes using supergravity calculations.
Since also the fractional
D3 branes are D5 branes wrapped on a vanishing two-cycle corresponding
to a fixed point of an orbifold we can start from  the Born-Infeld
action of a D5 brane that in the string frame is given by:
\begin{equation}
S  = -\tau_5 \int d^{6} \xi {e}^{-( \phi - \phi_0)} \sqrt{- \det (
G_{ab} + B_{ab} + 2 \pi \alpha' F_{ab})}~~,~~ \tau_5 = \frac{1}{g_s
\sqrt{\alpha'} (2 \pi \sqrt{\alpha'})^5}
\label{boinfe45}
\end{equation}
We divide the $6$-dimensional world-volume into four flat
directions on which the gauge theory lives and $2$ directions on
which the brane is wrapped. Let us denote them with the indices $
a,b = ( \alpha, \beta; A,B)$ where $\alpha$ and $\beta$ denote the
flat four-dimensional ones, while $A$ and $B$ the wrapped ones.
Let us assume that the determinant in Eq. (\ref{boinfe45})
factorizes into a product of two determinants; one corresponding
to the four-dimensional flat directions where the gauge theory
lives and the other corresponding to the wrapped ones where we
only have  the metric and the NS-NS two-form field. By expanding
the first determinant and keeping only the quadratic term in the
gauge field we obtain:
\begin{equation}
- \tau_5 \cdot \frac{(2 \pi \alpha')^2}{8} \int d^{6} \xi
  {e}^{-(\phi - \phi_0)}
  \sqrt{-\det G_{\alpha \beta}} G^{\alpha \gamma} G^{\beta \delta}
  F_{\alpha \beta} F_{\gamma \delta} \sqrt{\det{( G_{AB} + B_{AB})}}
\label{binf74}
\end{equation}
We assume that  along the flat four-dimensional directions the metric
has only the warp factor, while along the wrapped ones, in addition to
the warp factor, there is also a nontrivial metric. This means that the
longitudinal part of the metric can be written as
\begin{equation}
ds^2 = H^{-1/2}\left(  d x_{3,1}^{2} + ds_{2}^{2} \right)~~,~~
{e}^{ (\phi - \phi_0)} = H^{-1/2}
\label{warp943}
\end{equation}
where we have also written the dilaton dependence on the warp
factor. Inserting this metric in Eq. (\ref{binf74}) we see that
the warp factor cancels in the Yang-Mills action and from it  we
can then extract the gauge coupling constant as the coefficient of
$-\frac{1}{4}F_{\mu\nu}F^{\mu\nu}$:
\begin{equation}
\frac{1}{g_{YM}^{2}} = \tau_5 \frac{( 2 \pi \alpha')^2}{2} \int
d^{2} \xi {e}^{- (\phi - \phi_0)} \sqrt{\det{( G_{AB} + B_{AB})}}
\label{gauge986}
\end{equation}
This formula is valid for both wrapped and fractional branes of the
orbifold $C^2 / Z_2$ and
can be rewritten as:
\begin{equation}
\frac{4 \pi}{g_{YM}^{2}} =  \frac{1}{g_s ( 2 \pi \sqrt{\alpha'})^{2}} \int
d^{2} \xi {e}^{- (\phi - \phi_0)} \sqrt{\det{( G_{AB} + B_{AB})}}
\label{volu87}
\end{equation}

The $\theta$ angle in the case of both fractional D3 branes and
wrapped D5 branes can be obtained extracting the coefficient of
$\frac{1}{32\pi^2}F_{\mu\nu}\tilde F^{\mu\nu}$ from the
Wess-Zumino-Witten  part
of the Born-Infeld action and is given by:
\begin{equation}
\theta_{YM} = \tau_5 (2 \pi \alpha')^2   (2 \pi)^2
\int_{{\cal{C}}_2} ( C_2 + C_0  B_2) = \frac{1}{2 \pi \alpha' g_s}
\int_{{\cal{C}}_2} ( C_2 + C_0  B_2)
\label{theta56}
\end{equation}

In the following subsections we will consider solutions of the
classical equations of ten-dimensional IIB supergravity and we
will insert them in the previously derived expressions for the
gauge couplings obtaining perturbative and non-perturbative
information on the gauge theory living on the world-volume of
these branes.

\subsection{Fractional branes}

In this subsection we will consider fractional D3 and D7 branes of
the orbifolds $C^2 /Z_2$ and $C^3 /(Z_2 \times Z_2)$ in order to
study the properties of respectively ${\cal{N}}=2$ and
${\cal{N}}=1$ supersymmetric gauge theories. We group the
coordinates of the directions $x^4, \dots x^9 $ transverse to the
world-volume of the D3 brane where the gauge theory lives, into
three complex quantities:
\begin{equation}
z_1 = x^4 + i x^5~~,~~z_2 = x^6+i x^7~~,~~
z_3 = x^8 + i x^9
\label{z1z2z3}
\end{equation}
In the case of the first orbifold the nontrivial generator $h$ of $Z_2$ acts
as
\begin{equation}
z_2  \rightarrow - z_2~~~,~~~z_3 \rightarrow  - z_3
\label{z2z3}
\end{equation}
while in the case of the second orbifold the three nontrivial
generators act as follows on the transverse coordinates:
\begin{eqnarray}
h_1\equiv h \times 1  \Rightarrow & z_1 \rightarrow z_1~,~  z_2 \rightarrow -
z_2 , & z_3 \rightarrow - z_3  \nonumber \\
h_2\equiv 1 \times h   \Rightarrow & z_1 \rightarrow - z_1~,~  z_2 \rightarrow
z_2 , & z_3 \rightarrow - z_3  \label{orb5} \\
h_3\equiv h \times h   \Rightarrow & z_1 \rightarrow - z_1 ~,~ z_2 \rightarrow
- z_2 , & z_3 \rightarrow  z_3  \nonumber
\end{eqnarray}
They are both non compact orbifolds with respectively one and
three fixed  points at the origin corresponding to the point $z_2
, z_3 =0$ and to the three points $z_1 , z_2 =0$, $z_1 , z_3 =0$
and $z_2 , z_3 =0$. Each fixed point corresponds to a vanishing
$2$-cycle. Fractional D3 branes are D5 branes wrapped on the
vanishing two-cycle and therefore are, unlike bulk branes, stuck
at the orbifold fixed point. By considering $N$ fractional D3 and
$M$ ($2M$) fractional D7 branes of the orbifold $C^2 / Z_2$ ($C^3
/(Z_2 \times Z_2)$) we are able to study  ${\cal{N}}=2$
(${\cal{N}}=1$) super QCD with $M$ hypermultiplets. In order to do
that we need to determine the classical solution corresponding to
the previous brane configuration. For the case of the orbifold
$C^2 /Z_2$ the complete classical solution has been found in
Ref.~\cite{d3d7}~\footnote{See also
Refs.~\cite{KLENE,d3,polch,grana2,marco} and Ref.~\cite{REV} for a
review on fractional branes.}. In the following we write it
explicitly for  a system of $N$  fractional D3 branes with
world-volume along the directions $x^0, x^1 , x^2 , $ and $ x^3$
and $M$ D7 fractional branes containing the D3 branes in their
world-volume and having the remaining four world-volume directions
along the orbifolded ones. The metric, the $5$-form field
strength, the axion  and the dilaton are given by~\footnote{We
denote with $\alpha$ and $\beta$ the four directions corresponding
to the world-volume of the fractional D3 brane, with $\ell$ and
$m$ those along the four orbifolded directions $x^6 , x^7, x^8 $
and $x^9$ and with $i$ and $j$ the directions $x^4$ and $x^5$ that
are transverse to both the D3 and the D7 branes.}:
\begin{eqnarray}
ds^2 &=& H^{-1/2}\, \eta_{\alpha\beta}\,d x^\alpha dx^\beta
+ H^{1/2} \,\left(\delta_{\ell m}\,dx^\ell dx^m + {\rm e}^{-\phi} \delta_{ij}
 dx^i dx^j\right)  ~~,
\label{met48} \\
{\widetilde{F}}_{(5)} &=&  d
\left(H^{-1} \, dx^0 \wedge \dots \wedge dx^3 \right)+ {}^* d
\left(H^{-1} \, dx^0 \wedge \dots \wedge dx^3 \right) ~~,
\label{f5ans}
\end{eqnarray}
\begin{equation}
\tau \equiv C_0 + i {\rm e}^{- \phi} =
{\rm i}\left(1 -\, \frac{Mg_s}{2\pi}\, \log
\frac{z}{\epsilon}\right)~~,~~ z \equiv x^4 + i x^5 = \rho {\rm e}^{i\theta}
\label{tausol}
\end{equation}
where the warp factor $H$ is a function of all coordinates that
are transverse to the D3 brane ($x^4,\ldots x^9$). The twisted fields
are instead given by  $B_2 = \omega_2 b$, $C_2 = \omega_2 c$
where $\omega_2$ is the volume form corresponding to the vanishing
$2$-cycle and
\begin{equation}
b {\rm e}^{-\phi} = \frac{(2 \pi \sqrt{\alpha'})^2}{2}
\left[ 1 + \frac{2N-M}{\pi} g_s \log \frac{\rho}{\epsilon} \right]~~,~~
c + C_0 b = - 2 \pi \alpha' \theta g_s (2N-M)
\label{bc63}
\end{equation}
It can be seen that the previous solution has  a naked singularity of
the repulson type at short distances. But, on the other hand, if we
probe it with a brane probe approaching the stack of branes
corresponding to the classical solution from
infinity, it can also be seen that  the tension of the probe vanishes
at a certain distance from the stack of branes that is larger than
that of the
naked singularity. The point where the probe brane becomes tensionless
is called in the literature enhan{\c{c}}on~\cite{enhancon} and at
this point the
classical solution cannot be used anymore to describe the stack of
fractional branes.

Now let us exploit the gauge/gravity correspondence to determine
the coupling constants of the world-volume theory from the supergravity
solution.
In the case of fractional D3 branes of the orbifold $C^2 / Z_2$ having
only one single vanishing two cycle Eq.(\ref{volu87}) becomes:
\begin{equation}
\frac{1}{g_{YM}^{2}} =  \frac{\tau_5 (2 \pi \alpha')^2}{2}
\int_{{\cal{C}}_2} {e}^{- \phi} B_2 = \frac{1}{4 \pi g_s (2 \pi
\sqrt{\alpha'})^2} \int_{{\cal{C}}_2} {e}^{- \phi} B_2
\label{coupli98}
\end{equation}

Inserting  in Eq.s (\ref{coupli98}) and (\ref{theta56}) the classical
solution we get the following expression for the gauge coupling constant and the $\theta$
angle~\cite{d3d7} :
\begin{equation}
\frac{1}{g_{YM}^{2}} = \frac{1}{8 \pi g_s} + \frac{2N-M}{8 \pi^2} \log
\frac{\rho}{\epsilon}~~,~~
\theta_{YM} =- \theta (2N-M)
\label{thetaym76}
\end{equation}
In the case of an ${\cal{N}}=2$ supersymmetric theory there is
also a complex scalar field $\Psi$ in the gauge multiplet that we
expect to find when deriving the Yang-Mills action from the
Born-Infeld one. In fact in this derivation we get a contribution
from the kinetic term of the brane coordinates $x^4$ and $x^5$
that are transverse to the ones on which the branes live and to
the orbifolded ones. This implies that the complex scalar field of
the gauge supermultiplet is related to the coordinate $z$ of
supergravity through the following gauge-gravity relation $\Psi
\sim \frac{z}{2 \pi \alpha'}$. This is a relation between a
quantity of the gauge theory living on the fractional D3 branes
and the coordinate $z$ of  supergravity. Such an identification
allows one to obtain the gauge theory anomalies from the
supergravity background. In fact, since we know how the anomalous
scale and $U(1)$ transformations act on $\Psi$, from the previous
gauge-gravity relation we can deduce how they act on $z$, namely
\begin{equation}
\Psi \rightarrow s {\rm e}^{2i \alpha} \Psi \Longleftrightarrow z
\rightarrow s  {\rm e}^{2 i \alpha} z   \Longrightarrow
\rho \rightarrow s
\rho~~,~~  \theta \rightarrow \theta + 2 \alpha
\label{gau78}
\end{equation}
Those transformations  do not leave invariant the supergravity background in
Eq.s (\ref{bc63}) and when we plug it in Eq.s
(\ref{coupli98}) and (\ref{theta56}), they generate the anomalies of the
gauge theory living on the fractional D3 branes. In fact acting with those
transformations on Eq.s (\ref{thetaym76}) we get:
\begin{equation}
\frac{1}{g_{YM}^{2}} \rightarrow \frac{1}{g_{YM}^{2}} +\frac{2N-M}{8
\pi^2} \log s~~,~~ \theta_{YM} \rightarrow \theta_{YM} - 2 \alpha (2N-M)
\label{tra38}
\end{equation}
The first equation implies that the $\beta$-function of ${\cal{N}}=2$
super QCD with $M$ hypermultiplets is given by:
\begin{equation}
\beta (g_{YM} ) = - \frac{2N-M}{16 \pi^2} g_{YM}^{3}
\label{beta34}
\end{equation}
while the second one reproduces the chiral $U(1)$
anomaly~\cite{KOW,ANOMA}.
In particular, if we choose $\alpha = \frac{2 \pi}{2(2N-M)}$, then
$\theta_{YM}$ is shifted by a multiple of $2 \pi$. But since $\theta_{YM}$
is periodic of $2 \pi$, this means that the subgroup $Z_{2(2N-M)}$ is
not anomalous in perfect agreement with gauge theory results.

{From} Eq.s (\ref{thetaym76}) it is easy to compute
the combination:
\begin{equation}
\tau_{YM} \equiv \frac{\theta_{YM}}{2 \pi} + i \frac{4
\pi}{g_{YM}^{2}} = i \frac{2N-M}{2 \pi} \log \frac{z}{ \rho_{e}}~~,~~
\rho_e = \epsilon {\rm e}^{\pi/(2N-M)g_s}
\label{tauym67}
\end{equation}
where $\rho_{\epsilon}$ is called in the literature
the enhan{\c{c}}on radius and corresponds in the gauge theory to
the dimensional scale $\Lambda$ generated by dimensional
transmutation. Eq. (\ref{tauym67})  reproduces the perturbative
moduli space of ${\cal{N}}=2$ super QCD, but not the instanton
corrections. This corresponds to the fact that the classical
solution is reliable for large distances in supergravity
corresponding to short distances in the gauge theory, while it
cannot be used below the enhan{\c{c}}on radius where
non-perturbative physics is expected to show up.

Indeed  in that corner of the moduli space the effective theory
will receive instanton corrections proportional to powers of the
one-instanton action \beq \label{dp1} \exp\left(-{8\pi^2\over g^2}
+\ii \theta_{YM}\right) = \exp \left(2\pi\ii\tau_{YM}\right) =
\left({\rho_e\over z}\right)^{2N-M}~. \eeq Thus the instantonic
contributions become quite suddenly important near  the enhan\c
con radius $|z|=\rho_e$. This means that in order to study
non-perturbative effects in the gauge theory we need to find a
classical solution free from enhan{\c{c}}ons and naked
singularities. In the next section we will see that this can be
done for ${\cal{N}}=1$ super Yang-Mills that lives in the
world-volume of a wrapped D5 brane described by the
Maldacena-Nu\~nez solution. Before passing to this solution let us
first extend the previous results to ${\cal{N}}=1$ super QCD that
can be obtained as a particular case  of the general one studied
in Ref.~\cite{NAPOLI}. In this case only the asymptotic behavior
for large distances of the  classical solution has been explicitly
obtained and this is sufficient for computing the gauge coupling
constant and the $\theta$ angle of ${\cal{N}}=1$ super QCD. As
explained in Ref.~\cite{NAPOLI}, together with $N$ fractional D3
branes of the same type, one must also consider two kinds of $M$
fractional D7 branes in order to avoid gauge anomalies

In the case of the orbifold $C^3/ (Z_2 \times Z_2 )$
the gauge coupling constant is related to the supergravity solution as follows:
\begin{equation}
\frac{1}{g_{YM}^{2}} = \tau_5 \frac{(2 \pi
 \alpha ')^2}{4}\left[\sum_{i=1}^{3}  \int_{{\cal{C}}_{2}^{(i)}}
{e}^{- \phi} B_2 - ( 2 \pi  \sqrt{\alpha'} )^2 \right]
\label{c3z2z2}
\end{equation}
while the $\theta_{YM}$ is given by:
\begin{equation}
\theta_{YM} = \tau_5 \frac{(2 \pi \alpha')^2}{2}   (2 \pi)^2
\sum_{i=1}^{3}  \int_{{\cal{C}}_{2}^{(i)}} ( C_2  + C_0 B_2)
\label{c3z2z2th}
\end{equation}
Notice that Eq.s (\ref{c3z2z2})  and (\ref{c3z2z2th}) differ
respectively from Eq.s (\ref{coupli98}) and (\ref{theta56}) by a
factor $1/2$ in the normalization. This is due to the fact that
the projector of this orbifold - which is
$P=\frac{1+h_1+h_2+h_3}{4}$ - has an additional factor $1/2$ with
respect  to the one of the orbifold $C_2/Z_2$ (which is
$P=\frac{1+h}{2}$). Using the explicit supergravity solution one
gets the following expressions for the gauge coupling constant and
the $\theta$ angle ( $z_i = \rho_i {\rm e}^{i
\theta_i}$)~\cite{ANOMA,FERRO,NAPOLI}:
\begin{equation}
\frac{1}{g_{YM}^{2}} = \frac{1}{16 \pi g_s} + \frac{1}{8 \pi^2} \left(
N \sum_{i=1}^{3} \log \frac{\rho_i}{\epsilon} - M \log
\frac{\rho_1}{\epsilon} \right)~~,~~\theta_{YM} =- N \sum_{i=1}^{3}
\theta_i + M \theta_1
\label{thetae4}
\end{equation}
As explained in Ref.~\cite{ANOMA,FERRO} the anomalous scale and $U(1)$
transformations act on $z_i$ as $z_i \rightarrow s {\rm
e}^{i2\alpha/3} z_i$. This implies that the gauge parameters are
transformed as follows:
\begin{equation}
\frac{1}{g_{YM}^{2}} \rightarrow \frac{1}{g_{YM}^{2}} +
\frac{3N-M}{8 \pi^2} \log s~,~\theta_{YM} \rightarrow \theta_{YM}
- 2 \alpha \left(N - \frac{M}{3} \right) \label{ano89}
\end{equation}
that reproduce the anomalies of ${\cal{N}}=1$ super QCD. The
difference between the anomalies in the ${\cal{N}}=2$
(Eq.(\ref{tra38})) and ${\cal{N}}=1$ (Eq.(\ref{ano89})) super QCD
can be easily understood in terms of the different structure of
the two orbifolds considered. If we consider the two gauge
coupling constants there is a factor $\frac{3}{2}$ between the
contributions coming from the pure gauge part, while the
contribution of the matter is the same. The factor $3$ follows
from the fact that the orbifold $C^3 / (Z_2 \times Z_2)$ has three
vanishing two-cycles instead of just one, while the factor
$\frac{1}{2}$ from the additional factor $\frac{1}{2}$ in the
orbifold projection for the orbifold $C^3 /(Z_2 \times Z_2)$ with
respect to the orbifold $C^2 /Z_2 $. This explains the factor
$\frac{3}{2}$ in the gauge field contribution to the
$\beta$-function. The matter part is the same because in the
orbifold $C^2 /Z_2$ we have only one kind of fractional branes,
while in the other orbifold, in order to cancel the gauge
anomaly~\cite{NAPOLI}, we need two kinds of fractional branes.
This factor $2$ cancels  the factor $\frac{1}{2}$  coming from the
orbifold projection. Similar considerations can also be used to
relate the two chiral anomalies.

In conclusion, by using the fractional branes we have reproduced
the one-loop perturbative behavior of both ${\cal{N}}=1$ and
${\cal{N}}=2$ super QCD, but, because of the enhan{\c{c}}on and
naked singularities, we are not able to enter the non-perturbative
region in the gauge theory corresponding to short distances in
supergravity. In order to do this we must find a classical
solution free of singularities. That is why in the next section we
turn to wrapped branes.

\subsection{Wrapped branes and topological twist}

In this section we consider D5  branes wrapped on some nontrivial
cycle of a Calabi-Yau space and by means of the topological twist
we construct the Maldacena-Nu\~nez solution. The topological twist
acts also on the gauge theory living on the world-volume of the
wrapped branes by reducing, in the case of the Maldacena-Nu\~nez
solution, the original ${\cal{N}}=4 $ to an ${\cal{N}}=1$
supersymmetry. We show that the gauge theory living on the
world-volume of the branes described by the Maldacena-Nu\~nez
solution is ${\cal{N}}=1$ super Yang-Mills.

If we consider a D5 brane wrapped on some nontrivial cycle of a
Calabi-Yau space in general we break completely supersymmetry because
 the Killing spinor equation:
\begin{equation}
D_{M} \epsilon = (\partial_M + \omega_M ) \epsilon =0
\label{killspi}
\end{equation}
does not in general admit any non trivial solution.
This means that it is not an
easy task to find a classical solution corresponding to a wrapped
D5 brane and preserving some supersymmetry  starting directly from the
action of 10-dimensional IIB
supergravity. As suggested in Ref.~\cite{MN} it is much more convenient
to start from the action of the 7-dimensional gauged supergravity that
corresponds to the 10-dimensional IIB supergravity on $R^{1,6} \times
S^3$. In this way one has an action that also contains the gauge
fields of $SO(4)$, that is the isometry group of $S^3$, and the
condition in Eq.(\ref{killspi}) becomes:
\begin{equation}
D_{M} \epsilon = (\partial_M + \omega_M  + A_M ) \epsilon =0
\label{kilspi2}
\end{equation}
In this case it is not difficult to keep some supersymmetry with a
constant spinor $\epsilon$ by requiring an identification of the spin
connection of the two-cycle $S^2$ around which we wrap the D5 brane
with a subgroup $U(1)$ of the gauge group $SO(4)$. This identification
is called the topological twist. In particular, if we write $SO(4) =
SU(2)_{L'} \times SU(2)_{R'}$, it is possible to see that one can
preserve four supersymmetries  corresponding to an ${\cal{N}}=1$
supersymmetric gauge theory if we identify the spin connection with
a $U(1)$ subgroup of $SU(2)_{L'}$.

The topological twist acts also on the gauge theory living in the
world-volume of the wrapped brane reducing the states of
${\cal{N}}=4$ super Yang-Mills to those of a gauge theory with
less supersymmetry. In particular, let us consider a D5 brane
wrapped on $S^2$ that breaks the ten-dimensional Minkowski
symmetry into:
\begin{equation}
SO(1,9) \rightarrow SO(1,5) \times SO(4)_R = SO(1,5) \times SU(2)_{L'}
\times SU(2)_{R'}
\label{deco98}
\end{equation}
According to this decomposition the vector and the four scalar fields
corresponding to the transverse coordinates of the brane, transform as
\begin{equation}
A_{\mu} \rightarrow (6; 1,1)~~~,~~~\Phi \rightarrow (1;2,2)
\label{tra34}
\end{equation}
while the fermions transform as follows
\begin{equation}
\Psi \rightarrow (4+;2,1) + (4-;1,2)
\label{tra46}
\end{equation}
where the index $\pm$ refers to the six-dimensional
chirality. Remember that, because of the GSO projection, the spinor
$\Psi$ has a definite (for instance negative) ten-dimensional
chirality. This is consistent with the assignment in Eq.(\ref{tra46})
because the state $(2,1)[ (1,2)]$ has negative (positive)
four-dimensional chirality and the product of the four- and
six-dimensional
chirality is equal to the ten-dimensional one.
When the brane is wrapped on $S^2$ there is a further breaking:
\begin{equation}
SO(1,9) \rightarrow SU(2)_L \times SU(2)_R \times SO(2)_{S^2}
\times SU(2)_{L'} \times SU(2)_{R'} \label{bre32}
\end{equation}
This means that the vector and scalar fields transform as follows:
\begin{equation}
A_{\mu} \rightarrow (2,2;1;1,1) + (1,1;2;1,1)~~~,~~~
\Phi \rightarrow (1,1;1;2,2)
\label{rta32}
\end{equation}
and the fermions as follows
\begin{equation}
\Psi \rightarrow (2,1;+;2,1) + (1,2;+;1,2) + (1,2;-;2,1) + (2,1;-;1,2)
\label{rta31}
\end{equation}
where now $\pm$ is the chirality in the two-dimensional space spanned
by $S^2$. Notice that the product of the four- and of two-dimensional
chirality gives the six-dimensional one.

Let us consider the case in which one of the two $SU(2)$
of the R-symmetry group is broken into $SO(2)$ and then this $SO(2)$
is identified with $SO(2)_{S^2}$. This means that:
\begin{equation}
SU(2)_{L'} \times SU(2)_{R'} \rightarrow SO(2)_{L'} \times SU(2)_{R'}
\label{bre49}
\end{equation}
that gives rise to the following decomposition:
\begin{equation}
(2,2) \rightarrow (+,2) + (-,2)~~,~~(2,1) \rightarrow (+,1) + (-,1)
\label{deco97}
\end{equation}
Since the massless states are singlets under the simultaneous action of
$SO(2)_{L'}$ and $SO(2)_{S^2}$ it is easy to see that we are left only
with the following states:
\begin{equation}
(2,2;1;1,1)~~~,~~~(2,1;+;-,1)~~~,~~~(1,2;-;+,1)
\label{rema34}
\end{equation}
The first one corresponds to a four-dimensional gauge field, while the
other two correspond to a Majorana spinor. This is the field content of
${\cal{N}}=1$ super Yang-Mills.

The gauged supergravity action containing only the fields that are
turned on is given in Eq.(2.33) of Ref.~\cite{DLM} and the classical
solution is obtained by introducing in the equations of motion that
follow from the gauged supergravity action, the following ansatz for
the metric:
\begin{equation}
d s_{7}^{2} ={\rm e}^{2f(r)} \left( d x_{1,3}^2 + dr^2 \right) +
\frac{1}{\lambda^2}\,{\rm e}^{2 g(r)}\,d\Omega_2^2
\label{anme}
\end{equation}
where $d x_{1,3}^2$ is the Minkowski metric on ${\rm I\!R}_{1,3}$,
$r$ is the transverse coordinate to the domain-wall, and
$d\Omega_2^2 = d \widetilde\theta^2 + \sin^2 \widetilde\theta\,
d\widetilde\varphi^2$ (with $0\leq \widetilde\theta\leq\pi$ and
$0\leq\widetilde\varphi \leq2\pi$) is the metric of a unit
2-sphere~\footnote{Notice that the factor of $\lambda^{-2}$ in
(\ref{anme}) is necessary for dimensional reasons and it turns out to
be equal to $\lambda^{-2}= N g_s \alpha'$.}.
We also add  the following ansatz for the gauge fields of $SU(2)_{L'}$:
\begin{equation}
A^{1} = -\,\frac{1}{2\lambda}\, a(r)\, d{\widetilde{\theta}}~~,~~
A^{2} =  \frac{1}{2\lambda}\,a(r) \sin {\widetilde{\theta}}\, d
{\widetilde{\varphi}}~~,~~ A^{3} = -\,\frac{1}{2\lambda}\,\cos
{\widetilde{\theta}}\, d
 {\widetilde{\varphi}}~~.
\label{vec45}
\end{equation}
The functions $a(r), f(r)$ and $g(r)$ are determined by the classical
equations of motion. Actually we have not turned on only a $U(1)$ gauge
field but all three gauge fields of $SU(2)_{L'}$ in order to have a
solution free from naked singularities at short distances.
Having found a classical solution in  7-dimensional gauged
supergravity one can use known formulas~\cite{CVETIC} that uplift it
to a ten-dimensional
solution of IIB supergravity. In this way one gets the following
ten-dimensional (string frame) metric~\cite{MN,CV}:
\begin{equation}
d s_{10}^{2}
={\rm e}^\Phi
\left[  dx_{1,3}^{2} +
  \frac{{\rm e}^{2h}}{\lambda^2} \left(d {\widetilde{\theta}}^2
+ \sin^2 {\widetilde{\theta}}\,d {\widetilde{\varphi}}^2 \right)
\right]
+\frac{{\rm e}^\Phi
}{\lambda^2} \left[ d \rho^2 +
\sum_{a=1}^3\left( \sigma^a - \lambda A^a \right)^2\right]~~,
\label{stri987}
\end{equation}
a ten-dimensional dilaton
\begin{equation}
{\rm e}^{2 \Phi} =
\frac{\sinh 2 \rho}{2 \,{\rm e}^h}~~,
\label{dila493}
\end{equation}
and the field strength corresponding to a R-R $2$-form given by:
\begin{equation}
F^{(3)} = \frac{2}{\lambda^2} \,\left(\sigma^1 - \lambda
A^{1}\right) \wedge \left(\sigma^2 - \lambda A^{2}\right) \wedge
\left(\sigma^3 -\lambda A^{3}\right) - \frac{1}{\lambda}\,
\sum_{a=1}^3F^{a} \wedge \sigma^a~~.
\label{h3f39}
\end{equation}
where
\begin{eqnarray}
{\rm e}^{2h} &=&  \rho \coth 2 \rho - \frac{\rho^2}{\sinh^2 2
\rho} - \frac{1}{4}~~, \label{solh6}\\ {\rm e}^{2k} &=&  {\rm
e}^{h}\,\frac{\sinh 2\rho}{2}~~, \label{solh61}\\
a &=& \frac{2
\rho}{\sinh 2 \rho} \label{solh62}
\end{eqnarray}
with $\rho \equiv \lambda \,r$, $ h \equiv g -f$ and $k \equiv
\frac{3}{2}f + g$. The left-invariant 1-forms of
$S^3$ are
\[
\sigma^1 =  \frac{1}{2}\Big[\cos \psi \,d\theta' + \sin \theta'
\sin \psi \,d \phi  \Big]~~,~~\sigma^2 = -\frac{1}{2} \Big[\sin
\psi \,d \theta' - \sin \theta' \cos \psi \,d\phi \Big]~~,
\]
\begin{equation}
\sigma^3 = \frac{1}{2} \Big[d \psi + \cos \theta' \,d \phi
\Big]~~,
\label{diffe49}
\end{equation}
with $0\leq \theta'\leq\pi$, $0\leq\phi\leq 2\pi$ and
$0\leq\psi\leq 4\pi$. Using the following formulas:
\begin{equation}
F^a = d A^a + \lambda \epsilon^{abc} A^b \wedge A^c~~,~~ d\sigma^a = -
\epsilon^{abc} \sigma^b \wedge \sigma^c
\label{equ32}
\end{equation}
it is possible to rewrite Eq.(\ref{h3f39}) as follows
\begin{equation}
F^{(3)}= \frac{1}{\lambda^2} \left[ 2 \sigma^1 \wedge \sigma^2 \wedge
  \sigma^3 + d \left(\sum_{a=1}^{3} \sigma^a \wedge \lambda A^a \right)\right]
\label{f367}
\end{equation}
and from it we can extract $C_2$
\begin{equation}
C_2 = \frac{1}{4 \lambda^2} \left[  \psi  \sin \theta' d \theta'
\wedge d \phi + 4 \sum_{a=1}^{3} \sigma^a \wedge \lambda A^a \right]
+ constant
\label{c298}
\end{equation}
that is equal to
\begin{eqnarray}
{C}^{(2)} &=& \frac{1}{4 \lambda^2} \left[ \psi\left(\sin
\theta'\, d \theta'
  \wedge d \phi - \sin {\widetilde{\theta}} \,d {\widetilde{\theta}}
  \wedge d {\widetilde{\varphi}}
\right) - \cos \theta' \,\cos
  {\widetilde{\theta}} \,
d \phi \wedge d {\widetilde{\varphi}}\right]\nonumber \\
&&+\,\frac{a}{2\lambda^2} \left[ d{\widetilde{\theta}} \wedge
\sigma^1 - \sin {\widetilde{\theta}}\,
 d{\widetilde{\varphi}} \wedge \sigma^2 \right] + constant
\label{f3exact}
\end{eqnarray}
when we insert in it the three gauge fields given in Eq. (\ref{vec45}).
In the next section we will use the Maldacena-Nu\~nez solution for
studying the properties of ${\cal{N}}=1$ super Yang-Mills.

\subsection{Gauge couplings from MN solution}

In Eq.s (\ref{volu87}) and (\ref{theta56})  we wrote the expression
of the gauge couplings in terms of the supergravity fields both for
fractional and wrapped branes.
In the case of wrapped branes having $B=0$
we see that the warp factor in Eq.(\ref{volu87}) cancels giving:
\begin{equation}
\frac{4 \pi}{g_{YM}^{2}} =  \frac{1}{g_s ( 2 \pi
\sqrt{\alpha'})^{2}} \int d^{2} \xi  \sqrt{\det{ {\hat{G}}_{AB} }}
\label{gauge56}
\end{equation}
where with ${\hat{G}}$ we have denoted the metric tensor in the
wrapped directions without the warp factor.

In order to get explicitly the gauge quantities from the supergravity solution we
have to identify the two-cycle. It is clear that in the
$7$-dimensional gauged supergravity the two-cycle is the one specified
by the coordinates ${\tilde{\theta}}$ and ${\tilde{\varphi}}$ keeping
the other variables fixed. But when we lift the solution up to ten
dimensions there is the topological twist that mixes
$({\tilde{\theta}},{\tilde{\varphi}} )$ with the variables
$(\theta', \phi, \psi)$
that describe $S^3$. In the literature two choices for the two-cycle
have been done. They are specified by:
\begin{enumerate}
\item{$({\tilde{\theta}},{\tilde{\varphi}} )$ keeping the other
      variables fixed}
\item{${\tilde{\theta}} = \pm \theta'$ and ${\tilde{\varphi}}= - \phi$
    keeping $\rho$ and $\psi$ fixed}
\end{enumerate}

In Ref.~\cite{DLM} the first choice for the two-cycle was made and one
found the following expression for the gauge coupling constant:
\begin{equation}
\frac{4 \pi^2}{N g_{YM}^2} = \frac{Y(\rho)}{4} \,
E\left(\sqrt{\frac{Y(\rho)-1}{Y(\rho)}}\right)~~,~~Y(\rho) =
4\rho\,\coth 2\rho -1
\label{gymfin}
\end{equation}
where
\begin{equation}
E(x) \equiv \int_{0}^{\pi/2}\!\! d\phi~ \sqrt{1-x^2\,\sin^2\!\phi}~~;
\label{elliptic}
\end{equation}
is the complete elliptic integral of second kind.
Using the properties of the elliptic integral, it is easy to see that
\begin{eqnarray}
\frac{1}{g_{\rm YM}^2}&\simeq&\frac{N\,\rho}{4\pi^2}~~~~~~~~~{\rm for}
~~\rho\to\infty~~,\label{gymUV}\\
\frac{1}{g_{\rm YM}^2}&\simeq&\frac{N}{32\pi}~~~~~~~~~{\rm for}~~\rho\to 0
~~.\label{gymIR}
\end{eqnarray}
implying that we get asymptotic freedom in the deep ultraviolet.
Putting together the ultraviolet behavior in Eq. (\ref{gymUV})
together with the relation connecting the supergravity variable
$\rho$ with the renormalization group scale $\mu$~\footnote{The
connection
  between the gaugino condensate and $a( \rho )$ was originally
  suggested in Ref.~\cite{MILANO}.}
\begin{equation}
a(\rho) = \frac{2 \rho}{\sinh 2 \rho} =\frac{\Lambda^3}{\mu^3}~~.
\label{holo1}
\end{equation}
allowed  the authors
of Ref.~\cite{DLM} to get the running coupling constant of
${\cal{N}}=1$ super Yang-Mills.
In fact from Eq. (\ref{holo1}) one can easily get:
\begin{equation}
\frac{\partial \rho}{\partial \log (\mu/\Lambda)} =
\frac{3}{2}\left[\frac{1}{1-(2\rho)^{-1}+2\,{\rm e
}^{-4\rho}\,\left(1-{\rm
e}^{-4\rho}\right)^{-1}}\right]~~.
\label{rholog}
\end{equation}
On the other hand from the ultraviolet behavior in
Eq.(\ref{gymUV}) one gets
\begin{equation}
\frac{\partial \rho}{\partial \log (\mu/\Lambda)} =
-\frac{8\pi^2}{Ng_{\rm YM}^3}\,\beta(g_{\rm YM}) \label{beta}
\end{equation}
where $\beta(g_{\rm YM})$ is the $\beta$-function.  Putting together
Eq.s (\ref{rholog}) and (\ref{beta}) we arrive at the following
$\beta$-function:
\begin{equation}
\beta(g_{\rm YM})=-\frac{3Ng_{\rm YM}^3}{16\pi^2} \left[
1-\frac{Ng_{\rm
YM}^2}{8\pi^2}+\frac{2\,\exp\Big({-\frac{16\pi^2}{N g_{\rm
YM}^2}}\Big)}{1- \exp\Big({-\frac{16\pi^2}{N g_{\rm
YM}^2}}\Big)}\right]^{-1}~~.
\end{equation}
This is precisely the complete perturbative NSVZ $\beta$-function
of the pure ${\cal{N}}=1$ SYM theory with gauge group $SU(N)$ in
the Pauli-Villars regularization \cite{NSVZ} with in addition
non-perturbative corrections due to fractional instantons.

This result was questioned in Ref.~\cite{OS} where it was shown
that, if one also includes the first non leading logarithmic
correction, one gets an extra contribution to the $\beta$-function
that modifies the one derived in Ref.~\cite{NSVZ} already at
two-loop level.  Then, in order to recover the correct two-loop
behavior, it was suggested in Ref.~\cite{OS} to add in
Eq.(\ref{holo1})  an extra  function $ f( g_{YM} )$ of the
coupling constant that can be fixed by requiring agreement with
the correct two-loop result. Of course it turns out that  $f (
g_{YM})$ must be singular at $g_{YM} \sim 0$ as the transformation
that is needed in going from the holomorphic to the wilsonian
$\beta$-function~\cite{SV}. But in this way the construction of
the NSVZ $\beta$-function  becomes not so direct and actually
rather involved.

Another problem that one encounters with the approach sketched so far
is that one gets different physical
properties if one uses  a gauge rotated vector field of gauged
supergravity in contradiction with the fact that a gauge
transformation cannot change physical properties. This can be seen
as follows. The $SU(2)$ gauge
field of 7-dimensional gauged supergravity is not vanishing but
becomes a pure gauge in the deep infrared at $\rho=0$. One can,
therefore, perform a $SU(2)$ gauge transformation that transforms it to
zero at $\rho=0$. In order to perform this gauge transformation it is
convenient to rewrite the gauge field of the Maldacena-Nu\~nez
solution as follows:
\begin{equation}
 A_{MN} = \frac{1}{2 \lambda} \left\{ (a-1) \left[ \sigma^2 \sin
{\tilde{\theta}} d {\tilde{\varphi}} - \sigma^1 d {\tilde{\theta}}
\right]
+ \left[ - \sigma^1 d {\tilde{\theta}} + \sigma^2 \sin {\tilde{\theta}}
d{\tilde{\varphi}} - \sigma^3 \cos {\tilde{\theta}} d{\tilde{\varphi}}
\right] \right\}
\label{mn87}
\end{equation}
The first term in the right hand side of the previous equation is
vanishing when $\rho =0$ because in this limit $a (\rho)$ becomes
equal to $1$, while the second term can be written as:
\begin{equation}
 A_{MN} ( \rho=0 ) = -\frac{i}{\lambda} d h h^{-1}~~,~~
h= {e}^{-i \sigma^1 \frac{{\tilde{\theta}}}{2}} {e}^{- i \sigma^3
\frac{{\tilde{\varphi}}}{2}}
\label{rho0}
\end{equation}
It is easy to see that the gauge field in Eq. (\ref{rho0}) can be
gauged to zero by performing the following gauge transformation:
\begin{equation}
 A_{MN} \rightarrow  A_{MN}' = h^{-1}  A_{MN} h +i \frac{1}{\lambda}
h^{-1} d h
\label{gautra34}
\end{equation}
where $h$ is given in Eq.(\ref{rho0}). On the other hand acting
with the  previous gauge transformation on the
entire field in Eq.(\ref{mn87}) one gets~\cite{OLESEN}:
\begin{equation}
A_{MN}^{1} {'} = \frac{1-a}{2 \lambda}\left[ d {\tilde{\theta}} \cos
{\tilde{\varphi}} - \sin {\tilde{\varphi}} \sin {\tilde{\theta}} \cos
{\tilde{\theta}} d {\tilde{\varphi}} \right]
\label{a189}
\end{equation}
\begin{equation}
 A_{MN}^{2} {'} = - \frac{1-a}{2 \lambda}\left[ d {\tilde{\theta}} \sin
{\tilde{\varphi}} + \cos {\tilde{\varphi}} \sin {\tilde{\theta}} \cos
{\tilde{\theta}} d {\tilde{\varphi}} \right]
\label{a289}
\end{equation}
\begin{equation}
 A_{MN}^{3} {'} = \frac{1-a}{2 \lambda}  \sin^2  {\tilde{\theta}}
 d {\tilde{\varphi}}
\label{a389}
\end{equation}
that is manifestly equal to $0$ at $ \rho=0$. We can now use
these gauge fields instead of the ones in Eq.(\ref{vec45}) in the
10-dimensional solution given in Eq.s (\ref{stri987}) and (\ref{c298})
and we
expect that the physical consequences are not modified. We will see
that this is not the case. In fact the term in Eq.(\ref{stri987}) for the
metric that is important
for determining the gauge coupling constant is equal to:
\begin{equation}
\sum_{i=1}^{3} (A^{i} {'})^2 = \frac{(a-1)^2}{4 \lambda^2} \left[  d
{\tilde{\theta}}^2 + \sin^2 {\tilde{\theta}} d {\tilde{\varphi}}   \right]
\label{asqu34}
\end{equation}
This means that the part of the metric relevant for computing the
gauge coupling constant is now given by:
\begin{equation}
ds_{10}^{2} = {e}^{\Phi} \left[ dx_{1,3}^{2} + \frac{1}{4 \lambda^2}
\left( 4{e}^{2h} + (a-1)^2    \right) ( d
{\tilde{\theta}}^2 + \sin^2 {\tilde{\theta}} d {\tilde{\varphi}})
\right] + \dots
\label{met453}
\end{equation}
and {from} it one obtains the following gauge coupling
constant~\cite{OLESEN}:
\begin{equation}
\frac{4 \pi^2}{N g_{YM}^{2}} = 4 {e}^{2h} + (a-1)^2 = \rho \tanh \rho
\label{gau63}
\end{equation}
that is totally different from the one obtained in
Eq.(\ref{gymfin}) although we have in the two cases used gauge
fields that differ just by a gauge transformation. It has the same
ultraviolet behavior as the expression given in Eq.(\ref{gymfin})
but a totally different infrared behavior, namely
\begin{equation}
\frac{4 \pi^2}{N g_{YM}^{2}} \sim  \rho~~.~~\rho \rightarrow
\infty~~~;~~~
\frac{4 \pi^2}{N g_{YM}^{2}} \sim   \rho^2~~,~~\rho \rightarrow 0
\label{uvifbe89}
\end{equation}
This means that now the gauge coupling constant is divergent for
$\rho \rightarrow 0$ and the
point $\rho =0$ corresponds in the gauge theory to the Landau pole.
One possible explanation of this mismatch is
that a change of gauge for the gauge fields corresponds in the gauge
theory living on the brane to a change of scheme of renormalization.
Moreover this change of scheme must be singular when $g_{YM}$
is small. But on the other hand, if we compute the vacuum angle
$\theta_{YM}$ after the gauge transformation one obtains a result that
is completely different from that found in Ref.~\cite{DLM}.
Inserting in Eq.(\ref{c298}) the solution in the new gauge given in
Eq.s (\ref{a189}), (\ref{a289}) and (\ref{a389}) we get
\[
C_2 = \frac{1}{4 \lambda^2} \left\{ \psi \sin \theta' d \theta' \wedge
  d \phi + 2 (1-a ) \left[ \sigma^1 \wedge \left(d {\tilde{\theta}} \cos
{\tilde{\varphi}} - \sin {\tilde{\varphi}} \sin {\tilde{\theta}} \cos
{\tilde{\theta}} d {\tilde{\varphi}}   \right)+
  \right. \right.
\]
\begin{equation}
\left. \left. - \sigma^2 \wedge \left( d {\tilde{\theta}} \sin
{\tilde{\varphi}} + \cos {\tilde{\varphi}} \sin {\tilde{\theta}} \cos
{\tilde{\theta}} d {\tilde{\varphi}} \right) + \sigma^3 \wedge \sin^2
    {\tilde{\theta}} d {\tilde{\varphi}} \right] \right\} + const.
\label{c254}
\end{equation}
Using this gauge transformed solution in Eq.(\ref{theta56}) one gets
that $\theta_{YM}$ is given by
\begin{equation}
\theta_{YM} = -N \psi_0
\label{theta65}
\end{equation}
instead of being proportional to $\psi$ as found in Ref.~\cite{DLM}.
Notice that in the previous equation we have taken the constant of
integration in Eq.(\ref{c254}) to be such to give Eq.(\ref{theta65}).

A natural and elegant way to solve the previous problems is
presented in Ref.s~\cite{BM,PM,MUE} and is based on the
observation that the correct cycle, i.e. the one that is
topologically nontrivial is not the cycle chosen in
Ref.~\cite{DLM}, but the one corresponding to the choice 2 at the
beginning of this section, namely the one specified by:
\begin{equation}
{\tilde{\theta}} = \pm \theta' ~~.~~{\tilde{\varphi}}= - \phi
\label{cycle2}
\end{equation}
keeping  $\rho$ and $\psi$ fixed. If we now compute the gauge
couplings on the cycle specified in the previous equation
we get~\cite{BM,PM,MUE}
\begin{equation}
\frac{4 \pi^2}{N g^{2}_{YM}} = \rho \coth 2 \rho \pm \frac{1}{2} a (
\rho) \cos \psi
\label{gaucou98}
\end{equation}
and
\begin{equation}
\theta_{YM} = \frac{1}{2 \pi g_s \alpha'}  \int C_2 =
 - N \left( \psi \pm a (\rho) \sin \psi  + \psi_0 \right)
\label{theta45}
\end{equation}
These two Eq.s must be considered together with the relation
between $\rho$ and the renormalization group scale given in Eq.
(\ref{holo1}), which in the following we are going to derive. The
derivation of eq(\ref{holo1})  is based on the fact that, as for
the fractional branes, there is a correspondence between the
symmetries of the classical supergravity solution and those of the
gauge theory living on the brane described by the classical
solution. If we look at the Maldacena-Nu\~nez solution it is easy
to see that the metric in Eq.(\ref{stri987}) is invariant under
the following transformations:
\begin{equation}
\left\{ \begin{array}{c}
                \psi \rightarrow \psi + 2 \pi~~~~~if~~ a \neq 0 \\
                \psi \rightarrow  \psi + 2 \epsilon~~~~~if~~ a =0
\end{array} \right.
\label{arr12}
\end{equation}
where $\epsilon$ is an arbitrary constant. On the other hand $C_2$ is
not invariant under the previous transformations but its flux, that is
exactly equal to $\theta_{YM}$ in Eq.(\ref{theta45}), changes
by an integer multiple of $2 \pi$. In fact one gets:
\begin{equation}
\theta_{YM} = \frac{1}{2 \pi \alpha' g_s} \int_{{\cal{C}}_2}  C_2 \rightarrow
\theta_{YM} +
\left\{ \begin{array}{c}
             - 2 \pi N~~,~~~ if~~ a \neq 0\\
             - 2 N \epsilon~~,~~if~~ a=0,\epsilon= \frac{\pi k}{N}
\end{array} \right.
\label{arr2}
\end{equation}
This changes $\theta_{YM}$ by a factor $2 \pi$ times an integer.
But since the physics is periodic in $\theta_{YM}$ under a
transformation $ \theta_{YM} \rightarrow \theta_{YM} + 2 \pi$ this
means that a change as in Eq.(\ref{arr2}) is an invariance. Notice
that also Eq.(\ref{gaucou98}) for the gauge coupling constant, is
invariant under the transformation in Eq.(\ref{arr12}). This means
that the classical solution and also the gauge couplings  are
invariant under the $Z_2$ transformation if $a \neq0$, while this
symmetry becomes $Z_{2N}$ if $ a$ is taken to be zero. This
implies that, since  in the ultraviolet $a (\rho)$ is
exponentially small, we can neglect it and we have a $Z_{2N}$
symmetry, while in the infrared where we cannot neglect $a (\rho)$
anymore, we have only a $Z_2$ symmetry left. It is on the other
hand well known that ${\cal{N}}=1$ super Yang-Mills has a non zero
gaugino condensate $<\lambda \lambda>$ that is responsible for the
breaking of $Z_{2N}$ into $Z_2$. Therefore it is natural to
identify the gaugino condensate with the function $ a(\rho)$ that
appears in the supergravity solution:
\begin{equation}
< \lambda \lambda> \sim \Lambda^3 = \mu^3 a( \rho )
\label{gauar78}
\end{equation}
This  gives the relation between the renormalization group scale $\mu$
and the supergravity space-time parameter $\rho$.

In the ultraviolet (large $\rho$) $a(\rho)$ is exponentially
suppressed and in Eq.s (\ref{gaucou98}) and (\ref{theta45})
we can neglect it obtaining:
\begin{equation}
\frac{4 \pi^2}{N g^{2}_{YM}} = \rho \coth 2 \rho ~~,~~
\theta_{YM} = - N \left( \psi  + \alpha \right)
\label{gautheta}
\end{equation}
The chiral anomaly can be obtained by performing the transformation
$\psi \rightarrow \psi + 2 \epsilon$ and getting:
\begin{equation}
\theta_{YM} \rightarrow \theta_{YM} - 2N \epsilon
\label{chi67}
\end{equation}
This implies that the $Z_{2N}$ transformations corresponding to
$\epsilon = \frac{\pi k}{N}$ are symmetries because they shift
$\theta_{YM}$ by multiples of $2 \pi$.

In general, however, Eq.s (\ref{gaucou98}) and (\ref{theta45}) are only
invariant under the $Z_2$ subgroup of $Z_{2N}$ corresponding
to the transformation:
\begin{equation}
\psi \rightarrow \psi + 2 \pi
\label{z2tra}
\end{equation}
that changes $\theta_{YM}$ in Eq.(\ref{theta45}) as follows
\begin{equation}
\theta_{YM} \rightarrow \theta_{YM} - 2 N \pi
\label{thtra}
\end{equation}
leaving invariant the gaugino condensate:
\begin{equation}
<\lambda^2 > = \frac{\mu^3}{3N g_{YM}^{2}} {e}^{- \frac{8 \pi^2}{N
    g_{YM}^{2}}}~~ {e}^{i \theta_{YM} /N}
\label{lala}
\end{equation}
Therefore the chiral anomaly and the breaking of $Z_{2N}$ to $Z_2$ are
encoded in Eq.s (\ref{gaucou98}) and (\ref{theta45}). Actually there
are $N$ vacua characterized by the value of the phase of the gaugino
condensate:
\begin{equation}
<\lambda^2 > \sim  \Lambda^3  {e}^{2i \pi k/N}~~ {e}^{i
\theta_{YM}} \label{vacua45}
\end{equation}
that are obtained by a shift of $\theta_{YM}$ by a factor $2 \pi k$ as
you can see in Eq.(\ref{lala}).

Let us turn now to the scale anomaly. It is easy to check that in the
ultraviolet (neglecting $a (\rho)$) from Eq.s (\ref{gautheta}) and
(\ref{gauar78}) one gets the NSVZ $\beta$-function. However, having
neglected $a(\rho)$ we cannot trust the contribution of the fractional
instanton. On the other hand, if we do not neglect $a(\rho)$ we get
also a dependence on $\psi$ for the gauge coupling constant and this
is not satisfactory. The
proposal formulated in Ref.~\cite{BM} has been to take the cycle that
has the minimal area. This forces $\psi$ to be equal to $(2k+1)\pi$ or
$2k \pi$ depending on the sign chosen in Eq.(\ref{cycle2}). In both
cases Eq. (\ref{gaucou98}) becomes:
\begin{equation}
\frac{4 \pi^2}{N g^{2}_{YM}} = \rho \coth 2 \rho - \frac{1}{2} a (
\rho) = \rho \tanh \rho
\label{gau84}
\end{equation}
precisely as in Eq.(\ref{gau63}). On the other hand if we insert in
Eq.(\ref{theta45}) the previous values of $\psi$ that minimize the
area of the two-cycle, we get again the result of Eq. (\ref{theta65})
apart from an irrelevant additional integer multiple of $2 \pi$.
This means that the choice of the
correct cycle depends on the gauge chosen for the gauge field of the
gauged supergravity~\cite{BM}. In the gauge where
the $SU(2)$ gauge field is vanishing at $\rho=0$, there is no mixing
between ${\tilde{\theta}}, {\tilde{\varphi}}$ and the variables
describing $S^3$ and in this case the correct cycle to be chosen is
the choice 1 at the beginning of this section, while in the other
gauge the correct cycle is the one specified in Eq.(\ref{cycle2}).

This brings us to the two following equations that determine the
running of the gauge coupling constant of ${\cal{N}}=1$ super Yang-Mills
as a function of the renormalization scale $\mu$:
\begin{equation}
\frac{4 \pi^2}{ N g_{YM}^2} = \rho \tanh \rho~~~~;~~~~
\frac{2 \rho}{\sinh 2 \rho}=
\frac{\Lambda^3}{\mu^3}
\label{fine34}
\end{equation}
It is easy to check that they imply the NSVZ $\beta$-function plus
corrections due to fractional instantons. In fact from the previous
two equations after some simple calculation one gets~\footnote{One of
  us (PdV) thanks M. Bertolini for pointing out a misprint in the
  previously published version~\cite{BERLIN} of this formula.}:
\begin{equation}
\frac{\partial g_{YM}}{\partial \log \frac{\mu}{\Lambda}} \equiv
\beta (g_{YM} ) = - \frac{3 N g_{YM}^{3}}{16 \pi^2}
\frac{1+ \frac{2 \rho}{\sinh 2 \rho}}{ 1 - \frac{N
g^{2}_{YM}}{8 \pi^2} + \frac{1}{2 \sinh^2 \rho}}
\label{beta23}
\end{equation}
This equation is exact and should be used together with the first
equation in (\ref{fine34}) in order to get the $\beta$-function as a
function of $g_{YM}$. It does not seem  possible, however, to trade $\rho$
with $g_{YM}$ in an analytic way. It can  be done in the
ultraviolet where,
from the first equation in (\ref{fine34}), it can be seen that
 $\rho$ can be approximated with
$\rho = \frac{4 \pi^2}{N g_{YM}^{2}} \coth \frac{4 \pi^2}{N
g^{2}_{YM}}$ obtaining the following $\beta$-function:
\begin{equation}
\beta (g_{YM} ) = - \frac{3 N g_{YM}^{3}}{16 \pi^2}
\frac{1+ \frac{4 \pi^2}{N g_{YM}^{2}} \sinh^{-2} \frac{4 \pi^2}{N g_{YM}^{2}}}{
1 - \frac{N
g^{2}_{YM}}{8 \pi^2} + \frac{1}{2} \sinh^{-2} \frac{4 \pi^2}{N g_{YM}^{2}} }
\label{betafin}
\end{equation}
that is equal to the NSVZ $\beta$-function plus non-perturbative
corrections due to fractional instantons.

\end{document}